\pdfoutput=1
\documentclass[a4paper]{article}
\usepackage[utf8]{inputenc}

\usepackage{hyperref}
\usepackage{amsmath,amsfonts,amssymb}
\usepackage{bbm}
\usepackage{amsthm}
\numberwithin{equation}{section}
\usepackage{color}
\usepackage{graphicx}
\usepackage{tikz}
\usepackage{tikz-cd}
\usepackage{tikz-3dplot}
\usetikzlibrary {decorations.pathmorphing,shadows,fadings}
\usepackage{pgfplots}
\usepackage{cite}
\usepackage{physics}
\usepackage{comment}
\usepackage{appendix}
\usepackage{youngtab}
\usepackage{float}
\usepackage{subcaption}
\usepackage[margin=1.8cm]{geometry}

\def\ds{\displaystyle}
\def\bea{\begin{array}{c}}
\def\ea{\end{array}}
\def\be{\begin{equation}\bea\ds}
\def\ee{\ea\end{equation}}
\def\bee{\begin{equation}\begin{array}{rcl}\ds}
\def\eee{\end{array}\end{equation}}

\def\Tr{{\rm Tr}\,}

\def\VP#1{{\color{violet} [VP: #1]}}

\title{States of 2D Yang-Mills and Large-Volume Entanglement}
\author{Dmitry Melnikov, Jefferson T. Oliveira, Valmir Peixoto and Marcia Tenser}
\date{}

\begin{document}

\maketitle

\begin{center}
\textit{\small International Institute of Physics, Federal University of 
Rio Grande do Norte, \\ Campus Universit\'ario, Lagoa Nova, Natal-RN  
59078-970, Brazil\\
~\\
Department of Theoretical and Experimental Physics, Federal University of Rio Grande do Norte,\\
Campus Universitário, Lagoa Nova, Natal-RN 59078-970, Brazil}

\vspace{2cm}

\end{center}

\vspace{-2cm}

\begin{abstract}
     We study entanglement in two-dimensional Yang-Mills theory, viewed as a quasi-topological model of emergent space. The most familiar class of states in this theory are states defined by Euclidean path integrals over Riemann surfaces. Bipartite states of this class have thermofield double structure, with entanglement consistently reducing with total area and the number of topological defects, turning separable in the infinite-area limit. In contrast, Wilson lines and loops generate rich non-monotonic behavior of the entanglement entropy. Most notably, we find that for a certain discrete set of configurations, entanglement remains finite at infinite area. The reduced density matrices, in such configurations, take the form of finite-dimensional projectors onto non-trivial vacuum sectors. We also discuss the implications of the large-volume effects for confinement and find that special asymptotic configurations are related to transitions in the confining force.
\end{abstract}

\tableofcontents

\section{Introduction}

The discovery of the geometrization of quantum entanglement~\cite{Ryu:2006bv}, in the context of a more general description of properties of quantum field theories by the holographic correspondence~\cite{tHooft:1993dmi,Susskind:1994vu,Maldacena:1997re,Gubser:1998bc,Witten:1998qj}, contributed to the development of the concept of space emerging from entanglement and quantum correlations~\cite{VanRaamsdonk:2010pw}. In this concept, the higher dimensional space attached to an ordinary nongeometric (nongravitating) system emerges as a convenient resource to encode the system's quantum correlations. A very similar holographic feature was somewhat known and used, before the advent of holography, in the axiomatic approach to topological quantum field theories (TQFT)~\cite{Atiyah:1989vu}, but received a revived and more focused interest after the appearance of its geometric analog.

TQFT axioms propose to view states in abstract Hilbert spaces as topological spaces attached to codimension-one boundaries, with boundaries themselves being invariants, representing the Hilbert space itself. The necessary properties of the quantum states can then be engineered by the choice of the topological details of the attached spaces. Similarly, gravitational systems, such as black holes, can encode quantum states living in flat asymptotic boundaries of curved spaces~\cite{Witten:1998zw,Maldacena:2001kr}. In other words, both topology and geometry can be viewed as a quantum resource.

Different aspects of the geometric presentation of quantumness were extensively studied in the holographic literature (see~\cite{VanRaamsdonk:2016exw} for a review). One popular slogan -- “ER = EPR” -- comparing the Einstein-Rosen bridge (a wormhole) with a pair of entangled particles~\cite{Maldacena:2013xja}, summarizes the broader efforts to understand black holes as ordinary quantum systems. In a more specific form, relevant for the later discussion, a double-sided black hole is understood as a thermofield double (TFD) state
\be
\label{eq:TFDstate}
|{\rm TFD}\rangle \ = \ \sum_n e^{-\beta E_n}|n\rangle |n\rangle\,.
\ee
Some other studies, focused on the general features of geometries and the properties of the multipartite states they describe include~\cite{Balasubramanian:2014hda,Bao:2015bfa,Balasubramanian:2024ysu}.

Most studies of the geometric quantification of entanglement are naturally limited by classical gravity backgrounds, which have only restricted access to possible forms of entanglement. This is particularly apparent in multipartite systems~\cite{Balasubramanian:2025hxg}. It seems that, to fully reflect the power of quantum correlations, the space also needs to be quantum. The topological encoding of entanglement is better placed in this respect, as TQFTs are usually integrable and one can easily reach the quantum regime. Some studies relevant to the present discussion can be found in \cite{Balasubramanian:2016sro,Balasubramanian:2018por,Melnikov:2018zfn,Melnikov:2022qyt,Melnikov:2025tui}. However, even in this case, it is not precisely known how topology can access arbitrary forms of entanglement.

In this work we take further steps in the development of the emerging space paradigm by looking at an intermediate example provided by the states of two-dimensional Yang-Mills theory (2D YM). This theory is quasi-topological since it has a very mild dependence on the geometry of the emerging spaces, via areas of their basic domains. Yet it remains integrable and, in principle, allows for the computation of arbitrary correlation functions~\cite{Migdal:1975zg,Witten:1991we}. 

States of 2D YM can be constructed as Euclidean path integrals over Riemann surfaces with circular boundaries. This approach defines states of the pure gauge theory on a collection of disjoint circles. On top of these, there is the possibility of adding particles, represented by endpoints of Wilson lines that cut the Riemann surfaces into domains. One may also consider adding closed Wilson lines (loops) as an additional feature to be added to the “emerging spaces”. In this paper, we will consider such configurations, beyond the well-studied cases of the pure YM states. Our primary goal is to develop general intuition about entanglement properties of the states encoded in each setup. 

We will start by reviewing the bipartite states defined on Riemann surfaces of different genus~\cite{Witten:1991we,Cordes:1994fc}. Entanglement properties of such states have already been discussed in the literature~\cite{Gromov:2014kia,Donnelly:2014gva,Donnelly:2019zde}. Our expectation, stemming from the study of TQFT states, is that topological defects should, in general, decrease the amount of entanglement because the defects consistently decrease the space's connectedness. Calculating the von Neumann entropy, we will see that this is indeed the case for this class of states. Moreover, we will recover the known exponential decrease of entanglement with the total area of the Riemann surface. In the bipartite case (cylinder) the state has the structure of the TFD state~(\ref{eq:TFDstate}), with the length of the cylinder playing the role of the inverse temperature. Consequently, the states are separable in the zero-temperature limit.  

Next, we will introduce Wilson loops and Wilson lines in the Riemann surface background. To our knowledge, such states have been much less explored (recent works where such states appear include~\cite{Blommaert:2018oro}). We analyze a wide range of examples, including bipartite states and states with a single boundary. In the former case, we consider Wilson lines connecting the same or opposite boundaries, as well as loops with trivial and nontrivial homotopy, and intersecting loops. The main observations of our studies can be summarized as follows:
\begin{itemize}

    \item Defects in the form of contractible nonnested Wilson loops, in general, affect the bipartite entanglement in a nonmonotonous fashion. However, there is a special regime in which such defects consistently decrease entanglement;

    \item In general, contractible loops have optimal sizes for the purpose of the entanglement. That is, the latter may have a (local) maximum for certain ratios of the areas inside and outside the loops;
    
    \item Though intuitively one expects entanglement to vanish in the large area limit, there are bipartite states for which entanglement asymptotes to a positive value. These are precisely the infinite volume\footnote{Since the theory is two-dimensional, ``volume'' and ``area'' refer to the same geometric characteristic.} limits of the optimal ratio values;

    \item Reduced density matrices of the bipartite states with contractible Wilson loops are of TFD type. In the large area limit, they take the form of projectors on finite-dimensional subspaces of the Hilbert space. More precisely, they are equivalent to finite-dimensional thermal density matrices;

    \item In the large area limit, the proliferation of contractible loops consistently decreases entanglement entropy;
    
    \item Bipartite states with noncontractible loops, in general, are not of the TFD type. However, they asymptote to TFD states in the infinite volume limit. Local maxima and finite entropy values for infinite volumes also occur for certain ratios of the areas of the domains separated by the loops. Reduced density matrices of such states also take the form of finite-dimensional thermal density matrices, consistently decreasing the entropy;

    \item Similar behavior is observed for bipartite states with Wilson lines connecting the same or the opposite boundaries. For certain ratios of the areas of the domains cut by the Wilson lines, the entanglement entropy has nonzero limit for infinite areas. Although the density matrices have a more complex structure in the presence of Wilson line endpoints, they take a very similar asymptotic form in the large volume limit;
    
    \item Reduced density matrices of the Wilson lines themselves, that is, density matrices with traced out gauge degrees of freedom, reduce to simple separable maximally mixed density matrices of the endpoints. This form of the density matrices is gauge invariant, despite the noninvariance of the original states and the Hilbert space.

\end{itemize}

We believe that the existence of finite results for the entanglement entropy in the infinite volume limit is the most interesting outcome of our studies. It seems to indicate the existence of a special sector of the theory in this limit, which is equivalent to zero temperature or strong coupling. 

As a slightly complementary line, we also review the properties of some states from the point of view of confinement. We note that, despite the fact that the endpoints of Wilson lines must be interpreted as being in a separable mixed state, this doesn't affect their ability to exhibit confining behavior. Since confinement in 2D YM has been studied extensively, we focus on a few complementary points: large volume effects on the confining force and on the hadronic force between meson-like configurations of Wilson lines. 

The existence of optimal area ratios with finite, locally maximal entropy values results in a series of crossover transitions in the confining force. In the infinite volume limit, these should become sharp phase transitions. Note that this limit is also the zero-temperature limit of the theory. Meanwhile, for the analog of the hadron force between meson-like configurations, no large-area effects occur.

The paper is organized as follows. In Section~\ref{sec:quantizationYM2D} we briefly review the basics of the calculation of Euclidean partition functions in 2D Yang-Mills. We consider a few standard examples of partition functions in the pure gauge theory and review the entanglement properties of two examples of bipartite states: on Riemann surfaces with two $S^1$ boundaries and on a disk with a partition of a single boundary.

In Section~\ref{sec:WilsonLoops} we introduce Wilson loops. We start in Section~\ref{subsec:1CWL} with the case of a single contractible Wilson loop on a cylinder and make a detailed investigation of the behavior of the entanglement entropy in different regimes. Our method, in this and the following sections, is based on the numerical observation of the entropy, followed by analytical arguments that explain special features of the numerical results. In most cases, we do not provide full analytical proofs of the numerical results. We believe that such proofs are straightforward, at least in the $SU(2)$ case. For general $SU(N)$, proofs may be complicated due to the lack of general formulas for group-theoretic quantities and, especially, the multiplicities in the tensor products of representations. Besides the single contractible loop, Section~\ref{sec:WilsonLoops} also includes subsections with configurations of multiple loops (Section~\ref{subsec:manyloops}), intersecting loops (Section~\ref{sec:intersectingloops}), and noncontractible loops (Section~\ref{subsec:1NCWL}).

In Section~\ref{sec:WilsonLines} we connect the boundaries of a cylinder with Wilson lines and study the entanglement present in either of the boundaries. We investigate the cases of lines ending at the same (Section~\ref{sec:UUlines}) or at opposite boundaries~(Section~\ref{sec:IIlines}). In the last case, an example of entropy enhancement due to a nontrivial multiplicity effect is demonstrated.

Section~\ref{sec:MixedParticles} also studies Wilson lines on a cylinder and on a disk, but focuses on the entanglement of the Wilson lines with the gauge degrees of freedom. In Section~\ref{sec:disklines} we consider a single line connecting two endpoints on a single boundary. On a cylinder, we mark a pair of points on each boundary and consider three most natural ways of connecting them with lines: $\cup\cap$-configuration (Section~\ref{sec:UUWilsonLines}), ${\rm II}$-configuration (Section~\ref{sec:IIWilsonLines}) and ${\rm X}$-configuration (Section~\ref{sec:XWilsonLines}). We show that in either case the reduced density matrix has the structure of a separable maximally mixed state.

Section~\ref{sec:confinement} reviews the confining behavior of particles (quarks) represented by the endpoints of Wilson lines. We compute the free energy associated with several examples of partition functions and check its behavior at different quark separations. In Section~\ref{sec:ConfinementLargeVolume} we discuss the effects of large volume and the presence of the optimal area fractions on the confining force. In Section~\ref{sec:hadronforce} we discuss the analog of the hadronic force between meson-like configurations of quarks.

In Section~\ref{sec:conclusions} we draw our conclusions and discuss possible follow-up questions. Appendix~\ref{sec:usefulformulae} contains an additional list of group-theoretic formulae useful for calculations.

\section{States in pure 2D Yang-Mills theory}
\label{sec:quantizationYM2D}


We will consider Yang-Mills theory defined on a 2D Riemannian manifold $\mathcal{M}$, with a compact simple Lie group $G$ and its Lie algebra $\mathfrak{g}$. The theory is said to be quasi-topological and its partition function $Z$ depends on the metric solely via the surface area $\varrho$ of the manifold $\mathcal{M}$. In particular, on a compact orientable manifold of genus $g$, the partition function has the form of the following sum over irreducible representations $\alpha$ of $G$:
\begin{equation}\label{eq:partitionfunctiongenusg}
    Z(\varrho)=\sum_\alpha (d_\alpha)^{2-2g} e^{-C_2(\alpha)\varrho/2}\,.
\end{equation}
Here $d_\alpha$ is the dimension of $\alpha$ -- its exponent is the Euler characteristic of $\mathcal{M}$ -- and $C_2(\alpha)$ is the quadratic Casimir of $\alpha$.\footnote{In this work we will adopt the convention for the quadratic Casimir given by equation~(\ref{eq:quadratic_casimir_formula_su_n}).} 

In practice, for manifolds of $g>0$, one computes the partition function \eqref{eq:partitionfunctiongenusg} using a cell decomposition technique~\cite{Witten:1991we,Cordes:1994fc}. By decomposing a genus-$g$ Riemannian manifold in 2D in terms of $p$ plaquettes (these will be polygons of 4$g$ edges), one arrives at a set $L_p$ of $l$ links or edges composing each plaquette. Around them, the holonomy is $U_p=\prod_{l\in L_p} U_l$. One may then compute $Z$ integrating over the elements $dU_l$ -- the Haar measure on the gauge group $G$ -- as follows:
\begin{equation}
\label{eq:celldecomposition}
    Z(\varrho) = \int \prod_l dU_l \prod_p \Gamma(U_p,\varrho_p)\,, \qquad \Gamma(U_p,\varrho_p)\equiv \sum_\alpha d_\alpha\, \chi_\alpha(U_p)\,e^{-C_2(\alpha)\varrho/2} \,.
\end{equation}
The quantities $\chi_\alpha(U_p)$ are the characters of the group element $U_p$ in the irreducible representations $\alpha$ of $G$. According to the Peter-Weyl theorem, characters form an orthonormal basis for the conjugacy class functions. In particular, they satisfy
\begin{eqnarray}
    \int dU \,\chi_\alpha(VU)\chi_\beta(U^{-1}W)&=\delta_{\alpha\beta}\,(d_\alpha)^{-1} \, \chi_\alpha(VW)\,, \label{eq:teo_char_eq_1}\\
    \int dU\,\chi_\alpha(VUWU^{-1})&= (d_\alpha)^{-1} \chi_\alpha(V)\chi_\alpha(W)\,. \label{eq:teo_char_eq_2}
\end{eqnarray}
Moreover, for unitary representations, it follows that $\chi_\alpha(U^{-1}) = \overline{\chi_\alpha(U)}$. In the following sections we will make extensive use of these properties. 

When the manifold $\mathcal{M}$ has $n_b$ boundaries, the prescription is to consider holonomies $W_i$ in representation $\alpha_i$ along them. The partition function will depend on these boundary data and, in particular,
\begin{equation}
    Z(\varrho;W_i)= \int \prod_l dU_l  \prod_p \Gamma(W_i,U_p,\varrho_p)\,.
\end{equation}
Note that there will be plaquettes containing holonomies at the boundary, but these are not integrated over. Equivalently, provided that $\Gamma(W_i)$ is a class function, one may write the partition function in the representation basis. This is implemented by projection on the character basis $\chi_{\alpha_i}(W_j^{-1})$, via integration over $dW_j$:
\begin{equation}
\label{eq:partitionfunctionwithboundaries}
    Z(\varrho;\alpha_i)=\int \prod_j^{n_b} dW_j\, \chi_{\alpha_i}(W_j^{-1}) \int \prod_l dU_l  \prod_p \Gamma(W_k,U_p,\varrho_p)\,.
\end{equation}

One may also glue manifolds $M_i$ along their boundaries, such that the resulting manifold $\mathcal{M}$ has fewer boundaries: $\mathcal{M}=\cup \mathcal{M}_i$. Considering that the boundaries of $\mathcal{M}_i$ contain holonomies $W_{i_j}$ in representation $\alpha_{i_j}$, the gluing is done by {\it i)} taking the product of the partition functions, {\it ii)} identifying the representations living on coincident boundaries and {\it iii)} summing over all boundary representations:
\begin{equation}
    Z_{\mathcal{M}}(\varrho)=\prod_i\sum_{\alpha_{i_j}}Z_{\mathcal{M}_i}(\varrho_i,\{\alpha_{i_j}\})\,.
\end{equation}

Suppose we consider a 2D manifold with one boundary and we associate a holonomy $U$ to it. We refer to such a manifold as $\mathcal{M}_U$. A wavefunction $\psi$ may then be prepared by path integrating over $\mathcal{M}_U$ with $U$ as a boundary condition:
\begin{equation}
    \psi(U)=\int \prod_p dV_p \ \Gamma(U,V_p,\varrho_p)\,,
\end{equation}
where the integration is performed over all holonomies except the one on the boundary where the wavefunction is evaluated. By definition, $\psi(U)$ will be a class function.

Naturally, the Hilbert space may be equipped with a positive inner product by means of the Haar measure on $G$,
\begin{equation}
    \langle \psi_1|\psi_2\rangle = \int_G dU\, \overline{\psi_1(U)}\psi_2(U)\,.
\end{equation}
So the Hilbert space $\mathcal{H}$ consists of square-integrable class functions with respect to the Haar measure.

We will apply the cutting and gluing techniques described for the computation of partition functions for construction of wavefunctions and density matrices. The corresponding amplitudes will also often be referred to as partition functions, with some abuse of terminology.

In the rest of this section we will apply these techniques to discuss the entanglement properties of typical states of pure 2D Yang-Mills theory, focusing on the bipartite entanglement.

\subsection{The disk}
\label{sec:discpartition}

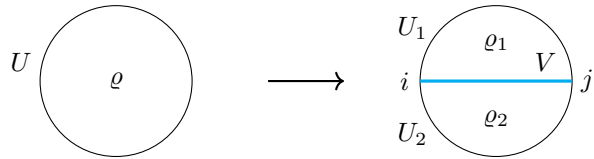
\begin{figure}[h!]
\centering
     \begin{tikzpicture}
        \fill[white] (0,0) circle (1);
         \draw (0,0) circle (1);
         \draw (-1.25,0.25) node {$U$};
         \draw (0,0) node {$\varrho$};

         \draw[->,thick] (2,0) -- (3,0);
     
         \fill[white] (5,0) circle (1);
         \draw (5,0) circle (1);
         \draw[line width=1.2,cyan] (4,0)  -- (6,0) ;
         \draw (5,0.5) node {$\varrho_1$};
         \draw (5,-0.5) node {$\varrho_2$};
         \draw (5.65,0.25) node {$V$};
         \draw (3.9,0.7) node {$U_1$};
         \draw (3.9,-0.7) node {$U_2$};
         \draw (3.8,0) node {$i$};
        \draw (6.2,0) node {$j$};

     \end{tikzpicture} 
\caption{Cell decomposition of the disk. The partition can also be thought as the insertion of a Wilson line, with the corresponding group representation indices shown. The empty disk partition function is recovered when the representation of the Wilson line is considered to be trivial.}
    \label{fig:diskoneline}
\end{figure}

The entanglement entropy associated with a single interval on the boundary of a disk may be computed using the replica trick as follows. The disk's partition function is given by the contribution of a single plaquette with holonomy $U$ at the boundary,
\begin{equation}\label{eq:Zdisk}
    Z(\varrho,U)=\sum_\alpha d_\alpha \,\chi_\alpha(U)\,e^{-C_2(\alpha)\varrho/2}\,.
\end{equation}
So we start by partitioning the system in terms of two plaquettes, of area $\varrho_1$ and $\varrho_2$. Their sum equals the total area of the disk $\varrho$. The decomposition is depicted in Figure~\ref{fig:diskoneline}. We may write
\begin{equation}\label{eq:Zdiskpart}
        Z(\varrho_1,\varrho_2;U_1,U_2) = \int dV\left(\sum_\alpha d_\alpha \,\chi_\alpha(U_1 V)\,e^{-C_2(\alpha)\varrho_1/2}\right)\left(\sum_\beta d_\beta\, \chi_\beta( V^{-1}U_2)\,e^{-C_2(\beta)\varrho_2/2}\right)\,.
\end{equation}
Naturally, if one performs the integration over $V$, one recovers \eqref{eq:Zdisk}, given that $U=U_1 U_2$ and $\varrho=\varrho_1+\varrho_2$.

The partition function \eqref{eq:Zdiskpart} may be seen as a state with two labels, $U_1$ and $U_2$, $\psi(U_1,U_2)$,
to which one can associate a density matrix. A (unnormalized) reduced density matrix $\rho$ may then be obtained via
\begin{align}
    \begin{split}
        \rho(U_1,U_3^{-1}) &= \int dU_2 \, Z(U_1,U_2)\overline{Z(U_2,U_3)}=\sum_\alpha d_\alpha \,\chi_\alpha(U_1 U_3^{-1}) e^{-C_2(\alpha)\varrho}
    \end{split}
\end{align}
The normalization factor is simply
\begin{equation}
    Z_1 \ = \ {\rm Tr}\,\rho(U_1,U_3^{-1}) \ = \ \int dU \,\rho(U,U^{-1})=\sum_\alpha (d_\alpha)^2 e^{-C_2(\alpha)\varrho/2}\,,
\end{equation}
which is nothing but the partition function of a sphere $S^2$, cf. equation~(\ref{eq:partitionfunctiongenusg}).

To use the replica trick, we need to compute $Z_n={\rm Tr}\left(\rho^n(U_1,U_3^{-1})\right)$, and it is not difficult to show that
\begin{equation}
    Z_n =\sum_\alpha (d_\alpha)^2 e^{-nC_2(\alpha)\varrho}\,.
\end{equation}
The entanglement entropy is then given by
\begin{equation}
    S=-\lim_{n\to 1}\frac{d}{dn}\frac{Z_n}{Z_1^n}\,.
\end{equation}
Introducing the notation
\begin{equation}
    p(\alpha)\equiv \frac{(d_\alpha)^2 \exp\left(-C_2(\alpha)\varrho\right)}{Z_1}\,,
\end{equation}
we may write the resulting entanglement entropy as
\begin{equation}
    S=-\sum_\alpha p(\alpha)\log p(\alpha) + 2\sum_\alpha p(\alpha)\log d_\alpha\,.
\end{equation}
This basic example reproduces the results of \cite{Donnelly:2014gva} via the replica trick applied to the cell decomposition approach. In particular, the second term appearing on the right-hand side originates from additional degrees of freedom living on the entangling surface, the so-called edge modes. They appear due to the impossibility of gauge invariant factorization of the Hilbert space of the gauge theory. The replica trick derives their contribution in a gauge invariant way.

As explained in \cite{Donnelly:2014gva}, if one requires gauge invariance of the reduced density matrix $\rho(U_1,U_3^{-1})$, it must commute with gauge transformations acting at both boundaries. It then follows from Schur’s lemma that, for each irreducible representation $\alpha$, the boundary degrees of freedom must form a maximally mixed state of dimension $d_\alpha$. This, in turn, leads to a contribution $2\log d_\alpha$ to the entropy, where the factor of two accounts for the two boundaries. In Section~\ref{sec:MixedParticles} we will see how gauge invariant reduced density matrices explicitly appear after integration over the gauge modes.


\subsection{The cylinder}

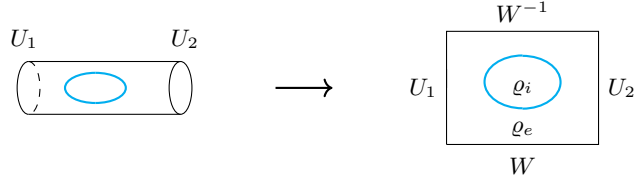
\begin{figure}[h!]
\centering
\begin{tikzpicture}[scale=1, every node/.style={font=\small}]

\begin{scope}[xshift=-0.5cm]
  \draw[dashed] (-1,1.35) arc[start angle=90, end angle=-90,x radius=0.15cm, y radius=0.35cm];
  \draw (-1,0.65) arc[start angle=270, end angle=90,x radius=0.15cm, y radius=0.35cm];
  \draw (1,1) ellipse (0.15 and 0.35);
  \draw (-1,1.35) -- (1,1.35);
  \draw (-1,0.65) -- (1,0.65);
  
  \draw[thick, color=cyan] (-0.12,0.8) arc [x radius=0.4, y radius=0.2, start angle=270, end angle=450];
  \draw[thick, color=cyan] (-0.12,1.2) arc [x radius=0.4, y radius=0.2, start angle=90, end angle=270];

  \node[left] at (-0.75,1.65) {$U_1$};
  \node[right] at (0.75,1.65) {$U_2$};
  
\end{scope}

\draw[->, thick] (1.75, 1) -- (2.5, 1);

\begin{scope}[xshift=4cm]
  \draw[thick, color=cyan] (1,0.725) arc [x radius=0.5, y radius=0.35, start angle=270, end angle=450];
  \draw[thick, color=cyan] (1,1.43) arc [x radius=0.5, y radius=0.35, start angle=90, end angle=270];
  \draw (0,1) -- ++(0,0.75) -- ++(2,0) -- ++(0,-1.5) -- ++(-2,0);
  \draw (0,0.25) -- (0,1.25) ;

  \node[left] at (0,1) {$U_{1}$};
  \node[above] at (1,1.75) {$W^{-1}$};
  \node[below] at (1,0.2) {$W$};
  \node[right] at (2,1) {$U_{2}$};
  \node at (1,1) {$\varrho_{i}$};
  \node at (1,0.45) {$\varrho_{e}$};
\end{scope}

\end{tikzpicture}
\caption{Cell decomposition of the cylinder. For later use we show the setup with one contractible Wilson loop.}
\label{fig:cylinder1contractWL}
\end{figure}

In the case of a cylinder, we have the set-up shown in Figure~\ref{fig:cylinder1contractWL}. Following the prescription outlined in \eqref{eq:partitionfunctionwithboundaries}, we compute the partition function by {\it i)} identifying the edges by gluing along $W$ and {\it ii)} introducing holonomies $U_1$ and $U_2$ at the boundaries. The resulting partition function will depend on the area and on the representation of the holonomies, say $\alpha_1$ and $\alpha_2$. In the representation basis we have
\begin{align}\label{eq:CylinderRepBasis}
\begin{split}
    Z(\varrho;\alpha_1,\alpha_2)&=\int dU_1 dU_2 dW \,\sum_\beta d_\beta\, \chi_\beta(U_1 W^{-1} U_2 W)e^{-C_2(\beta)\varrho/2} \overline{\chi_{\alpha_1}(U_1)} \,\overline{\chi_{\alpha_2}(U_2)}\\
    &= \int dU_1 dU_2 \,\sum_\beta \chi_\beta(U_1)\chi_\beta(U_2)e^{-C_2(\beta)\varrho/2}\chi_{\alpha_1}(U_1^{-1}) \,\chi_{\alpha_2}(U_2^{-1})\\
    &= \delta_{\alpha_1\alpha_2} \exp\left(-\frac{\varrho}{2}C_2(\alpha_1)\right).
\end{split}
\end{align}
Equivalently, in the holonomy basis,
\be\label{eq:Zcylinder}
Z(\varrho;U_1,U_2)=\sum_\alpha \chi_\alpha(U_1)\chi_\alpha(U_2)\,e^{-C_2(\alpha)\varrho/2}\,.
\ee
Note that one can obtain this result immediately by using a refined cell decomposition that sums over cells with given Euler characteristic, zero in this case, which amounts to omitting the dimension $d_\alpha$ factor from~(\ref{eq:celldecomposition}) and inserting a character for each boundary. 

The two-boundary partition function may be seen as coefficients of a state decomposed in the character basis as follows
\begin{equation}
    \label{eq:StateInRepBasis}
    \vert\psi\rangle=\sum_{\alpha_1,\alpha_2} \psi_{\alpha_1\alpha_2}|\chi_{\alpha_1}\chi_{\alpha_2}\rangle\,.
\end{equation}
So~\eqref{eq:CylinderRepBasis} makes it evident that the state is a TFD state (for now unnormalized):
\begin{equation}\label{eq:psicylcharacterbasis}
    |\psi\rangle=\sum_\alpha e^{-C_2(\alpha)\varrho/2}|\chi_\alpha\chi_\alpha\rangle\,.
\end{equation}

We may then consider the pure state with density matrix given by
\begin{equation}
    \rho = |\psi\rangle\langle\psi| = \sum_{\alpha,\beta} e^{-\left(C_2(\alpha)+C_2(\beta)\right)\varrho/2} |\chi_\alpha\chi_\alpha\rangle\langle\chi_\beta\chi_\beta|\,.
\end{equation}
The reduced density matrix obtained by tracing out the $U_1$ boundary, for instance, is
\begin{equation}
    \rho_2=\frac{1}{\mathcal{N}}\sum_\alpha e^{-C_2(\alpha)\varrho} |\chi_\alpha\rangle\langle\chi_\alpha|\,, \qquad \mathcal{N}\equiv\sum_\alpha e^{-C_2(\alpha)\varrho}\,.
\end{equation}

It is instructive to visualize how the von Neumann entropy is obtained from the replica trick,
\begin{equation}
    S=-\lim_{n\to 1}\frac{d}{dn} {\rm Tr}(\rho_2^n)=\left( 1-\varrho\frac{d}{d\varrho}\right)\log \mathcal{N}\,,
\end{equation}
where $\rho_2^n$ is the same partition function as $\rho$, but computed on a longer cylinder. Otherwise, the entropy is directly computed for the diagonal matrix via
\begin{equation}\label{eq:cylinder_no_wilson_entropy_log}
    \boxed{\quad S = -\sum_\alpha h_\alpha\log h_\alpha\,, \quad\quad h_\alpha = \dfrac{e^{-C_2(\alpha)\varrho}}{\mathcal{N}}\,.\quad }
\end{equation}

When $G=SU(2)$, for example, $C_2(j)=j(j/2+1)$ with $j=0,1,2,3,\dots$ and one has
\begin{equation}
    \mathcal{N}=\frac{e^{\varrho/2}}{2}\left( \vartheta_3(0,e^{-\varrho/2})-1\right)\,, \qquad \vartheta_3(z,q)=\sum_{n=-\infty}^{\infty} q^{n^2}\exp(2n iz)\,,
\end{equation}
where $\vartheta_3(z,q)$ are Jacobi theta functions on the torus.

\subsection{Higher genera}
\label{sec:highergenus}

Building on the intuition gained from considering the cylindrical case, we go a step further and introduce holes in such a topology. This will lead to two-boundary states defined on Riemann surfaces of genus $g$.

\begin{figure}[h!]
    \centering
\begin{subfigure}{0.6\textwidth}
\centering
    \begin{tikzpicture}
\begin{scope}
  \draw[dashed] (-4,0.5) arc[start angle=90, end angle=-90,x radius=0.25cm, y radius=0.5cm];
  \draw (-4,-0.5) arc[start angle=270, end angle=90,x radius=0.25cm, y radius=0.5cm];
  \draw (-1,1)   ellipse (0.25 and 0.5);   
  \draw (-1,-1)  ellipse (0.25 and 0.5);  

  \draw (-4,0.5)  .. controls (-3,0.6)  and (-2,1.2)  .. (-1,1.5);
  \draw (-4,-0.5) .. controls (-3,-0.6) and (-2,-1.2) .. (-1,-1.5);

  \draw (-1,0.5)  .. controls (-1.5,0.3)  and (-2,-0.3)  .. (-1,-0.5);

  \draw (-4.25,0) .. controls (-3,0.2)  and (-2,0.9)  .. (-1.25,1);
  \draw (-4.25,0) .. controls (-3,-0.2) and (-2,-0.9) .. (-1.25,-1);

  \node[left]  at (-4.4,0)     {$V$};
  \node[right] at (-0.7,1.1)   {$U$};
  \node[right] at (-0.7,-1.1)  {$W$};
  \node        at (-2.3,0.35)   {$X$};
  \node        at (-2.3,-0.35)  {$Y$};
\end{scope}
\hspace{0.3cm}
\begin{scope}[xscale=-1]
  \draw (-4,0)   ellipse (0.25 and 0.5);   
  \draw (-1,0.5) arc[start angle=-90, end angle=90,x radius=0.25cm, y radius=0.5cm];
  \draw[dashed] (-1,0.5) arc[start angle=270, end angle=90,x radius=0.25cm, y radius=0.5cm]; 
  \draw (-1,-1.5) arc[start angle=-90, end angle=90,x radius=0.25cm, y radius=0.5cm];
  \draw[dashed] (-1,-1.5) arc[start angle=270, end angle=90,x radius=0.25cm, y radius=0.5cm];   

  \draw (-4,0.5)  .. controls (-3,0.6)  and (-2,1.2)  .. (-1,1.5);
  \draw (-4,-0.5) .. controls (-3,-0.6) and (-2,-1.2) .. (-1,-1.5);

  \draw (-1,0.5)  .. controls (-1.5,0.3)  and (-2,-0.3)  .. (-1,-0.5);

  \node[right] at (-4.4,0)     {$T$};
  \node[left]  at (-0.7,1.1)   {$U^{-1}$};
  \node[left]  at (-0.7,-1.1)  {$W^{-1}$};
\end{scope}
\end{tikzpicture}
\subcaption{}
\label{subfig:genus1cell}
\end{subfigure}
\hfill
\begin{subfigure}{0.25\textwidth}
\centering
    \begin{tikzpicture}
\coordinate (A) at (-0.5, 0.3);
\coordinate (B) at ( 0.5, 0.3);
\coordinate (D) at ( 1,-0.35);
\coordinate (E) at ( 0.5,-1.0);
\coordinate (F) at (-0.5,-1.0);
\coordinate (G) at (-1,-0.7);
\coordinate (H) at (-1, 0.0);

\draw (A)--(B)--(D)--(E)--(F)--(G)--(H)--cycle;

\path (A) -- (B) node[midway,above] {$U$};
\path (B) -- (D) node[midway,above right] {$X^{-1}$};
\path (D) -- (E) node[midway,below right] {$Y$};
\path (E) -- (F) node[midway,below] {$W$};
\path (F) -- (G) node[midway,below left] {$Y^{-1}$};
\path (G) -- (H) node[midway,left] {$V$};
\path (H) -- (A) node[midway,above left] {$X$};

\node[left]  at (0.2,-0.4)   {$\varrho$};

\end{tikzpicture}
\subcaption{}
\label{subfig:plaquettepairofpants}
\end{subfigure}
\caption{A genus-1 cell may be built by gluing two pairs of pants as depicted in (\subref{subfig:genus1cell}). Each pair of pants may be cut open and represented as a heptagon. In (\subref{subfig:plaquettepairofpants}) we depict the plaquette obtained by cutting the left pair of pants along $X$ and $Y$.}
    \label{fig:2pairsofpants}
\end{figure}
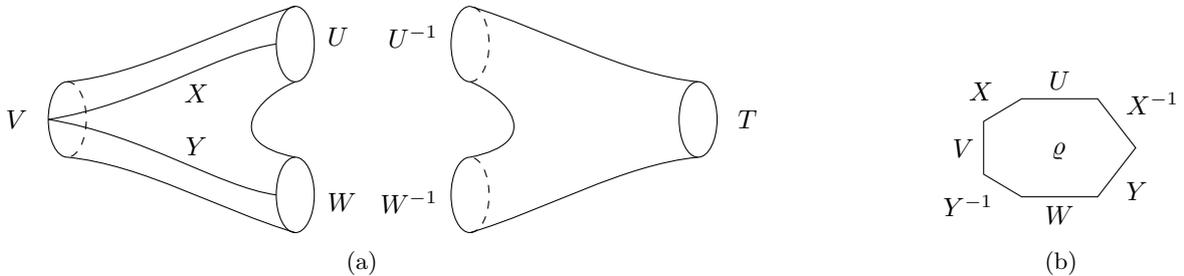

A hole may be introduced by gluing together two pairs of pants, as depicted in Figure \ref{subfig:genus1cell}. Each pair of pants, on the other hand, may be cut open and represented in terms of a plaquette with identified edges. In Figure \ref{subfig:plaquettepairofpants} we choose to cut the left pair of pants along the curves $X$ and $Y$.

Each pair of pants may be seen as a three-boundary state. Focusing on the pair of pants on the left, we may read off its representation in the holonomy basis directly from Figure \ref{subfig:plaquettepairofpants}. We have
\begin{align}
    \begin{split}
        \psi_l(V,U,W)&=\int dX dY \sum_\alpha d_\alpha\, \chi_\alpha(VX UX^{-1}Y WY^{-1})\,e^{-C_2(\alpha)\varrho/2}\\
        &=\int dX \sum_\alpha d_\alpha \, \frac{\chi_\alpha(VXUX^{-1})\chi_\alpha(W)}{d_\alpha}\,e^{-C_2(\alpha)\varrho/2}\\
        &=\sum_\alpha \frac{1}{d_\alpha}\, \chi_\alpha(V)\chi_\alpha(U)\chi_\alpha(W)\,e^{-C_2(\alpha)\varrho/2}\,.
    \end{split}
\end{align}
Analogously, the state associated with the pair of pants on the right is
\begin{align}
    \begin{split}
        \psi_r(T,U^{-1},W^{-1})&=\sum_\alpha \frac{1}{d_\alpha}\,\chi_\alpha(T)\chi_\alpha(U^{-1})\chi_\alpha(W^{-1})\,e^{-C_2(\alpha)\varrho/2}\,.
    \end{split}
\end{align}
The genus-one cell of Figure \ref{subfig:genus1cell} is the two-boundary state obtained by gluing $\psi_l$ and $\psi_r$ above along $U$ and $W$:
\begin{align}\label{eq:holecell}
    \begin{split}
        \psi_1(V,T)&\equiv\int dU dW \psi_l(V,U,W)\psi_r(T,U^{-1},W^{-1})\\
        &=\sum_\alpha \frac{1}{(d_\alpha)^2}\chi_\alpha(V)\chi_\alpha(T)\,e^{-C_2(\alpha)\varrho}\,.
    \end{split}
\end{align}

Successively gluing copies of states of the form \eqref{eq:holecell}, we build two boundary states of genus $g$. For instance, gluing two of them gives\footnote{From now on, and until the end of this section, instead of keeping track of how many factors of the area of a single pair of pants we have, we choose to write the area dependence in terms of the total area of the state under consideration. We keep the notation $\varrho$ nevertheless.}
\begin{equation}
    \psi_2(V,R)=\int dT\, \psi_1(V,T)\,\psi_1(T^{-1},R)= \sum_\alpha \frac{1}{(d_\alpha)^4}\,\chi_\alpha(V)\chi_\alpha(R)\,e^{-C_2(\alpha)\varrho/2}\,.
\end{equation}
It is easy to see that, in general, a bipartite genus-$g$ state will be described by\footnote{This agrees with the well-known results cited in~\cite{Witten:1991we,Cordes:1994fc}.}
\begin{equation}
    \psi_g(V,S)= \sum_\alpha \frac{1}{(d_\alpha)^{2g}}\,\chi_\alpha(V)\chi_\alpha(R)\,e^{-C_2(\alpha)\varrho/2}\,.
\end{equation}
Equivalently, in the character basis, we have
\begin{equation}
    \vert\psi_g\rangle= \sum_\alpha \frac{e^{-C_2(\alpha)\varrho/2}}{(d_\alpha)^{2g}} \vert \chi_\alpha \chi_\alpha\rangle\,,
\end{equation}
which is the genus $g$ version of \eqref{eq:psicylcharacterbasis}. We see that such states of pure Yang-Mills preserve the TFD structure, as seen in the cylinder case, for higher genera as well.

The von Neumann entropy is given by a generalization of \eqref{eq:cylinder_no_wilson_entropy_log}. In particular, we have
\begin{equation}\label{eq:entropy_cylinder_arbitrary_genus}
    \boxed{\quad S=-\sum_\alpha h_\alpha\log h_\alpha\,,\quad h_\alpha = \frac{e^{-C_2(\alpha)\varrho } (d_\alpha)^{-4g}}{\mathcal{N}}\,, \qquad\mathcal{N}=\sum_\alpha e^{-C_2(\alpha)\varrho } (d_\alpha)^{-4g}\,.\quad }
\end{equation}

\subsubsection{Entanglement entropy analysis}

\begin{figure}[h!]
    \centering

    \includegraphics[width=7cm]{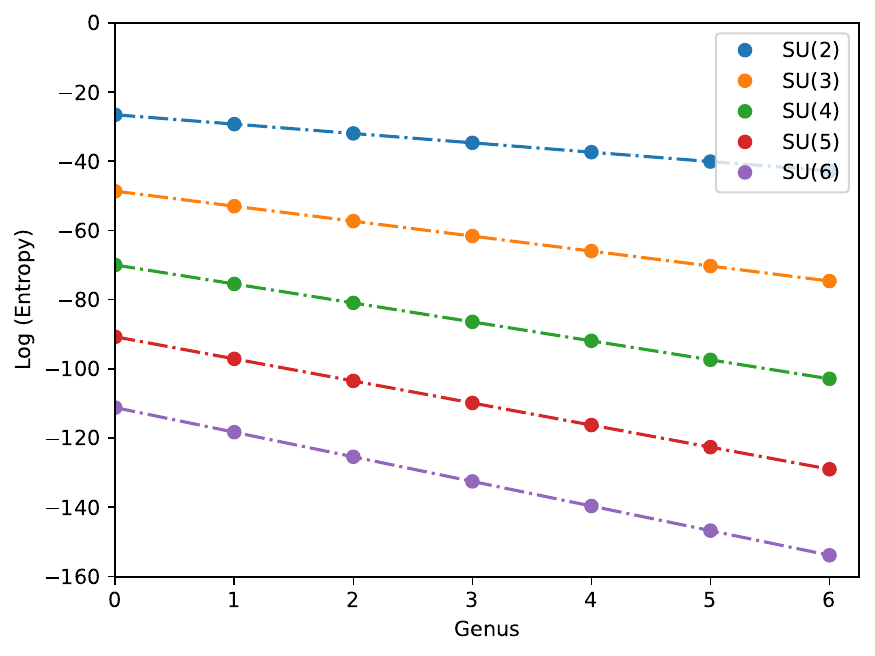}
\caption{Logarithm of the entropy as a function of genus for different gauge groups. The total area is fixed, $\varrho = 20$.}
 \label{fig:EmptyCylEntropyVsGenus}
\end{figure}

It is clear from \eqref{eq:entropy_cylinder_arbitrary_genus} that the entropy decreases exponentially with the area $\varrho$. We also see that the addition of holes increases the hierarchy among different eigenvalues of the density matrix, thereby further reducing the entropy. For $SU(2)$ this result is readily proven. All representations $\alpha$ are unequivocally labeled by only one number, the spin $j$, and we assume them to be related via $\alpha=2j$. The eigenvalues $h_\alpha$ of the reduced density matrix can be ordered by the value of $\alpha$. The ratio of two consecutive eigenvalues is
\be
\frac{h_{\alpha+1}}{h_\alpha} \ = \ \left(\frac{2\alpha+3}{2\alpha+1}\right)^{-4g}e^{-4(\alpha+1)\varrho }\,, 
\ee
confirming that the hierarchy among eigenvalues increases. For other groups, there is no unique way to label and order the representations, and we demonstrate the effect of entropy decay with genus numerically in Figure \ref{fig:EmptyCylEntropyVsGenus}. 

The reduction of the amount of entanglement with genus is expected in the space-emerging-from-entanglement paradigm discussed in the introduction. According to the paradigm, topological defects reduce the connectedness of the space emerging between the parties, which results in the loss of their correlation. In the limit of an infinite number of holes, the space essentially disconnects, breaking the correlations.


\section{Configurations with Wilson loops}
\label{sec:WilsonLoops}

Now we would like to turn on additional effects on the two-dimensional surfaces. In this section, we will focus on the example of the bipartite system of genus zero (cylinder) and study the effect of introducing basic Wilson loops to these states. We will consider contractible loops, with and without intersections, as well as non-contractible loops.


\subsection{Contractible loops without intersections}
\label{subsec:1CWL}

Let us add a contractible closed loop with no self-intersections on the surface of the cylinder; see Figure \ref{fig:cylinder1contractWL}. In this case, the procedure outlined in Section \ref{sec:quantizationYM2D} leads to a partition function written in terms of two boundary holonomies, $U_1$ and $U_2$, and the Wilson loop's representation $\gamma$. In particular, it has the form 
\begin{align}
\label{eq:1CWLpartfunction}
\begin{split}
Z(\varrho_i,\varrho_e;\gamma,U_1,U_2) \ &= \int dV \ \sum\limits_{\alpha,\beta} \frac{d_\beta}{d_\alpha} \chi_\alpha(U_1)\chi_\alpha(U_2)\chi_\alpha(V^{-1})\chi_\beta(V)\chi_\gamma(V)\ e^{-\left(C_2(\alpha)\varrho_e+C_2(\beta)\varrho_i\right)/2}\\
&=\ \sum\limits_{\alpha,\beta} \frac{d_\beta}{d_\alpha}N_{\beta\gamma}^{\alpha} \chi_\alpha(U_1)\chi_\alpha(U_2)\ e^{-\left(C_2(\alpha)\varrho_e+C_2(\beta)\varrho_i\right)/2}\,,
\end{split}
\end{align}
where $\varrho_i$ and $\varrho_e$ denote the internal and external areas bounded by the loop, respectively. Coefficients  $N_{\beta\gamma}^\alpha$ are the Littlewood-Richardson coefficients, or multiplicities of irreducible representation $\alpha$ appearing in the expansion of the tensor product of representations $\beta$ and $\gamma$:
\be
\label{eq:multiplicities}
\beta\otimes\gamma \ = \ \bigoplus_\alpha N_{\beta\gamma}^\alpha\, \alpha\,.
\ee
In equation~(\ref{eq:1CWLpartfunction}) we used equation~\eqref{eq:multiplicity_fusion_number_integral}, which expresses these coefficients in terms of a character integral. Note that for trivial $\gamma$, $N_{\beta 0}^\alpha=\delta_\beta^{\alpha}$, and one recovers the cylinder partition function~\eqref{eq:Zcylinder}.

We find it convenient to define the coefficients 
\be\label{eq:coeffC}
\mathcal{D}_\alpha(\gamma,\varrho) \ \equiv \ \sum\limits_{\beta} \frac{d_\beta}{d_\alpha}N_{\beta\gamma}^{\alpha} \,e^{-C_2(\beta)\varrho/2},
\ee
which allow us to write the partition function as 
\be\label{eq:partitionfunction1CWL}
Z(\varrho_i,\varrho_e;\gamma,U_1,U_2) \ = \ \sum\limits_{\alpha} \mathcal{D}_\alpha(\gamma,\varrho_i)e^{-C_2(\alpha)\varrho_e/2}\chi_\alpha(U_1)\chi_\alpha(U_2)\,.
\ee
This makes it more evident that the state has a similar diagonal (TFD) structure as the cylinder, but with renormalized coefficients. Moreover, note that
\be
\mathcal{D}_\alpha (\gamma,0) \ = \ \sum\limits_{\beta} \frac{d_\beta}{d_\alpha}N_{\beta\gamma}^{\alpha} \ = \ d_\gamma\,,
\ee
so that $\mathcal{D}_\alpha(\gamma,\varrho)$ may be interpreted as a ``deformation" of the dimension of the representation $\gamma$. This deformation preserves the reality of the dimension.

For $n$ nonintersecting loops, expression \eqref{eq:partitionfunction1CWL} is generalized by adding a $\mathcal{D}_\alpha(\gamma_k,\varrho_k)$ factor for each loop labeled by $k$, $k=1,\dots,n$:
\be
Z(\varrho_e,\{\varrho_k\};\{\gamma_k\},U_1,U_2) \ = \ \sum\limits_{\alpha} \left(\prod\limits_{k=1}^n \mathcal{D}_\alpha(\gamma_k,\varrho_k)\right)e^{-C_2(\alpha)\varrho_e/2}\chi_\alpha(U_1)\chi_\alpha(U_2)\,.
\ee
This pattern is similar to that of adding a topologically trivial Wilson loop in Chern-Simons partition functions, which amounts to inserting a factor of the ``quantum'' dimension~\cite{Witten:1988hf,Melnikov:2025tui}.

In the representation basis~(\ref{eq:StateInRepBasis}), such an amplitude has coefficients
\begin{equation}
\label{eq:manyloopstate}
    \psi_{\alpha\beta} = \delta_{\alpha\beta} \left(\prod\limits_{k=1}^n \mathcal{D}_\alpha(\gamma_k,\varrho_k)\right) e^{-C_2(\alpha)\varrho_e/2}\,.
\end{equation}
Then the reduced density matrix is
\begin{equation}
\label{eq:manyloopDM}
    \rho_1 = \mathcal{N}^{-1}\sum_\alpha \left(\prod^n_{k=1} \mathcal{D}_\alpha(\gamma_k, \varrho_k)^2\right) e^{-C_2(\alpha)\varrho_e}\,\ket{\chi_\alpha}\bra{\chi_\alpha}\,,
\end{equation}
where the normalization factor is given by
\begin{equation}
\label{eq:Nfor1CWL}
    \mathcal{N} = \sum_\alpha \left(\prod^n_{k=1} \mathcal{D}_\alpha(\gamma_k, \varrho_k)^2\right)e^{-C_2(\alpha)\varrho_e} \ = \ Z(T^2,\{\gamma_k\},\{\gamma_{k'}\})\,,
\end{equation}
which is equivalent to the partition function of a torus with $2n$ Wilson loops in their respective representations.

Consequently, the entropy takes the form
\begin{equation}
    \label{eq:1ConLoopEntropy}
    \boxed{\quad S = -\sum_\alpha h_\alpha \log h_\alpha\,, \quad\quad h_\alpha = \dfrac{e^{-C_2(\alpha)\varrho_e}}{\mathcal{N}}\prod^n_{k=1} \mathcal{D}_\alpha(\gamma_k,\varrho_k)^2\,.\quad}
\end{equation}


\subsubsection{Entanglement entropy analysis for a contractible loop}
\label{subsec:EE1CWL}

The qualitative behavior of the entropy in~(\ref{eq:1ConLoopEntropy}) can be understood with the help of numerical analysis. We start with the case of a single loop and investigate how its presence affects the entropy.

The first lesson learned is that entropy is not a monotonic function of the loop parameters. Let us look at the behavior of the entropy of a fixed-area state with a single loop as a function of the loop size. When the total area is sufficiently large in units set by the inverse of the Casimir of the loop's representation, a typical entropy plot reveals a series of peaks (local maxima) at specific values of the ratio between the area enclosed by the loop and the total area of the cylinder. The number of peaks is controlled by the loop representation; for instance, in the case of $SU(2)$, it coincides with its Dynkin label.\footnote{In the rest of the paper we will label the irreps of $SU(N)$ by Young diagrams, either drawing the diagram, or listing the corresponding $N$-integer box partition, e.g $\scalebox{0.3}{\yng(2,1)}\equiv[2,1,0]$ is the adjoint representation of $SU(3)$.} 

\begin{figure}[H]
    \centering
\begin{subfigure}{0.45\textwidth}
\centering
    \includegraphics[width=7cm]{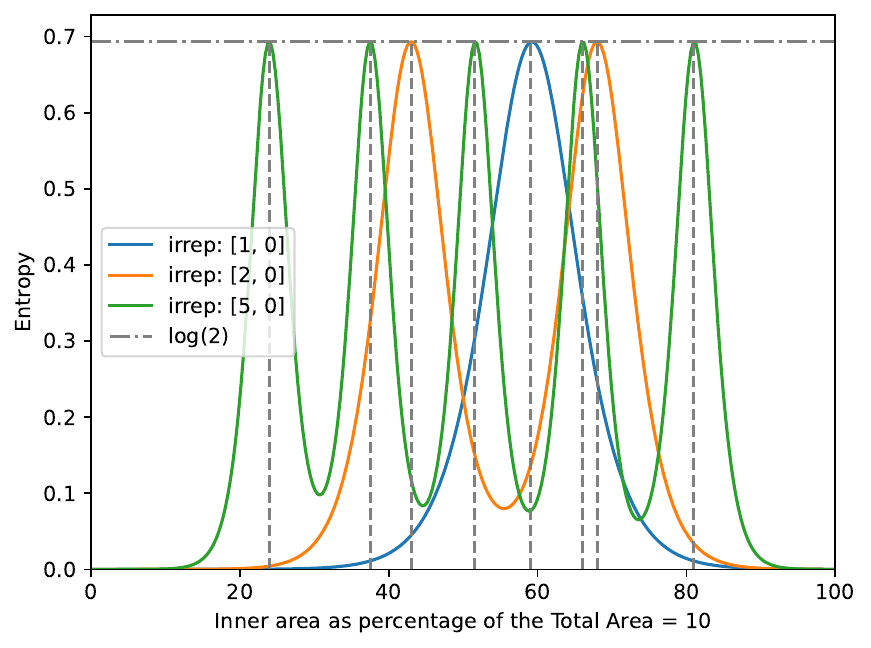}
\subcaption{}
\label{subfig:cylCWL_fixed_tot_su2a}
\end{subfigure}
\hfill
\begin{subfigure}{0.45\textwidth}
\centering
    \includegraphics[width=7cm]{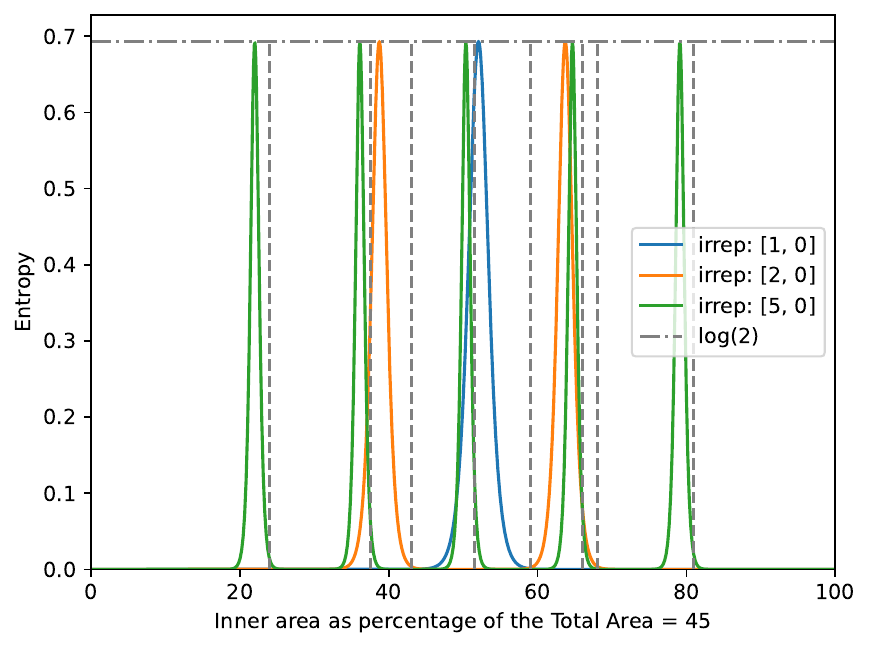}
\subcaption{}
\label{subfig:cylCWL_fixed_tot_su2b}
\end{subfigure}
\caption{Entropy~(\ref{eq:1ConLoopEntropy}) for varying area fraction for the loop in some $SU(2)$ representations calculated for intermediate (\subref{subfig:cylCWL_fixed_tot_su2a}) and large (\subref{subfig:cylCWL_fixed_tot_su2b}) total area $\varrho_t$. The vertical dashed lines correspond to the same fraction of the loop area and the total area on both plots.}
    \label{fig:cylCWL_fixed_tot_su2}
\end{figure}

A representative example is shown in Figure~\ref{fig:cylCWL_fixed_tot_su2}. In particular, Figures~\eqref{subfig:cylCWL_fixed_tot_su2a} and \eqref{subfig:cylCWL_fixed_tot_su2b} correspond to different values of the total area $\varrho_t$. They show that, as the total area increases, the peaks become sharply localized around certain “optimal’’ values of the area fraction. For example, the single blue peak associated with the fundamental representation localizes sharply at $50\%$ of $\varrho_t$ as its value increases.

\begin{figure}[h!]
\centering
    \includegraphics[width=7cm]{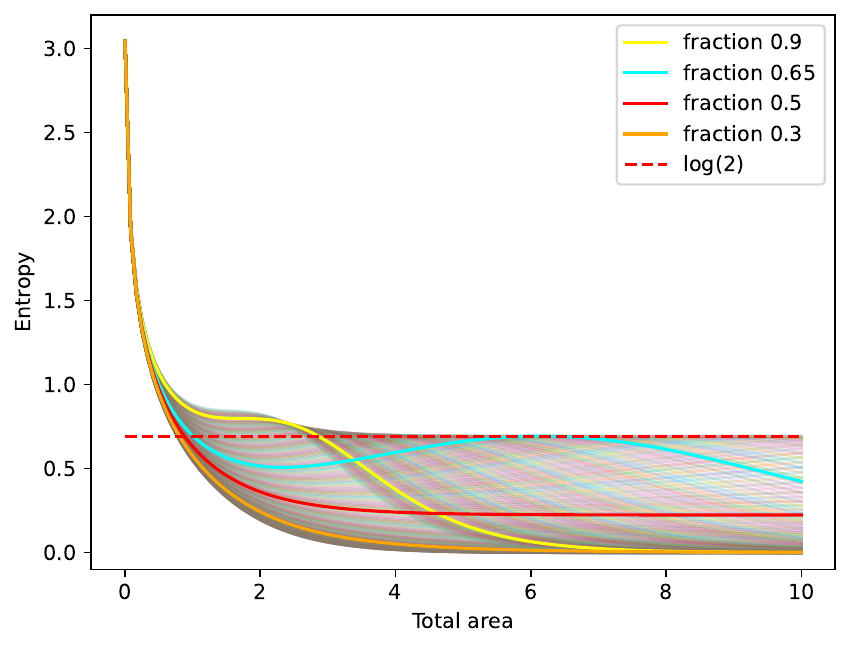}
\caption{Entropy~(\ref{eq:1ConLoopEntropy}) as a function of the total area $\varrho_t$ for the loop in the fundamental representation of $SU(2)$ for different fractions $r=\varrho_i/\varrho_t$ of the loop area and the total area. For fractions $0.5<r\lesssim 0.8$ the entropy curves have local maxima approaching $\log 2$. Approximately 200 curves plotted with selected ones labeled.}
    \label{fig:Ent_1CWL_su2_manycurves}
\end{figure}

The exact position of each peak is a function of the total area, and their heights are approximately equal to $\log 2$. Globally, the entropy is unbounded: It is divergent in the limit of vanishing total area, so the regime where this particular structure of the peaks is present corresponds to total areas above a threshold, as can be clearly seen from Figure~\ref{fig:Ent_1CWL_su2_manycurves}. The figure displays a collection of entropy curves for the fundamental representation, plotted as functions of the total area $\varrho_t=\varrho_i+\varrho_e$. Each curve corresponds to a fixed value of the area fraction $r=\varrho_i/\varrho_t$, with a few representative choices shown explicitly.

For $r<0.5$, the entropy curves decay monotonically to zero. For $0.5< r\lesssim 0.8$, the curves develop a local maximum of order $\log 2$; as $r\to0.5^+$, the position of this maximum is pushed to increasingly large values of $\rho_t$. This behavior suggests that states in this regime span subspaces of the Hilbert space that factorize into two two-dimensional components. For $r\gtrsim 0.8$, the maxima of the curves (if any) are notably above $\log 2$, but are confined to the region of small $\rho_t$. At large total area, the entropy again decays monotonically.

The value $r=0.5$ therefore separates two qualitatively distinct behaviors: curves with a local maximum and curves without it. As a result, the entropy at this optimal area fraction remains finite in the limit where the total area $\varrho_t$ is taken to infinity.

We can summarize the main observations in the $SU(2)$ case as follows:
\begin{itemize}
    \item For large total area, the entropy of one side of the cylinder in the presence of a Wilson loop lies in the range $0\leq S \lesssim \log 2$; 

    \item In the infinite-area limit, only a finite number of localized entropy peaks survive. This number equals the number of boxes in the Young diagram of the irreducible representation, or equivalently, twice its spin;

    \item For representations of integer spin, the peaks appear in pairs corresponding to reciprocal values of the area fraction. The same holds for half-integer spins, except that there is one additional peak at $50\%$;

    \item The heights of the infinite-area peaks depend on the representation and correspond to non-maximally entangled density matrices, with $S<\log2$.
    
\end{itemize}
These empirical observations can be confirmed by the analytical analysis presented below, which also allows one to derive the entropy values in the infinite-area limit.

The reason some entropies survive in the limit of infinite area is that there are two coefficients $h_\alpha$ with a finite limit. To see that for a loop in representation $\gamma$ of $SU(2)$, one first observes that, at leading order, there are $\gamma+1$ competing exponential terms in the normalization factor~(\ref{eq:Nfor1CWL}):
\begin{align}
\label{eq:Nfor1CWLexpansion}
\begin{split}
{\cal{N}} & =  d_\gamma^2 e^{-C_2(\gamma)\varrho_i} + \left(\frac{d_{\gamma-1}^2}{d_1^2}e^{-C_2(\gamma-1)\varrho_i-C_2(1)\varrho_e}+  \cdots\right) +  \cdots \\
&+ \left(\frac{d_{\gamma-n}^2}{d_n^2}e^{-C_2(\gamma-n)\varrho_i-C_2(n)\varrho_e}+  \cdots\right)
+  \cdots +  \left(\frac{1}{d_\gamma^2}e^{-C_2(\gamma)\varrho_e} +  \cdots\right) +  \cdots\,,
\end{split}
\end{align}
where every parenthesis corresponds to a choice of $\alpha = 0,\ldots, \gamma$ in~(\ref{eq:Nfor1CWL}), or to the numerator of $h_\alpha$. For a generic ratio $x=\varrho_i/\varrho_e$, only one of the $\alpha$-terms wins, there is a unique $h_\alpha=1$ and, hence, zero entropy in the infinite area limit.

Nonzero values for the entropy occur when $x$ is such that two exponential terms are of the same order. This happens when
\be
\label{eq:dominance}
C_2(\gamma-n)x + C_2(n) \  = \ C_2(\gamma-m)x+C_2(m)\,,
\ee
for some $m$ and $n$, $n\ne m$; that is, when two lines parameterized by $x$ intersect. Note that no more than two lines can intersect at the same point -- one can show that there are no solutions in this case -- so, indeed, there can be only two $h_\alpha$ contributing in the asymptotic limit.

Next, one notices that of all $\gamma(\gamma-1)/2$ intersections, only $\gamma$ of them correspond to the asymptotically dominant terms. In particular, one can show that the only relevant ones correspond to $m$ and $n$ in~(\ref{eq:dominance}) that differ by one unit. In such a case, for the optimal fractions $x_n$, the contributing components of the reduced density matrix are $h_{n}$ and $h_{n+1}$ where $n=1,\ldots,\gamma$.

Finally, solving $\gamma$ equations~(\ref{eq:dominance}) for $m=n-1$, and using $r=x/(1+x)$, one finds that the optimal ratios are given by
\be\label{eq:rnCasimir}
r_n=\frac{C_2(n)-C_2(n-1)}{C_2(n)-C_2(n-1) +C_2(\gamma-n+1)-C_2(\gamma-n)}=\frac{2n+1}{2\gamma+4}\,, \quad n=1,\dots,\gamma\,.
\ee
In the configuration of label $n$, the entropy approaches
\begin{align}
\label{eq:Sn1CWL}
\begin{split}
S_n(\gamma) & =  -\big(1-h_{n}(\gamma)\big)\log\big(1-h_n(\gamma)\big) - h_n(\gamma)\log\big(h_n(\gamma)\big)
\end{split}
\end{align}
with
\be\label{eq:hn1CWL}
h_n(\gamma)= \frac{1}{1+\left(\frac{d_n d_{\gamma-n+1}}{d_{\gamma-n}d_{n-1}}\right)^2} \,, \quad n=1,\dots,\gamma\,.
\ee

As stated earlier, there is a reflection symmetry in the distribution of ratios within the interval $[0,1]$, since
\be
1-r_n = r_{\gamma-n+1} \,.
\ee
This implies that a 50\% ratio is present only for odd representations -- indeed, $r=1/2$ appears when $n=(1+\gamma)/2$ and since $n$ is an integer, $\gamma$ needs to be odd. Moreover, reflection-symmetric ratios have the same entropy values, since
\be
S_n = S_{\gamma-n+1}\,.
\ee

\begin{figure}[h!]
\centering
    \includegraphics[width=0.5\linewidth]{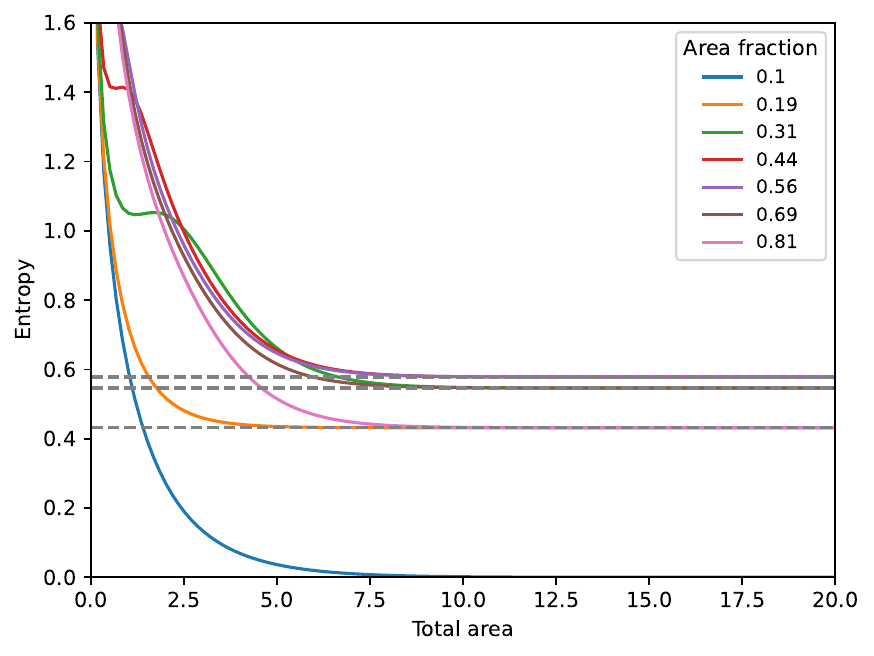}
\caption{Entropy~(\ref{eq:1ConLoopEntropy}) as a function of the total area $\varrho_t$ for a loop in representation $[6,0]$ of $SU(2)$ for different area fractions $r=\varrho_i/\varrho_t$. Continuous lines are obtained numerically while horizontal dashed lines are analytical predictions to the asymptotic limit.}
    \label{fig:cyl1CWLsu2gamma6}
\end{figure}

To illustrate the results of our analysis, we take $\gamma=6$ as an example. In this case, \eqref{eq:rnCasimir} implies that there are six values of ratios with a non-null asymptotic limit. In Figure \ref{fig:cyl1CWLsu2gamma6}, we compare our analytical and numerical results for this choice of representation, showing perfect agreement. We include a 0.1 ratio to explicitly show that the entropy vanishes when $r$ does not satisfy \eqref{eq:rnCasimir}. Furthermore, we see that complementary area fractions have the same asymptotic limit, even though they behave differently at small areas. For instance, the orange and pink curves are only distinguishable for $\varrho_t\lesssim7.5$.

We have thus established analytically the empirical observations concerning the infinite-area limit of the entropy for different representations of $SU(2)$. This also explains the bounded behavior of the entropy at large areas: the corresponding density matrices are diagonal, with only two dominant entries. As a result, they effectively behave as quasi-projectors onto two-dimensional subspaces of the Hilbert space, labeled by pairs of adjacent representations.

\begin{figure}[h!]
    \centering
\begin{subfigure}{0.45\textwidth}
\centering
    \includegraphics[width=7cm]{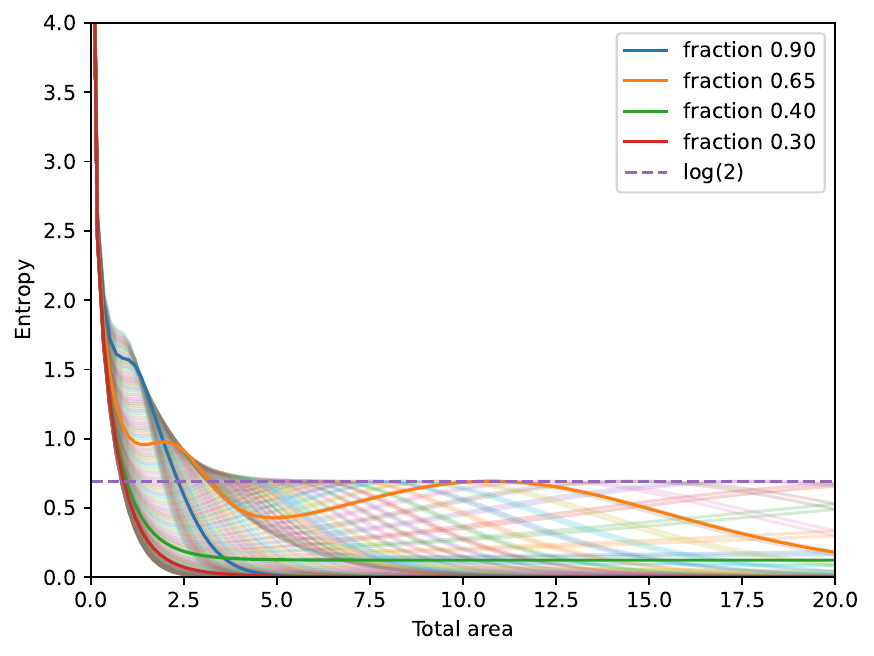}
\subcaption{}
\label{subfig:cylCWL_su3_manycurvesa}
\end{subfigure}
\hfill
\begin{subfigure}{0.45\textwidth}
\centering
    \includegraphics[width=7cm]{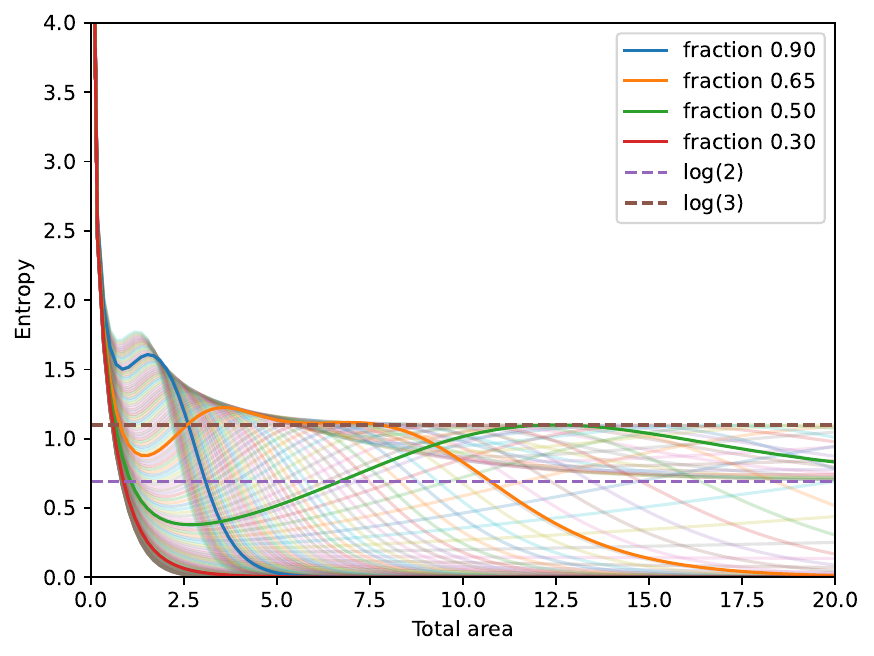}
\subcaption{}
\label{subfig:cylCWL_su3_manycurvesb}
\end{subfigure}
\caption{Entropy curves for a contractible loop in representation $[2,0,0]$ (\subref{subfig:cylCWL_su3_manycurvesa}) and $[2,1,0]$ (\subref{subfig:cylCWL_su3_manycurvesb}) of $SU(3)$.}
    \label{fig:cylCWL_su3_manycurves}
\end{figure}

For $SU(3)$ the structure of the entropy peaks becomes richer. There are two classes of representations in this case: those that can be viewed as $SU(2)$ representations embedded in $SU(3)$, and genuinely $SU(3)$ representations.\footnote{That is, by ``genuinely $SU(3)$" we mean representations whose Young diagrams are neither single-row diagrams nor conjugate to single-row diagrams.} For the first class, the area-entropy curves also show the $\log2$-bounded behavior at large areas. For the diagrams of the second class, the bound shifts to $\log3$. The two behaviors can be seen in Figure~\ref{fig:cylCWL_su3_manycurves}.

\begin{figure}[h!]
\centering
\begin{subfigure}{0.45\textwidth}
\centering
    \includegraphics[width=7cm]{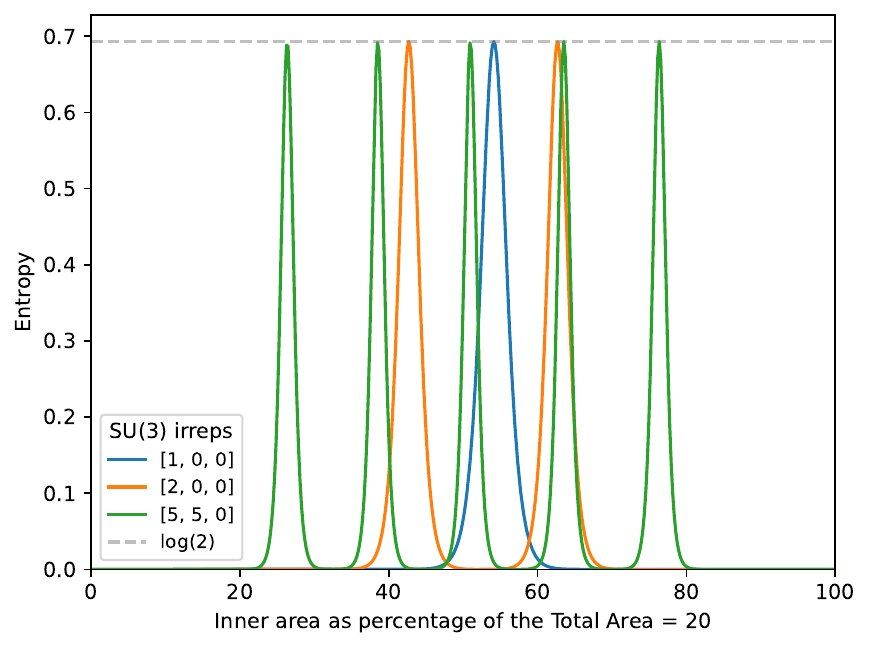}
\subcaption{}
\label{subfig:cylCWL_su3_peaksa}
\end{subfigure}
\hfill
\begin{subfigure}{0.45\textwidth}
\centering
    \includegraphics[width=7cm]{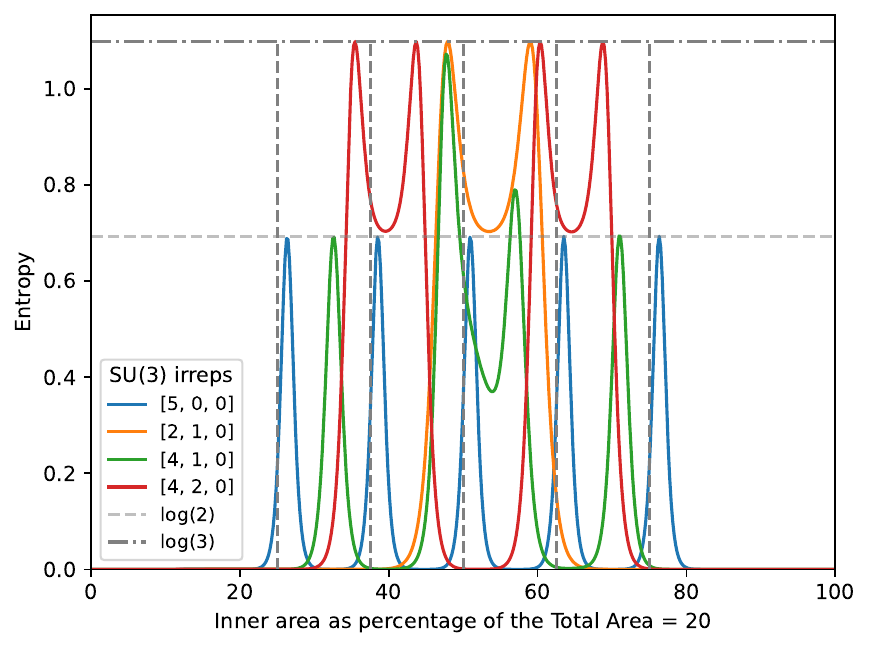}
\subcaption{}
\label{subfig:cylCWL_su3_peaksb}
\end{subfigure}
\caption{Peaks of $SU(3)$ entropy for fixed total area for selected symmetric irreps or their conjugates (\subref{subfig:cylCWL_su3_peaksa}) and genuinely $SU(3)$ representations (\subref{subfig:cylCWL_su3_peaksb}). Representation $[5,0,0]$ is shown as a reference in the second plot. For irreps $[2,1,0]$ and $[4,2,0]$ there are regions bounded by $\log 2$ from below. Vertical grid indicates the positions of asymptotic optimal fractions for $[5,0,0]$ given by equation~(\ref{eq:rnCasimirSU3}).}
    \label{fig:cyl1CWLsu3}
\end{figure}

Besides, for the irreps of the second class, the new feature is that there are finite regions of the fraction $r$ for which the entropy remains finite in the infinite area limit. This is manifest in the accumulation of curves between $\log2$ and $\log3$ in Figure~\ref{subfig:cylCWL_su3_manycurvesb}. Alternatively, we can analyze this behavior in terms of the entropy peaks. These are shown in Figure~\ref{fig:cyl1CWLsu3}. We see that, for symmetric representations or their conjugates, the regions between peaks have a null asymptotic limit. In contrast, for genuinely $SU(3)$ representations, there are regions between peaks that are bounded by $\log2$ from below.

Our empirical observations in the $SU(3)$ case can be summarized as follows: 
\begin{itemize}
    \item the number of peaks seems to be controlled by the number of boxes in the first row of the Young diagram; 
    
    \item at large areas, the entropy is asymptotically bounded by $\log 3$;
    
    \item in symmetric representations or their conjugates, the peaks are bounded by $\log 2$. These peaks are infinitely localized (that is, they have zero width) and assume finite values, typically below $\log2$, in the infinite area limit;
    
    \item for representations that are not symmetric or their conjugates, some of the peaks appear in the gap between $\log2$ and $\log3$;
    
    \item for such representations there may exist finite regions of fraction value $r$ with finite entropy in the infinite volume limit.
\end{itemize}

Here we will work out a few explicit examples illustrating some details analytically. We leave detailed proofs of the observations for future work. 

\begin{figure}[h!]
\centering
\begin{subfigure}{0.45\textwidth}
\centering
    \includegraphics[width=7cm]{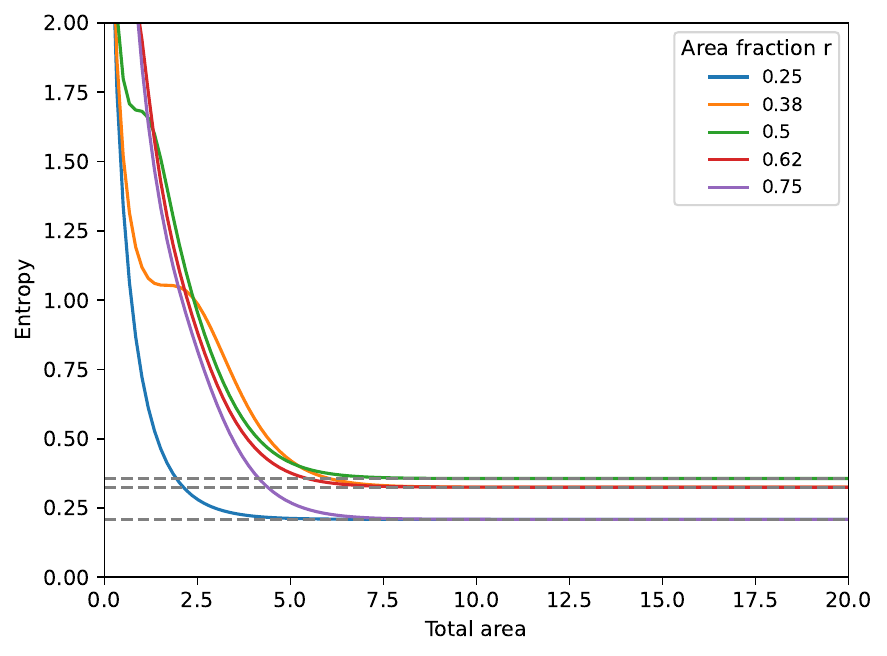}
\subcaption{}
\label{subfig:cylCWL_su3_a}
\end{subfigure}
\hfill
\begin{subfigure}{0.45\textwidth}
\centering
    \includegraphics[width=7cm]{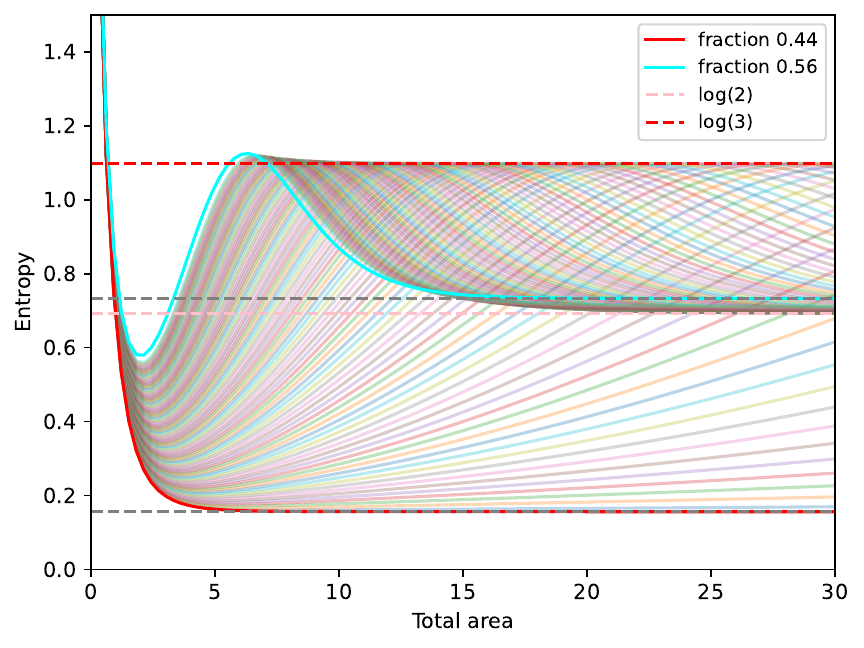}
\subcaption{}
\label{subfig:cylCWL_su3_b}
\end{subfigure}
\caption{Entropy versus total area $\varrho_t$ for a loop in $[5,0,0]$ (\subref{subfig:cylCWL_su3_a}) and $[2,1,0]$ (\subref{subfig:cylCWL_su3_b}) of $SU(3)$ for different area fractions $r=\varrho_i/\varrho_t$. In both plots, continuous lines are obtained numerically while gray horizontal lines are analytical results for the asymptotic limit. In (\subref{subfig:cylCWL_su3_b}) translucent curves correspond to area fractions within the interval $[4/9,5/9]$ while colored horizontal curves are upper and lower bounds for their asymptotic limit.}
    \label{fig:cyl1CWLSU3}
\end{figure}

For a loop in a symmetric representation, the infinite area limit appears to be dominated by the same set of exponential terms as in~(\ref{eq:Nfor1CWLexpansion}), where $\gamma$ again labels the number of boxes in the first row of the Young diagram. For generic values of the ratio $r=\varrho_i/\varrho_t$ (or $x=\varrho_i/\varrho_e$), only one of the exponential terms dominates. However, for a number $\gamma$ of discrete values, exactly two terms have the same order. These values are defined by equation~(\ref{eq:dominance}) and its solution~(\ref{eq:rnCasimir}), but this time with the $SU(3)$ Casimir
\be
 C_2([\gamma,0,0]) \ = \ \gamma(\gamma+2) - \frac{\gamma^2}{3}\,.
 \ee
That is,
\be\label{eq:rnCasimirSU3}
r_n=\frac{C_2(n)-C_2(n-1)}{C_2(n)-C_2(n-1) +C_2(\gamma-n+1)-C_2(\gamma-n)}=\frac{n+1}{\gamma+3}\,, \quad n=1,\dots,\gamma\,.
\ee
At these special configurations, the asymptotic behavior of the entropy is again captured by \eqref{eq:Sn1CWL} with the same $h_n(\gamma)$, modulo adjusting the dimension of the representation to its $SU(3)$ value: $d_\gamma=\frac{(\gamma+1)(\gamma+2)}{2}$. In Figure~\ref{subfig:cylCWL_su3_a} we show an example of these predictions at work for representation $[5,0,0]$.

On the other hand, when the loop is not in a symmetric representation, we observe non-null lower bounds for the entropy between peaks; see, for instance, the $\log2$ lower bound for irreps $[2,1,0]$ and $[4,2,0]$ in Figure~\ref{subfig:cylCWL_su3_peaksb}. In the infinite area limit, such regions saturate at exactly $\log 2$ values because of the degeneracy of the quadratic Casimir and dimensions for conjugate representations. To see this, let us consider the $[2,1,0]$ case in detail. For this representation, the relevant exponential terms are
\begin{align}
\label{eq:1CWLlog3}
\begin{split}
{\cal{N}} & =  d_{\scalebox{0.2}{\yng(2,1)}}^2 \,e^{-C_2(\scalebox{0.2}{\yng(2,1)})\varrho_i} + \left(e^{-C_2(\scalebox{0.2}{\yng(1)})\varrho_i-C_2(\scalebox{0.2}{\yng(1)})\varrho_e}+ \cdots\right)  +
\left(e^{-C_2(\scalebox{0.2}{\yng(1,1)})\varrho_i-C_2(\scalebox{0.2}{\yng(1,1)})\varrho_e}+  \cdots\right) \\
& + \left(\frac{1}{d_{\scalebox{0.2}{\yng(2,1)}}^2}e^{-C_2(\scalebox{0.2}{\yng(2,1)})\varrho_e}+ \cdots\right) +  \cdots \,,
\end{split}
\end{align}
where the dots indicate exponential terms that are subleading in comparison with the four relevant ones. The four terms compete for different choices of the fraction $x$, and in contrast to the case of symmetric irreps, two of them are completely degenerate. This fact has two consequences: First, there are triple intersection points defined by
\be
C_2(\scalebox{0.25}{\yng(1)})x_- + C_2(\scalebox{0.25}{\yng(1)}) \ = \ C_2(\scalebox{0.25}{\yng(1,1)})x_- +C_2(\scalebox{0.25}{\yng(1,1)}) \ = \ C_2(\scalebox{0.25}{\yng(2,1)}) x_-
\ee
and
\be
C_2(\scalebox{0.25}{\yng(1)})x_+ + C_2(\scalebox{0.25}{\yng(1)}) \ = \ C_2(\scalebox{0.25}{\yng(1,1)})x_+ +C_2(\scalebox{0.25}{\yng(1,1)}) \ = \ C_2(\scalebox{0.25}{\yng(2,1)}), 
\ee
yielding $x_- =4/5$ and $x_+ = 5/4$, or $r_- =4/9$ and $r_+=5/9$, and entropies $S_-\simeq 0.16$ and $S_+\simeq 0.73$. Second, for $4/9< r < 5/9$ the entropy should asymptote to the value of $\log 2$ at infinite area. This is precisely what is seen in Figure~\ref{subfig:cylCWL_su3_b}.

\begin{figure}[h!]
\centering
\begin{subfigure}{0.45\textwidth}
\centering
    \includegraphics[width=7cm]{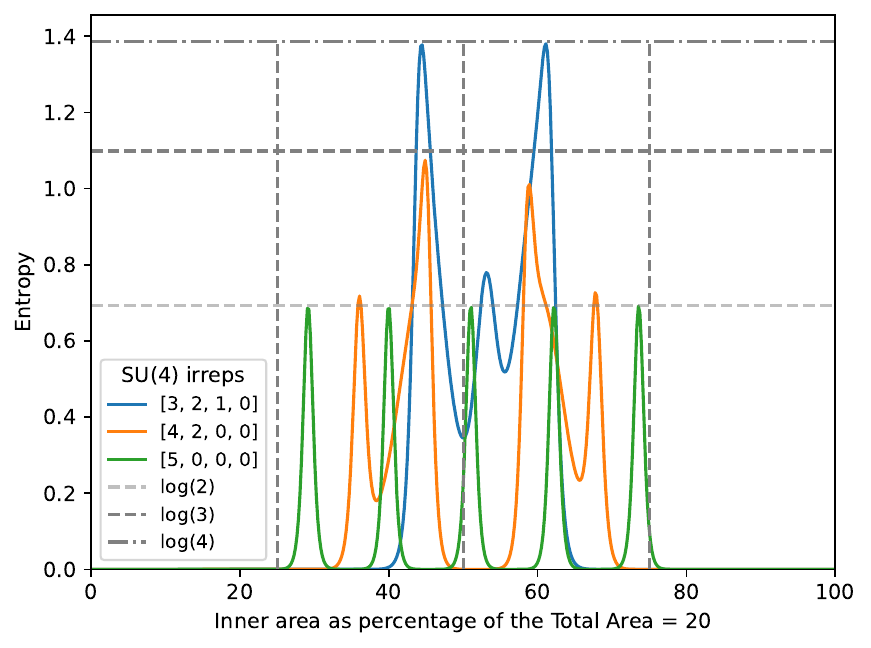}
\subcaption{}
\label{subfig:su4_peaks}
\end{subfigure}
\hfill
\begin{subfigure}{0.45\textwidth}
\centering
    \includegraphics[width=7cm]{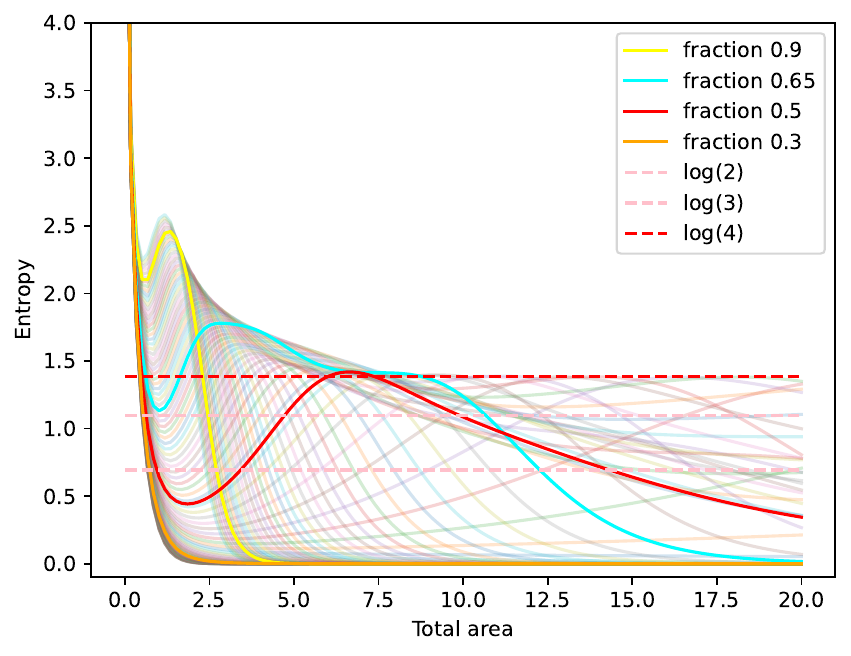}
\subcaption{}
\label{subfig:su4_manycurves}
\end{subfigure}
\caption{(\subref{subfig:su4_peaks}) Peaks of the $SU(4)$ entropy illustrating $\log(n+1)$ bounds for Young diagrams with $n$ rows. (\subref{subfig:su4_manycurves}) Asymptotically spanned entropy ranges by irrep $[3,2,1,0]$.}
    \label{fig:cyl1CWLSU4manycurves}
\end{figure}

We expect our main observations to hold for all $SU(N)$ groups. In particular, we expect to see entropy peaks with an asymptotic $\log n$ hierarchy, with $1\leq n\leq N$, established by the irreps, whose Young diagrams have $n$ rows. We expect that the number, positions, and heights of the peaks to be dependent on the details of representation theory, such as degeneracies and multiplicities. In the limit of infinite areas, we expect the peaks to be replaced by an infinitely localized set of lines, with the possibility of spaces between the lines filled by regions of finite entropy with discretized $\log n$ values. Samples of this behavior are illustrated in Figure~\ref{fig:cyl1CWLSU4manycurves}. Although the precise classification of the asymptotic large-area states requires a thorough group-theoretical analysis, the most important properties of these states must be independent of its result. By this, in particular, we understand the existence of a set of finite-dimensional bipartite density matrices of the TFD type.

\subsubsection{Entanglement entropy analysis for multiple contractible loops}
\label{subsec:manyloops}

With the above results it is straightforward to construct examples of large-area finite-entropy states with multiple loops. Consider state~(\ref{eq:manyloopstate}) with $n$ nonintersecting contractible loops in the same representation. Its reduced density matrix is given by equation~(\ref{eq:manyloopDM}). This density matrix can be obtained by concatenation of $n$ density matrices of the single-loop state. In particular one can concatenate infinitely large cylinders with identical loops occupying fractions of the total area given by equation~(\ref{eq:rnCasimir}). For example, for the fundamental irrep of $SU(2)$ there is a single optimal fraction corresponding to the density matrix
\be
\label{eq:10SU2DM}
\rho_{\scalebox{0.2}{\yng(1)}} \ = \ \left(
\begin{array}{cccc}
    1/17 & 0 & 0 & \cdots \\
    0 & 16/17 & 0 & \cdots \\
    0 & 0 & 0 & \cdots \\
    \vdots & \vdots & \vdots & \ddots
\end{array}
\right)\,,
\ee
where we assume the natural ordering of the $SU(2)$ representations. A two-loop state can have a reduced density matrix that is the normalized square of the above:
\be\label{eq:11SU2DM}
\rho_{(\scalebox{0.2}{\yng(1)},\scalebox{0.2}{\yng(1)})} \ = \ \left(
\begin{array}{cccc}
    1/257 & 0 & 0 & \cdots \\
    0 & 256/257 & 0 & \cdots \\
    0 & 0 & 0 & \cdots \\
    \vdots & \vdots & \vdots & \ddots
\end{array}
\right)\,,
\ee
and similarly, for any $n$. In fact, for a pair of loops in equal representations, the relative area fraction $\varrho_1/\varrho_2$ of the two loops does not affect the result: In the infinite-area limit, the density matrix has the same structure for any pair of loops with an arbitrary partition of the total inner area, provided that the latter is fixed at the optimal value $0.5$. That is, this holds for any $\varrho_1$ and $\varrho_2$ satisfying $\varrho_1+\varrho_2=\varrho_t/2$.

For a representation $[m,0]$ there will be $m$ density matrices and one can concatenate only compatible ones, i.e., with the same nonvanishing $2\times 2$ block. For the first symmetric representation, these are
\be
\label{eq:20SU2DM}
\rho_{\scalebox{0.2}{\yng(2)}}^{(1)} \ = \ \left(
\begin{array}{cccc}
    1/10 & 0 & 0 & \cdots \\
    0 & 9/10 & 0 & \cdots \\
    0 & 0 & 0 & \cdots \\
    \vdots & \vdots & \vdots & \ddots
\end{array}
\right) \qquad
\rho_{\scalebox{0.2}{\yng(2)}}^{(2)} \ = \ \left(
\begin{array}{cccc}
    0 & 0 & 0 & \cdots \\
    0 & 1/10 & 0 & \cdots \\
    0 & 0 & 9/10 & \cdots \\
    \vdots & \vdots & \vdots & \ddots
\end{array}
\right),
\ee
both with entropy $S\simeq\ 0.33$. Consequently, there are two series of finite-entropy states obtained from powers of the matrices above. The same happens for other identical irreps. In each case the density matrix is independent from the partition of the total area enclosed by the loops. In other words, one can understand the multiplication of loops as a splitting of a single loop that costs entanglement entropy. 

A similar pattern hold for loops of $SU(N)$ in identical representations. Recall that, for general representations, one can obtain density matrices of rank up to $N$. This increases the number of options, since now one can multiply/concatenate matrices corresponding to different optimal ratios, producing lower rank matrices. Consider the case of the $[2,1,0]$ irrep, which has two associated rank-three matrices:
\be
\rho_{\scalebox{0.2}{\yng(2,1)}}^{(1)} \ = \ \left(
\begin{array}{ccccc}
    32/33 & 0 & 0 & 0 & \cdots \\
    0 & 1/66 & 0 & 0 & \cdots \\
    0 & 0 & 1/66 & 0 & \cdots \\
    0 & 0 & 0 & 0 & \cdots \\
    \vdots & \vdots & \vdots & \vdots & \ddots
\end{array}
\right), \quad
\rho_{\scalebox{0.2}{\yng(2,1)}}^{(2)} \ = \ \left(
\begin{array}{ccccc}
    0 & 0 & 0 & 0 & \cdots \\
    0 & 64/129 & 0 & 0 & \cdots \\
    0 & 0 & 64/129 & 0 & \cdots \\
    0 & 0 & 0 & 1/129 & \cdots \\
    \vdots & \vdots & \vdots & \vdots & \ddots
\end{array}
\right).
\ee
Besides rank-three powers of matrices $\rho_{\scalebox{0.2}{\yng(2,1)}}^{(1)}$ and $\rho_{\scalebox{0.2}{\yng(2,1)}}^{(2)}$, one can also consider their product, which yields a maximally entangled rank-two density matrix. Such a product should correspond to a state with two loops of different size. The numerical analysis shows that this happens for any pair of loops with the total enclosed area ranging between $r_-=4/9$ and $r_+=5/9$ of the total area. This is compatible with the observation of Section~(\ref{subsec:EE1CWL}) that single loops with the enclosed area within the same range have maximally entangled rank-two density matrices.

The situation is more complex for nonequal representations. First, only single-loop states whose density matrices are compatible can be joined. However, there is no longer a unique optimal inner area fraction. Instead, the available optimal total areas enclosed by the loops range between the minimal of the two optimal single-loop fractions and the maximal one. For each asymmetry ratio $A=\varrho_1/\varrho_2$, there is a single optimal value of the parameter $I=(\varrho_1+\varrho_2)/\varrho_t$. For the smallest or the largest $I$, one of the loops must have zero area. The remaining loop is precisely the one whose representation defines the optimal size. 

Here we consider the simple example of a pair of loops in representations $[1,0]$ and $[2,0]$. Single-loop states for these irreps have optimal fractions $1/2$ for the former and the pair $3/8$ and $5/8$ for the latter. The $5/8$-loop is not compatible with the $1/2$-loop, as evident from density matrices~(\ref{eq:10SU2DM}) and~(\ref{eq:20SU2DM}). Thus, there are finite-entropy configurations of $3/8$ and $1/2$ loops with total enclosed areas $I$ varying in the range between the two fractions. All of these configurations have the same entropy fixed by the product of the two density matrices:
\be
\rho_{(\scalebox{0.2}{\yng(1)},\scalebox{0.2}{\yng(2)})} \ = \ \left(
\begin{array}{cccc}
    1/145 & 0 & 0 & \cdots \\
    0 & 144/145 & 0 & \cdots \\
    0 & 0 & 0 & \cdots \\
    \vdots & \vdots & \vdots & \ddots
\end{array}
\right)\,,
\ee
yielding $S\simeq 0.041$. Empirically, we find that within the interval $3/8\leq I\leq 1/2$ the optimal fractions are related via
\be
I(A) \ = \ \frac{3(1+A)}{2(4+3A)}\,.
\ee

\subsection{Contractible loops with intersections}
\label{sec:intersectingloops}

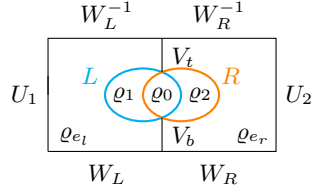
\begin{figure}[h!]
\centering
\begin{tikzpicture}[scale=1, every node/.style={font=\small}]

\begin{scope}
  \draw[thick, color=cyan] (1.75,1) arc [x radius=0.5, y radius=0.35, start angle=0, end angle=360];
  \draw[thick, color=orange] (2.25,1) arc [x radius=0.5, y radius=0.35, start angle=0, end angle=360];
  \draw (0,1) -- ++(0,0.75) -- ++(3,0) -- ++(0,-1.5) -- ++(-3,0);
  \draw (0,0.25) -- (0,1.25) ;
  \draw (1.5,0.25) -- (1.5,0.7) ;
  \draw (1.5,1.3) -- (1.5,1.75) ;

  \node[left] at (0,1) {$U_{1}$};
  \node[above] at (0.775,1.75) {$W_L^{-1}$};
  \node[above] at (2.25,1.75) {$W_R^{-1}$};
  \node[below] at (0.775,0.2) {$W_L$};
  \node[below] at (2.25,0.2) {$W_R$};
  \node[right] at (3,1) {$U_{2}$};
  \node[right] at (1.5,0.45) {$V_b$};
  \node[right] at (1.5,1.5) {$V_t$};
  \node at (1,1) {$\varrho_{1}$};
  \node at (2,1) {$\varrho_{2}$};
  \node at (1.5,1) {$\varrho_{0}$};
  \node at (0.35,0.45) {$\varrho_{e_l}$};
  \node at (2.7,0.45) {$\varrho_{e_r}$};
  \node at (0.55,1.25) {\textcolor{cyan}{$L$}};
  \node at (2.4,1.25) {\textcolor{orange}{$R$}};
\end{scope}

\end{tikzpicture}
\caption{Cell decomposition of the cylinder with two intersecting loops.}
\label{fig:intersectLoops}
\end{figure}

Another possibility is to study how entropy changes when a pair of loops on a cylinder intersects; see Figure \ref{fig:intersectLoops} for its plaquette decomposition.

In this case, the amplitude is obtained by computing
\begin{align}
\begin{split}
        \psi(U_1,U_2;\gamma_1,\gamma_2) = &\int dL\, dR\, dW\, dV\bigg[\chi_{\gamma_1}(L_iL_e)\chi_{\gamma_2}(R_iR_e)\bigg(\sum_\alpha d_\alpha e^{- C_2(\alpha)\varrho_{e_l}/2}\chi_\alpha(U_1 W_L V_b L_e V_t W_L^{-1})\bigg)\\
        &\times \bigg(\sum_\sigma d_\sigma e^{-C_2(\sigma)\varrho_{e_r} /2}\chi_\sigma(U_2 W_R^{-1} V_t^{-1} R_e V_b^{-1} W_R)\bigg)\bigg(\sum_{\beta} d_{\beta} e^{-C_2({\beta})\varrho_{1}/2}\chi_{\beta}(L_e R_i^{-1}) \bigg)\\
        &\times \bigg(\sum_{\alpha'} d_{\alpha'} e^{-C_2({\alpha'})\varrho_{0}/2}\chi_{\alpha'}(R_i L_i)\bigg)\bigg(\sum_{\beta'} d_{\beta'} e^{-C_2({\beta'})\varrho_{2}/2}\chi_{\beta'}(R_e L_i^{-1})\bigg)\,,
\end{split}
\end{align}
where subscripts $e$ and $i$ are assigned to the external and internal segments of the decomposition, respectively.

For simplicity we will consider the $SU(2)$ case. Using properties~\eqref{eq:ortho3j} and \eqref{6jfrom3j}, we may assemble the final result for the amplitude in terms of 6j symbols as follows\footnote{It is, in fact, known that an intersection of a pair of lines corresponds to the insertion of a 6j symbol (Racah coefficients), so the amplitudes can be constructed directly, bypassing the group integrals, e.g.~\cite{Witten:1991we,Blommaert:2018oro}.}
\be
\psi(U_1,U_2;\gamma_1,\gamma_2) = \sum\limits_{\alpha} f_\alpha(\gamma_1,\gamma_2;\varrho_0,\varrho_1,\varrho_2) \chi_\alpha(U_1)\chi_\alpha(U_2)e^{-C_2(\alpha)\varrho_e /2}\,,
\ee
where $\varrho_e = \varrho_{e_l} + \varrho_{e_r}$ and
\begin{equation}\label{eq:f_alpha_crossing_loops}
f_\alpha(\gamma_1,\gamma_2;\varrho_0,\varrho_1,\varrho_2) \ = \ \sum\limits_{\beta,\alpha',\beta'} \frac{d_{\alpha'} d_\beta d_{\beta'} }{d_\alpha}
\left\{
\begin{array}{ccc}
   \gamma_1  & \alpha& \beta \\
   \gamma_2  & \alpha' & \beta'
\end{array}
\right\}^2
e^{-(C_2(\alpha')\varrho_0 +C_2(\beta)\varrho_1 +C_2(\beta')\varrho_2 )/2}\,.
\end{equation}

In the character basis, tracing out the boundary holonomy $U_2$ yields a reduced density matrix that is diagonal in the representation basis,
\begin{equation}
    \rho_1 = \dfrac{1}{\mathcal{N}}\sum_\alpha f_\alpha^{~2} \,e^{-C_2(\alpha)\varrho_e} \ket{\chi_\alpha}\bra{\chi_\alpha}\,, \qquad \mathcal{N} = \sum_\lambda f_\lambda^{~2} e^{-C_2(\lambda)\varrho_e}\,.
\end{equation}
The straightforward application of the von Neumann formula gives
\begin{equation}\label{eq:2XLoopsEntropy}
\boxed{
    \quad S = -\sum_\alpha h_\alpha \log h_\alpha\,, \qquad h_\alpha = f_\alpha^{~2} \dfrac{e^{-C_2(\alpha)\varrho_e}}{\mathcal{N}}\,.\quad}
\end{equation}

At zero overlap ($\varrho_0 = 0$), the orthogonality relation of the 6j symbols~\eqref{eq:3j_orthogonality_1} allows us to rewrite \eqref{eq:f_alpha_crossing_loops} as
\begin{equation}\label{eq:zero_overlap_simplification}
    f_\alpha(\gamma_1,\gamma_2;0,\varrho_1,\varrho_2) \ = \mathcal{D}_\alpha(\gamma_1,\varrho_1)\, \mathcal{D}_\alpha(\gamma_2,\varrho_2)\,.
\end{equation}
In this limit, the expression factorizes, and we consistently recover the results of Section~\ref{subsec:1CWL} for two contractible, non-intersecting loops.

\subsubsection{Entanglement entropy analysis}

The numerical analysis of the entropy reveals a similar behavior compared to the non-intersecting case studied in Section~\ref{subsec:manyloops}. First of all, in the current case, the entropy \eqref{eq:2XLoopsEntropy} depends on several area fractions in addition to the internal area. For simplicity, we restrict ourselves to the case where $\varrho_1=\varrho_2$. Numerically, we see that for a fixed value of the overlap area, there is always a value of $I=\varrho_i/\varrho_t$ that renders the entropy finite at large areas. We shall refer to it as the optimal total inner fraction, and for concreteness, we focus on the case of two intersecting loops in the fundamental representation of $SU(2)$.

\begin{figure}[h!]
    \centering
\begin{subfigure}{0.45\textwidth}
\centering
    \includegraphics[width=7cm]{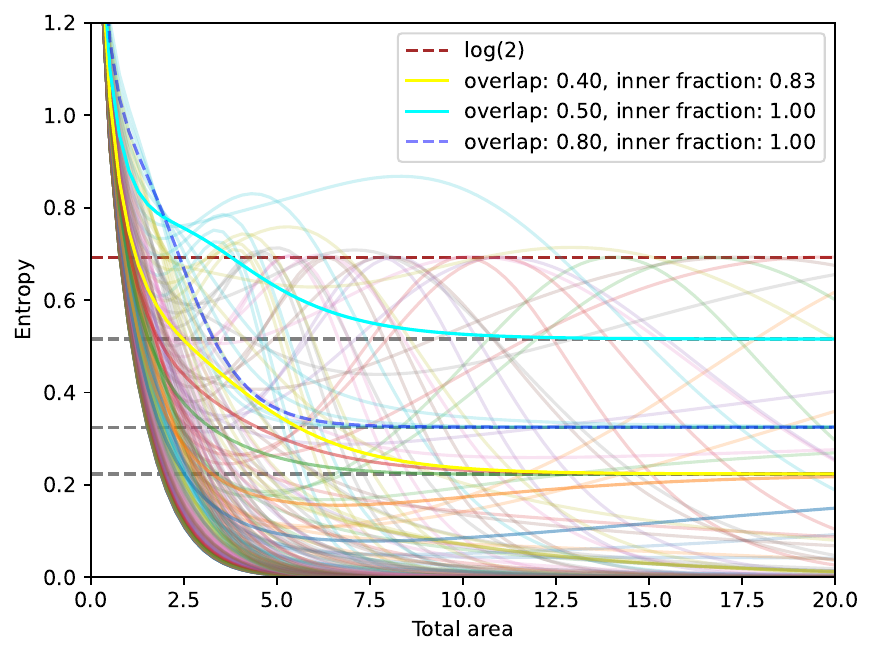}
\subcaption{}
\label{subfig:Ent_2XWL_su2_allcurves}
\end{subfigure}
\hfill
\begin{subfigure}{0.45\textwidth}
\centering
    \includegraphics[width=7cm]{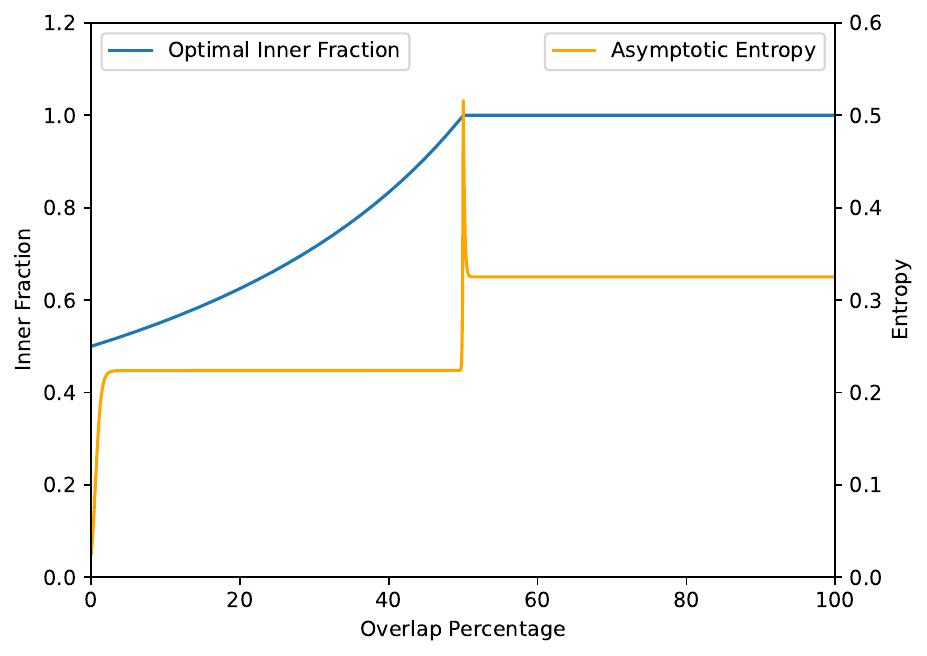}
\subcaption{}
\label{subfig:2xwl_optimal_fractions_and_entropy}
\end{subfigure}

\caption{(\subref{subfig:Ent_2XWL_su2_allcurves}) Entropy~(\ref{eq:2XLoopsEntropy}) as a function of the total area $\varrho_t$ for two crossing loops, both in the fundamental representation of $SU(2)$, for different fractions of the loop area and the total area as well as for different overlap fractions between the two loops areas. (\subref{subfig:2xwl_optimal_fractions_and_entropy}) Asymptotic values for the entropy for different combinations of overlap and total inner area fractions.}
    \label{fig:2xwl_fund_fund_su2}
\end{figure}

As in previous $SU(2)$ studies, the entropy is asymptotically bounded from above by $\log 2$. This can be seen directly from Figure~\ref{subfig:Ent_2XWL_su2_allcurves}, where we plot the entropy for different values of $\varrho_i$ and $\varrho_0$ as a function of the total area. In this plot, we also highlight three representatives of curves with different asymptotic limits. 

To understand these limits and, more generally, the relationship between entropy, overlap percentage ($O = \varrho_0 / \varrho_i$), and optimal inner area fraction it is convenient to look at Figure~\ref{subfig:2xwl_optimal_fractions_and_entropy}.
For small $O$, $I$ approaches 0.5, in agreement with the non-intersecting loop analysis. Furthermore, the entropy approaches $S\simeq 0.025$, which is the value obtained from the density matrix of two separate loops in the fundamental representation \eqref{eq:11SU2DM}.

When the overlap percentage is below $0.5$, the optimal total inner area fraction is related to it via
\be\label{eq:overlap_area_as_function_of_inner_area}
I(O)= \frac{1}{2(1-O)}\,.
\ee
In this case, the asymptotic value of the entropy is that obtained from the density matrix \eqref{eq:10SU2DM}, namely $S\simeq0.22$. This suggests that, in this regime, the system composed of two spin-$1/2$ intersecting loops behaves similarly to a single spin-$1/2$ loop at its optimal ratio $r=0.5$. 

At $O=0.5$, the optimal inner fraction reaches a plateau, as evident in the blue curve in Figure~\ref{subfig:2xwl_optimal_fractions_and_entropy}. At this point, the entropy value peaks and reaches its maximum, $S\simeq 0.52$. Though this value is numerically robust, indicating the existence of a specific analog configuration, we were unable to identify it.

Finally, when $I>0.5$, the optimal inner area fraction becomes constant and equal to 1. That is, the entropy will be finite in the large area limit only when the intersecting loops entirely cover the cylinder. In particular, its asymptotic value is $S \simeq 0.33$, which corresponds to the entropy derived from~\eqref{eq:20SU2DM}. That is, in this regime, the system behaves like a single spin-$1$ contractible loop at optimal ratio.

\subsection{Non-contractible loops}
\label{subsec:1NCWL}

\begin{figure}[h!]
\centering
\begin{tikzpicture}[scale=1, every node/.style={font=\small}]

\begin{scope}[xshift=-0.5cm]
 
  \draw[dashed] (-1,1.35) arc[start angle=90, end angle=-90,x radius=0.15cm, y radius=0.35cm];
  \draw (-1,0.65) arc[start angle=270, end angle=90,x radius=0.15cm, y radius=0.35cm];
  \draw (1,1) ellipse (0.15 and 0.35);
  \draw (-1,1.35) -- (1,1.35);
  \draw (-1,0.65) -- (1,0.65);
  
  \draw[color=cyan,dashed,thick] (0,1.35) arc[start angle=90, end angle=-90,x radius=0.15cm, y radius=0.35cm];
  \draw[color=cyan,thick] (0,0.65) arc[start angle=270, end angle=90,x radius=0.15cm, y radius=0.35cm];

  \node[left] at (-0.75,1.65) {$U_1$};
  \node[right] at (0.75,1.65) {$U_2$};
  \node[left] at (-0.2,1) {$\varrho_l$};
  \node[right] at (0.2,1) {$\varrho_r$};
  
\end{scope}

\end{tikzpicture}
\caption{Cylinder with non-contractible loop separating left and right regions with areas $\varrho_l$ and $\varrho_r$.}
\label{fig:cylinder1noncontractWL}
\end{figure}
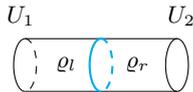

Now let us add a non-contractible loop, that is, a loop wrapping around the cylinder; see Figure \ref{fig:cylinder1noncontractWL}. In this case, the cylinder is split into left and right regions of areas $\varrho_l$ and $\varrho_r$, respectively. The amplitude, or partition function, depends on the boundary holonomies, $U_1$ and $U_2$, and on the representation $\gamma$ of the loop:
\be
Z(\varrho_l,\varrho_r;\gamma,U_1,U_2) \ = \ \sum\limits_{\alpha,\beta} N_{\alpha\gamma}^{\beta} e^{-\left(C_2(\alpha)\varrho_l +C_2(\beta)\varrho_r\right)/2}\chi_\alpha(U_1)\chi_\beta(U_2)\,.
\ee
In the representation basis, the components of the state are 
\be
C_{\alpha\beta} \ = \ N_{\alpha\gamma}^{\beta} e^{-\left(C_2(\alpha)\varrho_l +C_2(\beta)\varrho_r\right)/2}.
\ee
We see that, although the coefficient matrix of such a state is not diagonal, in general, its components are located at a finite distance from the diagonal, controlled by $\gamma$. As expected, for trivial $\gamma$, one returns to the diagonal result of the empty cylinder \eqref{eq:Zcylinder}.

In order to compute the entropy, we write the reduced density operator, say, of the left subsystem. In the representation basis, we find
\begin{equation}\label{eq:reduced_density_mtrx_nc_loop}
    \rho_l = \mathcal{N}^{-1}\sum_{\alpha,\alpha',\beta} N_{\alpha\gamma}^\beta\, N_{\beta\gamma^\ast}^{\alpha'}\,e^{-(C_2(\alpha)+C_2(\alpha'))\varrho_l/2-C_2(\beta)\varrho_r} \ket{\chi_\alpha}\bra{\chi_{\alpha'}}\,.
\end{equation}
Here $\mathcal{N}$ is the normalization factor, given by
\begin{equation}
\label{eq:ZofT2w2WLoops0}
    \mathcal{N} = \sum_{\alpha,\beta} (N_{\alpha\gamma}^\beta)^2\,e^{-C_2(\alpha)\varrho_l - C_2(\beta)\varrho_r}\,,
\end{equation}
where we used that $N_{\beta\gamma^\ast}^\alpha=N_{\alpha^\ast\gamma^\ast}^{\beta^\ast} = N_{\alpha\gamma}^\beta$.

Computation of the entropy using the replica trick requires analytical expression for the contraction of $2n$ multiplicity coefficients, weighted by exponential factors, which is probably unknown. Instead, a valid approximation can be obtained using the standard von Neumman formula,
\begin{equation}
    S = -\Tr(\rho_l\ln\rho_l) = -\Tr(\rho_r\ln\rho_r)\,,
\end{equation}
at least for sufficiently large values of the total area. The presence of exponential factors allows truncating the representation basis in order to obtain a finite matrix that can be diagonalized numerically. This enables us to perform a consistent numerical analysis, as presented below.

\subsubsection{Entanglement entropy analysis for one non-contractible loop}

For a Wilson loop wrapping the cylinder, we find some similarities compared to the contractible case analyzed in Section~\ref{subsec:EE1CWL}. Using the ratio between left and total areas $\varrho_l/(\varrho_l+\varrho_r)$ to characterize the system, when the gauge group is $SU(2)$, we find that:
\begin{itemize}
    \item there are ratios for which the entropy asymptotes to a non-zero value in the large area limit;
    \item for a loop colored with irrep $\gamma$, there are $\gamma$ of such fractions, determined by \eqref{eq:rnCasimir}; 
\end{itemize}
Now, in contrast to the state containing a contractible loop,
\begin{itemize}
    \item the exchange between the two areas involved, $\varrho_l$ and $\varrho_r$, represents a symmetry of the system and, therefore, also of the entropy;
    \item for the optimal ratios, the entropy asymptotes to $\log 2$ in the large area limit, regardless of the representation of the loop.
\end{itemize}

Let us further explain these statements. First, note that the normalization factor \eqref{eq:ZofT2w2WLoops0} has a structure that is similar to that of \eqref{eq:Nfor1CWL}. The first difference between them is the presence of a single sum over $\beta$ in the former, while the latter includes two sums, over $\beta_1$ and $\beta_2$. This difference, however, does not change the analysis of the ``optimal" ratios. Note that the analysis for \eqref{eq:Nfor1CWL} involves choosing representations with the lightest Casimirs, such that the minimization of $C_2(\beta_1)$ automatically minimizes $C_2(\beta_2)$. This implies that, in the end of the day, relevant contributions arise only for $\beta_1=\beta_2$. For this reason, the analysis outlined for \eqref{eq:Nfor1CWL} extends entirely to the \eqref{eq:ZofT2w2WLoops0} case. Thus, \eqref{eq:rnCasimir} appropriately captures the optimal ratios of the current setup.

Moreover, note that the procedure of choosing the lightest of the Casimirs also renders $\rho_l$ \eqref{eq:reduced_density_mtrx_nc_loop} diagonal in the large area limit: for a fixed $\beta$, minimization of $C_2(\alpha)$ automatically minimizes $C_2(\alpha')$ and selects $\alpha'=\alpha$. This allows us to repeat the analysis of entropy in the large area limit along the lines of Section~\ref{subsec:EE1CWL}.

We find that, while the ratios are the same, the value of the entropy in the asymptotic limits differs. This happens due to the second difference between \eqref{eq:ZofT2w2WLoops0} and \eqref{eq:Nfor1CWL}: the absence of factors of dimension in the former. Due to this fact, whenever we reach an optimal ratio, two elements will contribute to the entropy, but this time $h_n(\gamma)$, which in the contractible case was given by \eqref{eq:hn1CWL}, will simply be equal to $1/2$. Therefore, the entropy will always asymptote to $\log2$ for such states.

\begin{figure}[h!]
    \centering
\begin{subfigure}{0.45\textwidth}
\centering
    \includegraphics[width=7cm]{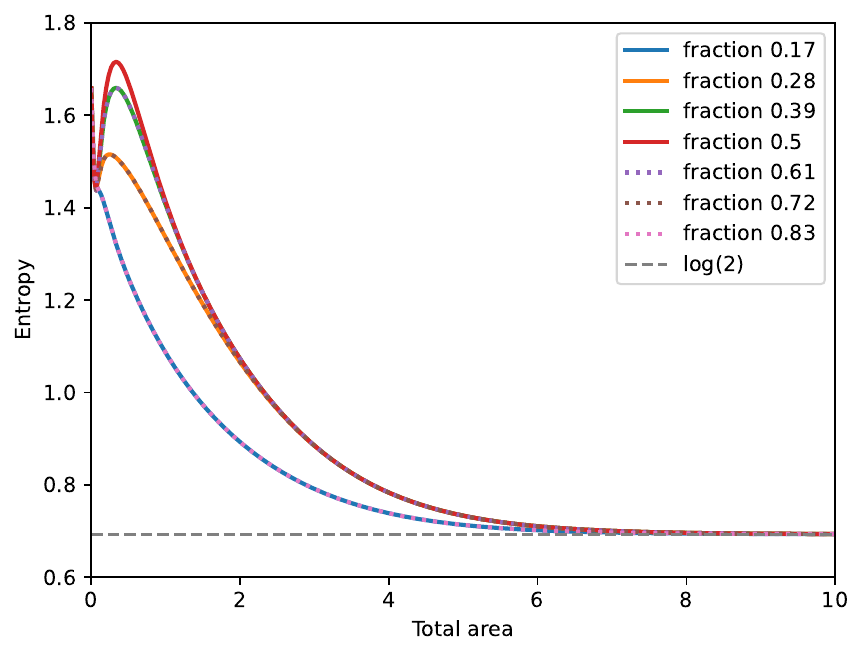}
\subcaption{}
\label{subfig:CylNonTriv1WL}
\end{subfigure}
\hfill
\begin{subfigure}{0.45\textwidth}
\centering
    \includegraphics[width=7cm]{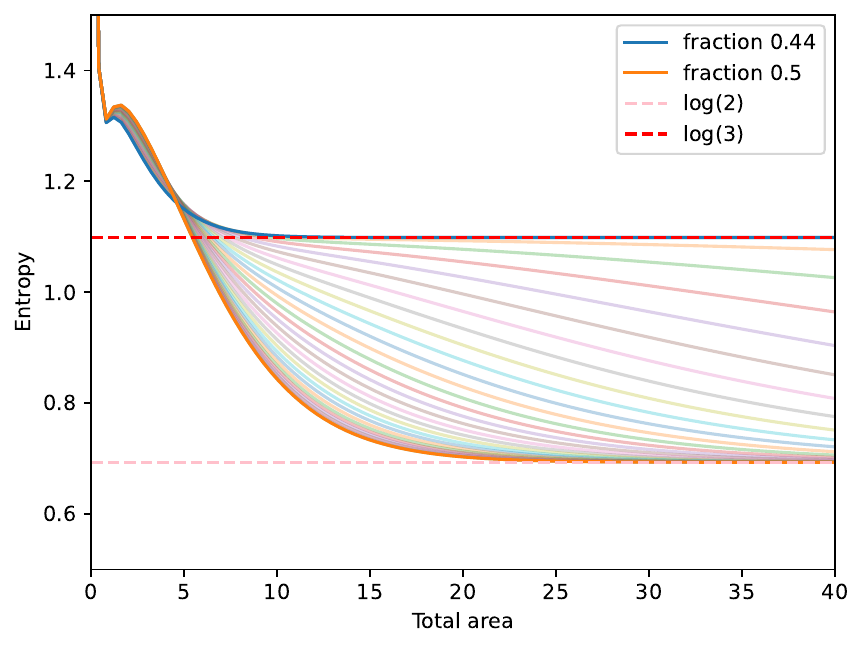}
\subcaption{}
\label{subfig:CylNonTriv1WLfixedarea}
\end{subfigure}
\caption{Entropy as a function of the total area using different fractions between left and right areas (\subref{subfig:CylNonTriv1WL}), and entropy as a function of the left area holding the total area $\varrho_l+\rho_r = 20$ fixed for different Wilson loop $SU(3)$ representations (\subref{subfig:CylNonTriv1WLfixedarea}).}
    \label{fig:CylNonTriv1WL}
\end{figure}

In Figure~\ref{subfig:CylNonTriv1WL} we present the case of $\gamma=[7,0]$ as an example. The asymptotic limit is equal to $\log2$ for all of them, as claimed. We choose dashed curves to illustrate fractions that are greater than $1/2$ in order to highlight the ($\varrho_l\leftrightarrow\varrho_r$)-symmetry of the entropy.

Finally, since there is a direct parallel between the analysis of peaks (or, equivalently, of the optimal ratios) in the present setup and that of the contractible loop, it is natural to examine the case in which the gauge group is $SU(3)$. For the representation $[2,1,0]$, we expect two central peaks located at $\varrho_l/(\varrho_l+\varrho_r)=4/9$ and $5/9$, corresponding to the two central maxima of the $[2,1,0]$ curve shown in Figure~\ref{subfig:cylCWL_su3_peaksb}. In Figure~\ref{subfig:CylNonTriv1WLfixedarea}, we display the entropy as a function of the total area for this representation, considering ratios in the interval between these two values.

At $4/9\simeq 0.44$, the entropy approaches the asymptotic value $\log 3$. As the ratio increases, the asymptotic limit of the curves descends toward $\log 2$, with the $0.5$ curve reaching $\log 2$ faster than the others. For ratios between $0.5$ and $5/9\simeq 0.56$, the curves ascend eventually changing the asymptotic value to $\log 3$ at the $5/9$ fraction. Thus, the lower and upper bounds coincide with those in Figure~\ref{subfig:cylCWL_su3_b}; however, in the present case they attain the exact value $\log 3$ at the $4/9$ and $5/9$ fractions.

\section{Configurations with Wilson lines}
\label{sec:WilsonLines}

In this section we will enrich the spatial topology including open Wilson lines ending on the boundaries. In the cell decomposition framework, the inclusion of Wilson line operators is accomplished via the insertion of the respective representation matrices along appropriate links of the lattice~\cite{Witten:1991we}, instead of closed loop characters. For instance, suppose $V$ is an edge of the cell decomposition of $\mathcal{M}$ along which a Wilson line is supported. We shall denote it as 
\begin{equation}
    \gamma(V)_{ij}\,,
\end{equation}
where $\gamma$ stands for its representation and indices $i$ and $j$ are representation indices.
Wilson lines are natural partition cuts and the partition function integrates the inserted operator over the holonomy $V$ associated with the cut.

The addition of Wilson line endpoints to the boundaries enhances the corresponding Hilbert spaces. They are no longer spanned by the gauge invariant basis of representations $|\chi_\gamma\rangle$, but by a more general basis of matrix elements $|\gamma;i,j\rangle$, so that
\be
\gamma(V)_{ij} \ = \ \langle \gamma;i,j|V\rangle\,.
\ee

In the following, we shall apply this framework by considering configurations with several Wilson lines anchored on the same or opposite boundaries of a cylinder. We will see that, as in the case of Wilson loops, configurations with open lines can also lead to nonvanishing entropies in the infinite area limit.


\subsection{Wilson lines anchored on the same boundary of a cylinder}
\label{sec:UUlines}

\begin{figure}[h!]
\centering
\begin{tikzpicture}[scale=1, every node/.style={font=\small}]

\begin{scope}[xshift=-0.5cm]

  \draw[dashed] (-1,1.35) arc[start angle=90, end angle=-90,x radius=0.15cm, y radius=0.35cm];
  \draw (-1,0.65) arc[start angle=270, end angle=90,x radius=0.15cm, y radius=0.35cm];
  \draw (1,1) ellipse (0.15 and 0.35);
  \draw (-1,1.35) -- (1,1.35);
  \draw (-1,0.65) -- (1,0.65);
  
  \draw[thick, color=cyan] (-1.12,0.8) arc [x radius=0.4, y radius=0.2, start angle=270, end angle=450];
  \draw[thick, color=orange] (0.87,1.2) arc [x radius=0.4, y radius=0.2, start angle=90, end angle=270];

  \node[left] at (-0.75,1.65) {$U_1$};
  \node[right] at (0.75,1.65) {$U_2$};
  \node[left] at (-1.15,1.2) {$i$};
  \node[left] at (-1.15,0.7) {$i'$};
  \node[right] at (1.15,1.2) {$j'$};
  \node[right] at (1.15,0.7) {$j$};
\end{scope}

\draw[->, thick] (1.75, 1) -- (2.5, 1);

\begin{scope}[xshift=4cm]
  \draw[thick, color=cyan] (0,1.025) arc [x radius=0.5, y radius=0.35, start angle=270, end angle=450];
  \draw[thick, color=orange] (2,1.73) arc [x radius=0.5, y radius=0.35, start angle=90, end angle=270];
  \draw (0,1) -- ++(0,0.75) -- ++(2,0) -- ++(0,-1.5) -- ++(-2,0);
  \draw (0,0.25) -- (0,1.25) ;

  \node[left] at (0,1.4) {$U_{1i}$};
  \node[above] at (1,1.75) {$W^{-1}$};
  \node[below] at (1,0.2) {$W$};
  \node[right] at (2,1.4) {$U_{2i}$};
  \node[right] at (2,0.6) {$U_{2o}$};
  \node[right] at (0.4, 1.2) {$V_1$};
  \node[left] at (1.65, 1.2) {$V_2$};
  \node[left] at (0,1.75) {$i$};
  \node[left] at (0,1) {$i'$};
  \node[right] at (2,1.75) {$j'$};
  \node[right] at (2,1) {$j$};
  \node[left] at (0,0.6) {$U_{1o}$};
  \node at (0.25,1.4) {$\varrho_{1}$};
  \node at (1,0.6) {$\varrho_{e}$};
  \node at (1.75,1.4) {$\varrho_{2}$};
\end{scope}

\end{tikzpicture}
\caption{Cell decomposition of the cylinder with two Wilson lines, each anchored on a boundary.}
\label{fig:cylindertwolines_diff_boundary}
\end{figure}
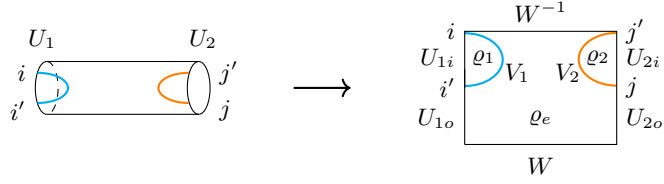

Let us insert a pair of Wilson lines, each anchored on a different boundary in the cylinder topology. The corresponding cell decomposition consists of three plaquettes; see Figure \ref{fig:cylindertwolines_diff_boundary}. We denote the inner areas enclosed by each line as $\varrho_1$ and $\varrho_2$, while the area external to the lines is $\varrho_e$. 

The wavefunction, in this case, will depend not only on the boundary holonomies, which the line insertion splits into $U_{1i}$, $U_{1o}$ and $U_{2i}$, $U_{2o}$, but also on $\gamma_1(V_1)_{ii'}$ and $\gamma_2(V_2)_{jj'}$, with indices $i,i'$ and $j,j'$ labeling the internal state of two pairs of particles on a circle in the Euclidean preparation picture. The density matrix will be a function of the full set of data: \footnote{In this and the following sections, pairs of indices separated by a semicolon label the endpoints of the same Wilson line.}
\begin{align}
\label{eq:UUsetup}
\begin{split}
    \psi(U_{1i},U_{2i}, U_{1o}, U_{2o}; \gamma_1,\gamma_2)_{ii';jj'} = \int &dV_1 dV_2 dW  \Bigg[ \bigg( \sum_\alpha d_\alpha\, e^{-C_2(\alpha)\varrho_{1}/2}\,\chi_\alpha(U_{1i}V_1)\bigg)\\
    &\times\bigg(\sum_\beta d_\beta\, e^{-C_2(\beta)\varrho_e/2}\,\chi_\beta(U_{1o}V_{1}^{-1}W^{-1}V_2^{-1}U_{2o}W)\bigg)\\
    &\times \bigg(\sum_{\sigma} d_{\sigma}\, e^{-C_2(\sigma)\varrho_{2}/2}\,\chi_{\sigma}(U_{2i}V_2)\bigg)\,\gamma_1(V_1)_{ii'}\,\gamma_2(V_2)_{jj'}\Bigg]
\end{split}
\end{align}
Identifying the edges to form the cylinder, that is, integrating over $W$, gives
\begin{align}
\begin{split}
    \psi(U_{1i},U_{2i}, U_{1o}, U_{2o}&; \gamma_1,\gamma_2)_{ii';jj'} = \sum_{\alpha,\beta,\sigma} d_\alpha d_{\sigma}\, e^{-(C_2(\alpha)\varrho_1 + C_2(\beta)\varrho_e + C_2(\sigma)\varrho_2)/2}\\
    &\times \int dV_1\ dV_2\ \chi_\alpha(U_{1i}V_1)\chi_{\beta}(U_{1o}V_1^{-1})\chi_\beta(V_2^{-1}U_{2o})\chi_{\sigma}(U_{2i}V_2)\gamma_1(V_1)_{ii'}\,\gamma_2(V_2)_{jj'}\,
\end{split}
\end{align}

Let us trace out one half of the system~(\ref{eq:UUsetup}). For the reduced density matrix of the left subsystem, labeled by holonomy $U_1$ and Wilson line with indices $i$ and $i'$, one obtains
\begin{align}
\label{eq:DM_2UUlines}
\begin{split}
    \rho(U_{1i},U_{3i},U_{1o},U_{3o};\gamma_1,\gamma_2)_{ii';kk'} \ = \  \mathcal{N}^{-1} \sum_{\alpha,\beta,\alpha'} d_\alpha d_{\alpha'}\, e^{-C_2(\beta)\varrho_e-(C_2(\alpha) + C_2(\alpha'))\varrho_1/2} \mathcal{D}_\beta(\gamma_2,2\varrho_2&)\\
    \times \int dV\ dW\ \chi_\alpha(VU_{1i})\chi_{\beta}(V^{-1}U_{1o})\chi_\beta(U_{3o}^{-1}W)\chi_{\alpha'}(U_{3i}^{-1}W^{-1})\gamma_1(V)_{ii'}\,\gamma_1(W^{-1})_{k'k}&\,,
\end{split}
\end{align}
where the normalization factor is the partition function of a torus with a pair of closed Wilson loops, cf.~(\ref{eq:Nfor1CWL}):
\be
\label{eq:torus2loops}
\mathcal{N} \ = \ \Tr\rho(\gamma_1,\gamma_2) \ = \ \sum_{\beta}\mathcal{D}_\beta(\gamma_1,2\varrho_1)\mathcal{D}_\beta(\gamma_2,2\varrho_2)\, e^{- C_2(\beta)\varrho_e}\,.
\ee

Expression~(\ref{eq:DM_2UUlines}) has a straightforward generalization to $\rho^n$:
\begin{align}
\begin{split}
    \rho^n(U_{1i},U_{3i},U_{1o},U_{3o};\gamma_1,\gamma_2)_{ii',kk'}  =   \mathcal{N}^{-n} \sum_{\alpha,\beta,\alpha'} d_\alpha d_{\alpha'}\, e^{-nC_2(\beta)\varrho_e-(C_2(\alpha)\varrho_1 + C_2(\alpha')\varrho_1/2} \left(\mathcal{D}_\beta(\gamma_2,2\varrho_2)\right)^n\\
    \times\left(\mathcal{D}_\beta(\gamma_1,2\varrho_1)\right)^{n-1} \int dV\ dW\ \chi_\alpha(VU_{1i})\chi_{\beta}(V^{-1}U_{1o})\chi_\beta(U_{3o}^{-1}W)\chi_{\alpha'}(U_{3i}^{-1}W^{-1})\gamma_1(V)_{ii'}\,\gamma_1(W^{-1})_{k'k}&\,,
\end{split}
\end{align}
Note that just by comparison with~(\ref{eq:torus2loops}) the trace of the above can be cast as
\be
\Tr \rho^n(\gamma_1,\gamma_2) \ = \ \sum_\beta h_\beta(\gamma_1,\gamma_2;\varrho_1,\varrho_2, \varrho_e)^n\,,
\ee
so that the entanglement entropy of the left half of the system is
\begin{equation}
\label{eq:2UULineEntropy}
    \boxed{S = -\sum_\beta h_\beta\log h_\beta\,, \qquad h_\beta(\gamma_1,\gamma_2,\varrho_1,\varrho_2, \varrho_e) = \dfrac{1}{\mathcal{N}}\,{\mathcal{D}_\beta(\gamma_1,2\varrho_1) \mathcal{D}_\beta(\gamma_2, 2\varrho_2) e^{-C_2(\beta)\varrho_e}}\,,}
\end{equation}
where $\mathcal{N}$ is the same as in~\eqref{eq:torus2loops}.

Finally, we can also consider a somewhat intermediate scenario between lines and loops. Namely, let us assume that only one cylinder boundary has a line attached. Although the entropy is independent of which side is traced out, if we trace out the side with the line, the reduced density matrix will essentially be the state shown in Figure~\ref{fig:cylinder1contractWL}.  The entropy may then be obtained by taking one of the lines above to be in the trivial representation:
\begin{equation}\label{eq:S1linecyl}
    \boxed{S = -\sum_\beta h_\beta\log h_\beta\,, \quad h_\beta(\gamma,\varrho_i,\varrho_e) = \dfrac{1}{\mathcal{N}}\,{\mathcal{D}_\beta(\gamma,2\varrho_1) e^{-C_2(\beta)\varrho_e}}\,, \quad \mathcal{N} = \sum_{\alpha}{\mathcal{D}_\alpha(\gamma,2\varrho_1) e^{-C_2(\alpha)\varrho_e}}\,.}
\end{equation}

\subsubsection{Entanglement entropy analysis}

Note that the formula for the coefficients in the expression~(\ref{eq:2UULineEntropy}) is a variation of formula~(\ref{eq:1ConLoopEntropy}), which looks like a ``square root" of the configuration with two loops in representations $\gamma_1$ and $\gamma_2$, so the analysis is very similar to the case of a cylinder with contractible loops. To compare, we quote here the normalization factor~\eqref{eq:Nfor1CWL}, which had the structure
\be
\mathcal{N} = \sum_\alpha \left(\prod^n_{k=1} \mathcal{D}_\alpha(\gamma_k, \varrho_k)^2\right)e^{-C_2(\alpha)\varrho_e}\,.
\ee
In the large area limit, this fact implies the same hierarchy of the exponential terms and the same optimal ratios as in the case of a single loop. What will change are the values of the entropy because the factors of dimensions in $h_\beta$ will not be squared. For $SU(2)$ this results in the following values, cf.~(\ref{eq:hn1CWL}):
\be\label{eq:hn1WLB}
h_n(\gamma)= \frac{1}{1+\frac{d_n d_{\gamma-n+1}}{d_{\gamma-n}d_{n-1}}}\,.
\ee

\begin{figure}[h!]
\centering
\begin{subfigure}{0.45\textwidth}
\centering
    \includegraphics[width=7cm]{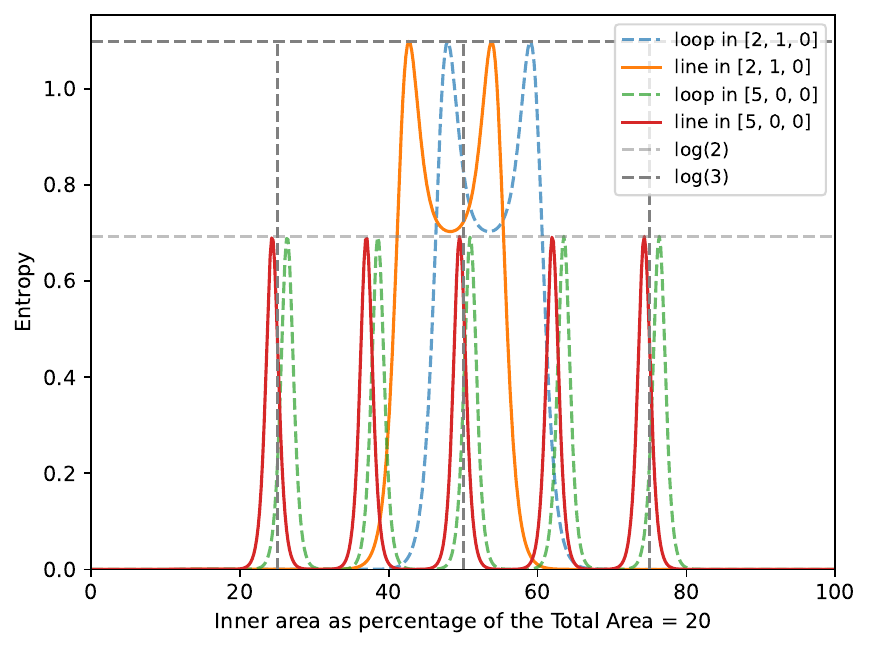}
\subcaption{}
\label{subfig:cyl1WLsu2gamma3a}
\end{subfigure}
\hfill
\begin{subfigure}{0.45\textwidth}
\centering
    \includegraphics[width=7cm]{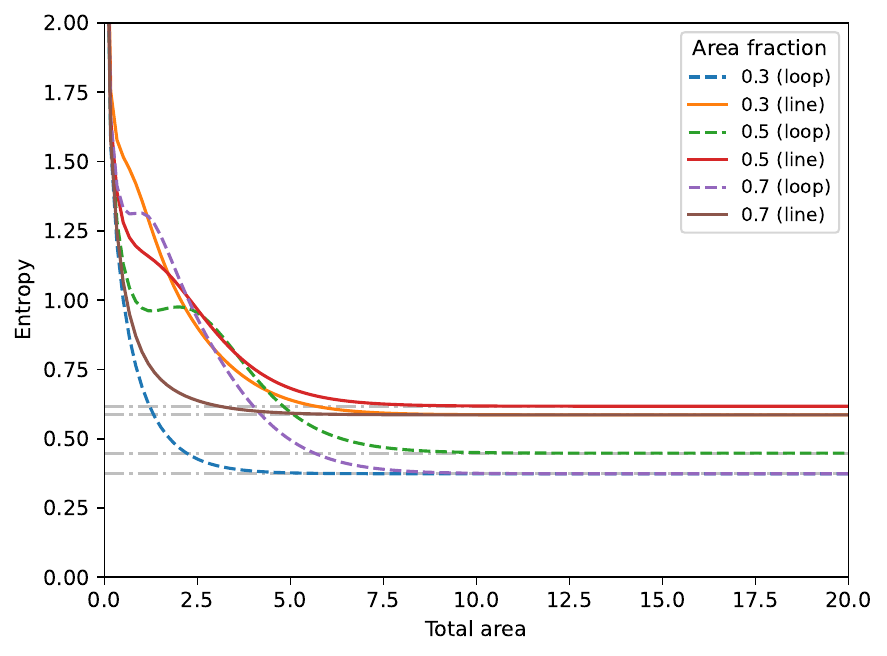}
\subcaption{}
\label{subfig:cyl1WLsu2gamma3b}
\end{subfigure}
\caption{Entropy~\eqref{eq:S1linecyl} of Wilson lines anchored on one of the cylinder's boundaries as a function of the area fraction for irreps $[5,0,0]$ and $[2,1,0]$ of $SU(3)$ (\subref{subfig:cyl1WLsu2gamma3a}) and as a function of the total area $\varrho_t$ in irrep $[3,0]$ of $SU(2)$ (\subref{subfig:cyl1WLsu2gamma3b}). Dashed curves show similar qualitative results for the case of Wilson loops.}
    \label{fig:cyl1WLsu2gamma3}
\end{figure}

To illustrate the similarity and the difference between the two cases, we plot the entropies~\eqref{eq:S1linecyl} and~(\ref{eq:1ConLoopEntropy}) for Wilson lines and Wilson loops with compatible representations and areas in Figure \ref{fig:cyl1WLsu2gamma3}. For Wilson lines, Figure~\ref{subfig:cyl1WLsu2gamma3a} and similar plots  show that the entropy profiles have similar shapes, but there is a consistent shift of the entropy peaks towards smaller area fractions. In the infinite area limit, the entropy peaks occur at the same area fractions, as illustrated in Figure~\ref{subfig:cyl1WLsu2gamma3b}. The entropy of a single Wilson line is higher in the infinite area limit, as expected from the perspective that the line is only half of the loop.

\subsection{Wilson lines anchored on opposite boundaries of a cylinder}
\label{sec:IIlines}

\begin{figure}[h!]
\centering
\begin{tikzpicture}[scale=1.2, every node/.style={font=\small}]

\begin{scope}[xshift=-0.5cm]
    \draw[dashed] (-1,-0.4) arc[start angle=90, end angle=-90,x radius=0.15cm, y radius=0.35cm];
  \draw (-1,-1.1) arc[start angle=270, end angle=90,x radius=0.15cm, y radius=0.35cm];
  \draw (1,-0.75) ellipse (0.15 and 0.35);
  \draw (-1,-0.4) -- (1,-0.4);
  \draw (-1,-1.1) -- (1,-1.1);
  
  \draw[cyan, thick] (-1.15,-0.65) -- (0.85,-0.65);
    \draw[orange, thick] (-1.15,-0.85) -- (0.85,-0.85);
  
  \node[left] at (-1.15,-0.34) {$U_1$};
  \node[right] at (1.15,-0.34) {$U_2$};
  \node[left] at (-1.15,-0.61) {$i$};
  \node[right] at (0.83,-0.64) {$i'$};
    \node[left] at (-1.15,-0.88) {$j$};
  \node[right] at (0.83,-0.86) {$j'$};
\end{scope}

\draw[->, thick] (1.5, -0.75) -- (2.5, -0.75);

\begin{scope}[xshift=3.5cm]
  \draw (0,0) rectangle (3,-0.9);
  \draw (0,-1.1) rectangle (3,-2);
  
  \node[above] at (1.5,0) {$W_1^{-1}$};
  \node[below] at (1.5,-1.05) {$W_2^{-1}$};
  \node[above] at (1.5,-0.95) {$W_2$};
  \node[below] at (1.5,-2.15) {$W_1$};
  
  \node[right] at (1.25,-0.25) {$\varrho_1$};
  \node[right] at (1.25,-1.75) {$\varrho_2$};
  
  \draw[cyan, thick] (0,-1) -- (3,-1);
  \draw[orange, thick] (0,-2.1) -- (3,-2.1);
  
  \node[left] at (0,-0.5) {$U_{1t}$};
  \node[left] at (0,-1) {$i$};
  \node[right] at (3.05,-0.5) {$U_{2t}$};
  \node[right] at (3.05,-1) {$i'$};
  \node[left] at (0,-1.5) {$U_{1b}$};
  \node[left] at (0,-2.1) {$j$};
  \node[right] at (3.05,-1.5) {$U_{2b}$};
  \node[right] at (3.05,-2.1) {$j'$};
\end{scope}

\end{tikzpicture}
\caption{Cylinder with a pair of parallel Wilson lines and its cell decomposition.}
\label{fig:cylinderparallel}
\end{figure}
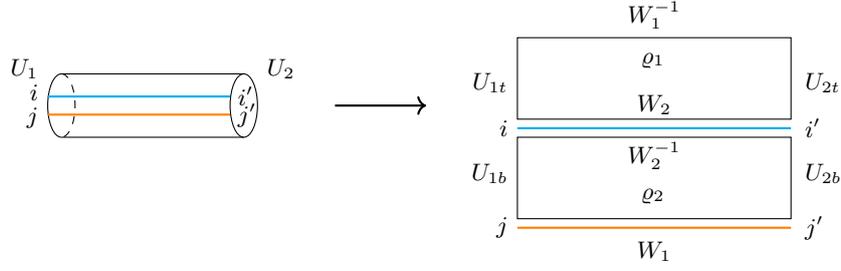

If one inserts two nonintersecting lines stretched between the opposite boundaries of the cylinder, the following cell decomposition can be employed. We denote the respective cell areas as $\varrho_1$ and $\varrho_2$, see Figure~\ref{fig:cylinderparallel}. The corresponding wavefunction in the holonomy basis is written as:
\begin{align}
\label{IIsetupState}
\begin{split}
    \psi(U_{1t},U_{2t}, U_{1b}, U_{2b}&; \gamma_1,\gamma_2)_{ii',jj'} = \sum_{\alpha,\beta} d_\alpha d_\beta\,e^{-(C_2(\alpha)\varrho_1+C_2(\beta)\varrho_2)/2}\\
    &\times\int dW_1\ dW_2\ \chi_\alpha(U_{1t}W_2U_{2t}W_1^{-1})\,\chi_\beta(U_{1b}W_1U_{2b}W_2^{-1})\,\gamma_1(W_1)_{ii'}\,\gamma_2(W_2)_{jj'}
\end{split}
\end{align}
It's not hard to see that by making another copy of the cylinder and gluing both together to form a density operator, one obtains the same result, except for doubled area factors in the exponential.

As before, we compute the entropy associated with the left/right subsystems. By the symmetry of the state matrix~(\ref{IIsetupState}), the left/right reduced density matrix and all its powers are given by essentially the same expression, with the only difference being the amount of area covered by the cylinder. Similarly to the calculation in Section~\ref{sec:UUlines}, one can find the trace of the $n$-th power of the reduced density matrix of the left subsystem as a sum over weighted ``probabilities'':
\begin{equation}
    \Tr \rho^n \ = \ \sum\limits_{\alpha,\beta} N^{\alpha}_{\gamma_1\beta} N^{\beta}_{\alpha\gamma_2} \,p_{\alpha\beta}^n\,, \qquad p_{\alpha\beta} \ = \ \frac{1}{\cal{N}}\,e^{-(C_2(\alpha)\varrho_1 + C_2(\beta)\varrho_2)}\,,
\end{equation}
where the normalization factor ${\cal N}$ is the partition of the torus $T^2$ with a pair of Wilson loops along the longitude of the torus:\footnote{This partition function does not distinguish between the longitude and the meridian and can be compared with the expression previously encountered in~(\ref{eq:ZofT2w2WLoops0}).} 
\be
\label{eq:ZofT2w2WLoops}
{\cal N} \ = \ Z(T^2;\gamma_1,\gamma_2) \ = \ \sum_{\lambda,\sigma}N^{\lambda}_{\gamma_1\sigma} N^{\sigma}_{\lambda\gamma_2}e^{-(C_2(\lambda)\varrho_1 + C_2(\sigma)\varrho_2)}\,.
\ee
The entropy is then
\begin{equation}\label{eq:entropy2parallelWLs}
    \boxed{\quad S = -\sum_{\alpha,\beta} N^{\alpha}_{\gamma_1\beta}  N^{\beta}_{\alpha\gamma_2} \,p_{\alpha,\beta}\log p_{\alpha,\beta}\,. \quad}
\end{equation}

\subsubsection{Entanglement entropy analysis}

As in previous examples, we find particular configurations for which the entropy does not vanish in the large area limit, which we characterize by the ratio between $\varrho_1$ and the total area $\rho_t=\varrho_1+\varrho_2$: $r\equiv \varrho_1/(\varrho_1+\varrho_2)$. 

Specializing to the case of $SU(2)$, we see that representations $\gamma_1$ and $\gamma_2$ are compatible if and only if they sum to an even number. Assuming, for simplicity, that $\gamma_2\geq\gamma_1$, we see that the leading contribution to the partition function \eqref{eq:ZofT2w2WLoops} is dominated by
\begin{align}
\label{eq:AsymptEntropyII}
    \begin{split}
        Z & \simeq \sum_{n=0}^{\gamma_1}\exp\left\{-\left[C_2\left(\frac{\gamma_2+\gamma_1}{2}-n\right)\varrho_1+C_2\left(\frac{\gamma_2-\gamma_1}{2}+n\right)\varrho_2\right]\right\}.
    \end{split}
\end{align}
In particular, the lowest values of the Casimirs are attained on the following representations: $\lambda=\frac{\gamma_2+\gamma_1}{2}-n$ and $\sigma=\frac{\gamma_2-\gamma_1}{2}+n$ for $n=0,\dots,\gamma_1$.

Writing $\varrho_2=x\varrho_1$, we see that optimal ratios appear when exactly two terms in $Z$ simultaneously dominate. In particular, whenever there are two dominating terms, they are neighbors in $n$. That is, they satisfy
\bee
C_2\left(\frac{\gamma_2-\gamma_1}{2}+n\right)x+C_2\left(\frac{\gamma_2+\gamma_1}{2}-n\right)=C_2\left(\frac{\gamma_2-\gamma_1}{2}+n+1\right)x+C_2\left(\frac{\gamma_2+\gamma_1}{2}+\gamma_1-n-1\right)\,.
\ee
Therefore, optimal fractions $r=x/(1+x)$ are given by
\begin{align}\label{eq:rngamma1gamma2}
\begin{split}
r_n(\gamma_1,\gamma_2)&=\frac{C_2(\frac{\gamma_2-\gamma_1}{2}+n-1)-C_2(\frac{\gamma_2-\gamma_1}{2}+n)}{C_2(\frac{\gamma_2+\gamma_1}{2}-n)-C_2(\frac{\gamma_2+\gamma_1}{2}-n+1)+C_2(\frac{\gamma_2-\gamma_1}{2}+n-1)-C_2(\frac{\gamma_2-\gamma_1}{2}+n)}\\
&=\frac{1+2n+\gamma_2-\gamma_1}{2\gamma_2+4}\,,
\end{split}
\end{align}
for $n=1,\dots,\gamma_1$. Note that these correspond to $\gamma_1$-many central peaks of the loop case \eqref{eq:rnCasimir} for representation $\gamma_2$. Moreover, for $\gamma_1=\gamma_2$ the result matches precisely~\eqref{eq:rnCasimir}. As in that case, there is a reflection symmetry in the distribution of ratios between $[0,1]$. 

At a particular fraction $r_n$, there are two asymptoticaly non-zero contributions $p_{\alpha\beta}$: one with $\alpha=\frac{\gamma_2+\gamma_1}{2}-n$, $\beta=\frac{\gamma_2-\gamma_1}{2}+n$ and another with $\alpha=\frac{\gamma_2+\gamma_1}{2}-n+1$, $\beta=\frac{\gamma_2-\gamma_1}{2}+n-1$. Each of them is equal to $1/2$, since there will be two contributing exponential terms from the normalization factor ${\cal N}$ in the denominator. Therefore, the entropy asymptotes to that of a Bell pair in the large area limit,
\be\label{eq:S2parallelWLs}
S=\log 2\,.
\ee
Note that the set of asymptotic density matrices constructed with parallel Wilson lines is different from the case of states with Wilson loops. These density matrices have much more complicated structure, cf.~(\ref{IIsetupState}), defined in a larger basis of representation matrices.

\begin{figure}[h!]
\centering
    \begin{subfigure}{0.45\textwidth}
\centering
    \includegraphics[width=7cm]{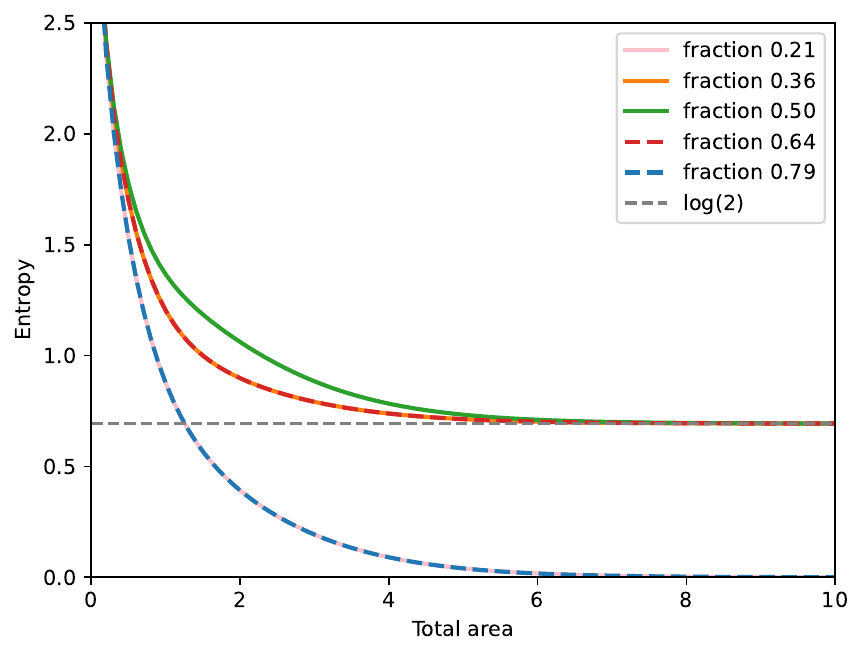}
\subcaption{}
\label{subfig:cyl2parallelWLsu2gamma3gamma7}
\end{subfigure}
\hfill
\begin{subfigure}{0.45\textwidth}
\centering
    \includegraphics[width=7cm]{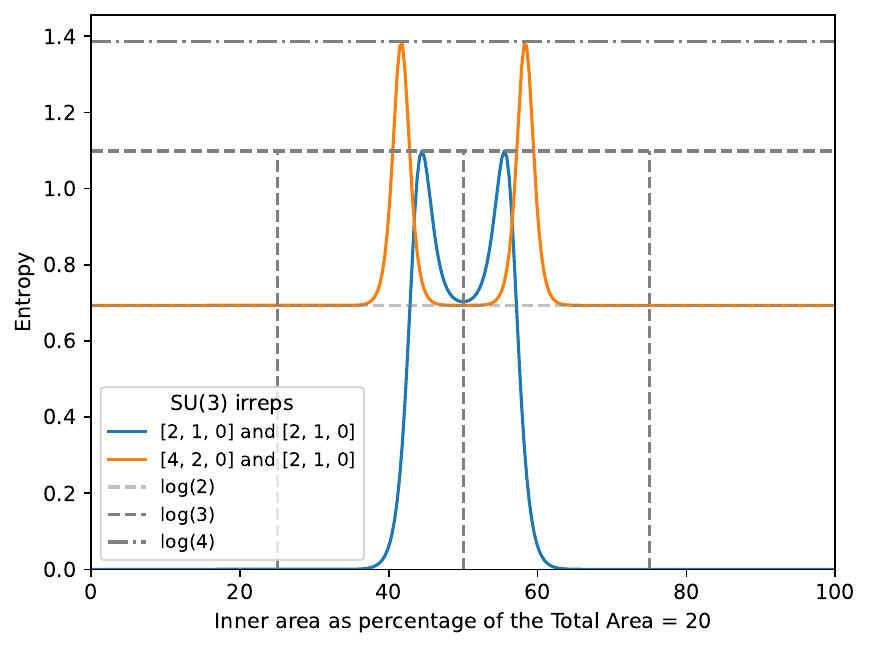}
\subcaption{}
\label{subfig:cyl2parallelWLsu3}
\end{subfigure}
\caption{(\subref{subfig:cyl2parallelWLsu2gamma3gamma7}) Entropy versus total area for a cylinder with two parallel lines in irreps $\gamma_1=[3,0]$ and $\gamma_2=[5,0]$ of $SU(2)$. Colored lines are obtained numerically while the gray horizontal line is the analytical result to the asymptotic limit. (\subref{subfig:cyl2parallelWLsu3}) Peaks of the entropy function for some choices of $SU(3)$ irreps.}
    \label{fig:cyl2parallelWL}
\end{figure}

To illustrate our results, in Figure \ref{subfig:cyl2parallelWLsu2gamma3gamma7} we consider the case $\gamma_1=[3,0]$ and $\gamma_2=[5,0]$. For this choice, \eqref{eq:rngamma1gamma2} yields three ratios with non-vanishing entropy in the large area limit, namely $5/14$, $1/2$ and $9/14$. As expected, the entropy associated with each of these fractions asymptotes to $\log 2$. These values precisely correspond to the three central peaks predicted by \eqref{eq:rnCasimir} for the representation $[5,0]$. 

The remaining candidate fractions, $3/14$ and $11/14$, are excluded once the second line in the irrep $[3,0]$ is introduced. This exclusion is visible in the plot. Moreover, complementary ratios produce coincident entropy curves, indicating that the entropy is insensitive to the direction from which the two lines approach each other.

For $SU(3)$ we restrict ourselves to examples giving $\log3$ and higher entropy in the infinite area limit. In this case, for $\gamma_1$ and $\gamma_2$ to be compatible representations, they should satisfy the ``triality'' condition; that is, the sum of the number of boxes of the corresponding diagrams must be a multiple of three.\footnote{For $SU(N)$ this is the $N$-ality condition that requires the sum of the number of boxes of the Young diagrams of $\gamma_1$ and $\gamma_2$ is an integer mutliple of $N$.} In order to reach the $\log3$ value, both representations must be genuine $SU(3)$ ones, so the minimal choice is $\gamma_1=\gamma_2=[2,1,0]$. In this case, one has the following dominant contribution to the partition function~(\ref{eq:ZofT2w2WLoops}):
\be
Z\ \simeq \ e^{-C_2(\scalebox{0.2}{\yng(2,1)})\varrho_2} + e^{-C_2(\scalebox{0.2}{\yng(1)})\varrho_1-C_2(\scalebox{0.2}{\yng(1)})\varrho_2 } + e^{-C_2(\scalebox{0.2}{\yng(1,1)})\varrho_1-C_2(\scalebox{0.2}{\yng(1,1)})\varrho_2 } + e^{-C_2(\scalebox{0.2}{\yng(2,1)})\varrho_1}\,.
\ee
As one can see, the two central terms are degenerate, which corresponds to a situation similar to the one described in~(\ref{eq:1CWLlog3}) for a contractible Wilson loop in the same representation. The optimal fractions are again $r_-=4/9$ and $r_+=5/9$. At these fractions, exactly three terms of the above expression contribute, and the entropy is $\log 3$. For $r_-< r<r_+$ the entropy is exactly $\log 2$ in the infinite area limit.

Let us consider another choice: $\gamma_1=[2,1,0]$ and $\gamma_2=[4,2,0]$. In this case, the dominant terms are
\begin{align}
\begin{split}
Z\ \simeq & \ e^{-C_2(\scalebox{0.2}{\yng(1)})\varrho_1-C_2(\scalebox{0.2}{\yng(3,1)})\varrho_2 } + e^{-C_2(\scalebox{0.2}{\yng(1,1)})\varrho_1-C_2(\scalebox{0.2}{\yng(3,2)})\varrho_2 } + 2e^{-C_2(\scalebox{0.2}{\yng(2,1)})\varrho_1-C_2(\scalebox{0.2}{\yng(2,1)})\varrho_2 } \\
& + e^{-C_2(\scalebox{0.2}{\yng(3,2)})\varrho_1-C_2(\scalebox{0.2}{\yng(1,1)})\varrho_2 } + e^{-C_2(\scalebox{0.2}{\yng(3,1)})\varrho_1-C_2(\scalebox{0.2}{\yng(1)})\varrho_2 }\,.
\end{split}
\end{align}
This example not only has two pairs of degenerate terms due to the equivalence of the Casimirs of conjugate representations, but also a case of nontrivial multiplicity. This leads to two possible combinations of maximal entropy for $r_-=5/12$ and $r_+=7/12$. The entropy at these points is $\log 4$, while for any other value of $r\in [0,1]$, it is $\log2$. We illustrate these $SU(3)$ examples in Figure~\ref{subfig:cyl2parallelWLsu3}. One may expect even higher values of the entropy for representations with higher multiplicities.

In comparison with the situation with closed loops, the entropy enhancement occurs since equation~(\ref{eq:entropy2parallelWLs}) has a different structure. In particular, $p_{\alpha\beta}$ do not have the exact meaning of probabilities unless the group is $SU(2)$ -- they only sum to one when weighted with the multiplicity factors. As a result, the entropy reaches $\log4$ value, even though only three terms contribute. Similarly, for $r_-<r<r_+$ the entropy is $\log2$ with only one contributing $p_{\alpha\beta}$. An alternative explanation of this effect is that the density matrices, in this case, live in a larger Hilbert space, augmented by the degrees of freedom associated with Wilson lines.


\section{Particles in the bath of gauge modes}
\label{sec:MixedParticles}

In this section, we will explore a different perspective on the configurations with Wilson lines. Endpoints of the Wilson lines can be understood as particles living in closed one-dimensional spaces. These particles may interact with each other via the interaction mediated by the gauge modes. Wilson lines ending at the same boundary, in this perspective, are pairs of entangled particles immersed in an external bath. Wilson lines connecting different boundaries, on the other hand, are particles in a mixed state exposed to the same bath. We would like to investigate how these naive entanglement properties of the particles change as a result of the interaction with the bath.


\subsection{Wilson line in a disk topology}
\label{sec:disklines}

Let us first consider a Wilson line in representation $\gamma$ added to the disk topology, as depicted in Figure~\ref{fig:diskoneline}. We would like to see how the particles, represented by the endpoints of the Wilson line, are correlated in the presence of the gauge modes. 

Generalizing equation~(\ref{eq:Zdiskpart}), the (unnormalized) reduced density matrix obtained by tracing out gauge variables $U_i$ will be given by
\be\label{eq:reddiskline}
    \rho(i,j;k,l) \ = \ \int dV dW \sum_{\alpha,\beta} d_\alpha\,d_\beta\,\chi_\alpha(V W^{-1})\,\chi_\beta(V^{-1}W)e^{-C_2(\alpha)\varrho_1 -C_2(\beta)\varrho_2 }\,\gamma_{ij}(V) \gamma_{kl}(W^{-1}) \,.
\ee
The integration over $V$ and $W$ in \eqref{eq:reddiskline} may then be performed using the expansion of the product of characters:
\begin{equation}
\chi_\alpha(VW^{-1})\chi_\beta((VW^{-1})^{-1})=\sum_\sigma N^\sigma_{\alpha\beta^*}\chi_\sigma(VW^{-1})\,,
\end{equation}
leading to
\begin{align}
\begin{split}
     \rho(i,j;k,l)
     &  =  \sum_{\alpha,\beta} d_\alpha\,d_\beta\,e^{-C_2(\alpha)\varrho_1 -C_2(\beta)\varrho_2 } \sum_\sigma N^{\sigma}_{\alpha\beta^*}\int dV\, \sigma_{pq}(V)\,\gamma_{ij}(V) \,\int dW\,  \sigma_{qp}(W^{-1})\,\gamma_{kl}(W^{-1}).
\end{split}
\end{align}
Writing the remaining character as $\chi_\sigma(VW^{-1})=\sigma_{pq}(V)\sigma_{qp}(W^{-1})$ and performing each of the integrations separately, e.g.
\begin{align}
\begin{split}
    \int dV\ \sigma_{pq}(V)\gamma_{ij}(V) = \int dV\ \sigma_{pq}(V)\,(\gamma_{ij}(V^{-1}))^{-1} &= \int dV\ \sigma_{pq}(V)\,(\gamma_{ij}(V^{-1}))^\dagger\\
    &= \int dV\ \sigma_{pq}(V)\,\gamma^*_{ji}(V^{-1})\\
    &= \dfrac{\delta_{\sigma,\gamma^*}}{d_{\gamma^*}}\delta_{pi}\delta_{qj}\,,
\end{split}
\end{align}
we arrive at
\begin{equation}
\label{eq:diskWL}
    \rho(i,j;k,l) = \frac{\delta_{il}\delta_{jk}}{(d_\gamma)^2}\sum_{\alpha,\beta} {d_\alpha d_\beta} \,N^{\beta}_{\alpha\gamma} \,  \,\exp\left(-(C_2(\alpha)\varrho_1 + C_2(\beta)\varrho_2)\right)\,.
\end{equation}
The sum in this expression is precisely the normalization factor, given by 
\begin{equation}
    \label{eq:ZofS2wWLoop}
     {\cal N} = \sum_{i,j}\rho(i,j;j,i)= \sum_{\alpha} d_\alpha^2\, {\cal D}_\alpha(\gamma,\rho_1)  \exp\left(-C_2(\alpha)\varrho_2\right)\,,
\end{equation}
which is nothing but the partition function $Z(S^2;\gamma)$ of a 2-sphere with the insertion of a closed loop.

The relevant part of the density matrix is just the tensor product of two identity matrices, each of dimension $d_\gamma$. Therefore, the entropy is
\begin{equation}
   \boxed{\quad S \ = \
   \log(d_\gamma^2)\,.\quad}
\end{equation}

The interpretation of this result is that the initial pair of particles connected by a Wilson line becomes completely separable when exposed to the bath of gauge degrees of freedom. Each of the particles assumes a maximally mixed state. We also note that, by being just the identity matrix, the particle's density matrices are gauge invariant.

\subsection{Pairs of Wilson lines in the cylinder topology}

\subsubsection{Lines anchored on the same boundary}
\label{sec:UUWilsonLines}

Let us now look at the entanglement properties of a similar pair of particles but with gauge modes in a mixed state. In other words, we anchor a Wilson line at one side of a cylinder in Figure~\ref{fig:cylindertwolines_diff_boundary}, say, the left side. We will consider a purified state~(\ref{eq:UUsetup}) that has a second Wilson line anchored on the opposite boundary, as in the figure. The reduced density matrix of the left Wilson line can be cast as
\begin{align}
\begin{split}
    \rho(\gamma_1,\gamma_2)_{ii';kk'} &= \sum_{\alpha,\beta,\lambda}\dfrac{d_\alpha d_\lambda}{(d_\beta)^2}\, e^{-(C_2(\alpha)\varrho_1 + C_2(\beta)\varrho_e + C_2(\lambda)\varrho_2)}\, I_{ii',kk'}\,,
\end{split}
\end{align}
where the integral over holonomies $I_{ii',kk'}$ can be evaluated as follows:
\begin{align}
\begin{split}
    I_{ii';kk'} &= \int dV_1\ dV_2\ dW_1\ dW_2\ \chi_\alpha(V_1W_1)\chi_\beta((V_1W_1)^{-1})\chi_\beta((V_2W_2)^{-1})\chi_\lambda(V_2W_2)\\
    &\quad\quad\times \gamma_1(V_1)_{ii'}\gamma_1(W_1)_{kk'} \chi_{\gamma_2}(V_2W_2) 
    = \dfrac{N_{\lambda\gamma_2}^\beta}{d_{\gamma_1}}\delta_{ki'} \dfrac{N^{\gamma_1^*}_{\alpha\beta^*}}{d_{\gamma_1}}\delta_{ik'}
\end{split}
\end{align}

Hence, we obtain the following reduced density matrix of the left Wilson line:
\begin{equation}
    \label{ZinUUconfig}
    \rho(\gamma_1,\gamma_2)_{ii';kk'} = \dfrac{\delta_{ik'}\delta_{ki'}}{(d_{\gamma_1})^2}\sum_{\alpha,\beta,\lambda}\dfrac{d_\alpha d_\lambda}{(d_\beta)^2}\, e^{-(C_2(\alpha)\varrho_1 + C_2(\beta)\varrho_e + C_2(\lambda)\varrho_2)}\, N^{\beta}_{\alpha\gamma_1}N_{\lambda\gamma_2}^\beta\,,
\end{equation}
where we used the fact that $N^{\gamma_1^*}_{\alpha\beta^*} = N^\beta_{\alpha\gamma_1}$. Note that the result can also be cast in terms of coefficients $\mathcal{D}_\beta(\gamma,\varrho)$~(\ref{eq:coeffC}) associated with a closed Wilson loop of area $\varrho$ carrying representation $\gamma$ as follows:
\be
\rho(\gamma_1,\gamma_2)_{ii';kk'} = \dfrac{\delta_{ik'}\delta_{ki'}}{(d_{\gamma_1})^2}\sum_{\beta}\mathcal{D}_\beta(\gamma_1,2\varrho_1)\mathcal{D}_\beta(\gamma_2,2\varrho_2)\, e^{- C_2(\beta)\varrho_e}\,.
\ee
As in the example of the disk~(\ref{eq:diskWL}), the result factorizes into two pieces: this time, the product of the torus partition function $Z(T^2;\gamma_1,\gamma_2)$, with two nonintersecting Wilson loops, and the same product of identity matrices for the particles representing each endpoint. The torus partition function~(\ref{eq:torus2loops}) is the normalization factor. The endpoints of the Wilson line are in a separable, maximally mixed state.

\subsubsection{Lines anchored on opposite boundaries}
\label{sec:IIWilsonLines}

Now we turn to state~\eqref{IIsetupState}, shown in Figure~\ref{fig:cylinderparallel}. This case is aimed at representing a pair of particles in a mixed state exposed to a mixed state of gauge modes. As in the previous example, we first trace out completely the gauge field degrees of freedom and the endpoints of the Wilson lines on the right hand side. For this we glue $U_{1t}$ with $U_{2t}$ and $U_{1b}$ with $U_{2b}$. The resulting partition function is labeled by the open ends of the Wilson lines:
\begin{align}
\begin{split}
    \rho(\gamma_1&,\gamma_2)_{ii';kk'} 
    = \sum_{\alpha,\beta} d_\alpha d_\beta\, e^{-(C_2(\alpha)\varrho_1 + C_2(\beta)\varrho_2)}\,\int dW_1dW_2\ \gamma_1(W_1)_{ii'}\,\gamma_2(W_2)_{kk'} \\
    &\hspace{5cm}\times \Bigg[\int dU_t\ dU_b\ \chi_\alpha(U_{t}W_2U_{t}^{-1}W_1^{-1})\,\chi_\beta(U_{b}W_1U_{b}^{-1}W_2^{-1})\Bigg]\\
    &= \sum_{\alpha,\beta} e^{-(C_2(\alpha)\varrho_1 + C_2(\beta)\varrho_2)}\, \int dW_1\ dW_2\ \chi_\alpha(W_2)\,\chi_{\alpha^*}(W_1)\,\chi_\beta(W_1)\,\chi_{\beta^*}(W_2)\, \gamma_1(W_1)_{ii'}\,\gamma_2(W_2)_{kk'}\,. 
\end{split}
\end{align}
The number of characters in this expression may be reduced by introducing the multiplicities:
\begin{align}
\begin{split}
    \rho(\gamma_1,\gamma_2)_{ii';kk'} &= \sum_{\alpha,\beta} e^{-(C_2(\alpha)\varrho_1 + C_2(\beta)\varrho_2)}\,\sum_{\sigma,\lambda} N^\sigma_{\alpha^*\beta} N^\lambda_{\alpha\beta^*} \int dW_1\ dW_2\ \chi_\sigma(W_1)\,\chi_\lambda(W_2)\, \gamma_1(W_1)_{ii'}\,\gamma_2(W_2)_{kk'}
\end{split}
\end{align}    
and each integration may be carried out using the orthogonality relation for matrix elements~(\ref{eq:orthogonality_mat_elem}).

The final expression for the density matrix of the pair of parallel Wilson lines becomes
\begin{align}
\label{eq:IIsetupDM}
\begin{split}
     \rho(\gamma_1,\gamma_2)_{ii';kk'} 
    \ = \ \frac{\delta_{ii'}\delta_{kk'}}{d_{\gamma_1}d_{\gamma_2}} \sum_{\alpha,\beta} e^{-(C_2(\alpha)\varrho_1 + C_2(\beta)\varrho_2)}\,  {N^{\alpha}_{\gamma_1\beta} N^{\beta}_{\alpha\gamma_2}}\,.
\end{split}
\end{align}
Again, the density matrix factorizes into the product of identity matrices for the endpoints of the Wilson lines and the torus partition function (here torus with the Wilson lines winding along the non-contractible cycle). This result has the same structure as the density matrices of endpoints in Sections~\ref{sec:disklines} and~\ref{sec:UUWilsonLines}. In either case, the particles on the same circle are separable and maximally mixed, and their density matrices are gauge invariant.

\subsubsection{Intersecting lines}
\label{sec:XWilsonLines}

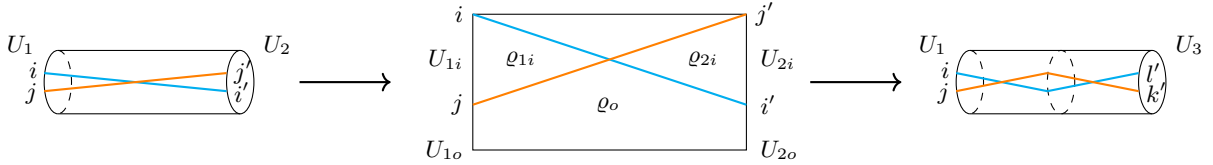
\begin{figure}[h!]
\centering
\begin{tikzpicture}[scale=1.2, every node/.style={font=\small}]

\begin{scope}[xshift=-0.5cm]

    \draw[dashed] (-1,-0.4) arc[start angle=90, end angle=-90,x radius=0.15cm, y radius=0.35cm];
  \draw (-1,-1.1) arc[start angle=270, end angle=90,x radius=0.15cm, y radius=0.35cm];
  \draw (1,-0.75) ellipse (0.15 and 0.35);
  \draw (-1,-0.4) -- (1,-0.4);
  \draw (-1,-1.1) -- (1,-1.1);
  
  \draw[cyan, thick] (-1.15,-0.65) -- (0.85,-0.85);
  
    \draw[orange, thick] (-1.15,-0.85) -- (0.85,-0.65);
  
  \node[left] at (-1.15,-0.34) {$U_1$};
  \node[right] at (1.15,-0.34) {$U_2$};
  \node[left] at (-1.12,-0.62) {$i$};
  \node[right] at (0.83,-0.64) {$j'$};
    \node[left] at (-1.12,-0.88) {$j$};
  \node[right] at (0.83,-0.86) {$i'$};
\end{scope}

\draw[->, thick] (1.15, -0.75) -- (2.15, -0.75);

\begin{scope}[xshift=3.05cm]
  \draw (0,0) rectangle (3,-1.5);

  \node[right] at (1.25,-1) {$\varrho_o$};
  \node[right] at (0.25,-0.5) {$\varrho_{1i}$};
  \node[right] at (2.25,-0.5) {$\varrho_{2i}$};
  
  \draw[cyan, thick] (0,0) -- (3,-1);
  \draw[orange, thick] (0,-1) -- (3,0);
  
  \node[left] at (0,-0.5) {$U_{1i}$};
  \node[left] at (0,0) {$i$};
  \node[right] at (3.05,-0.5) {$U_{2i}$};
  \node[right] at (3.05,-1) {$i'$};
  \node[left] at (0,-1.5) {$U_{1o}$};
  \node[left] at (0,-1) {$j$};
  \node[right] at (3.05,-1.5) {$U_{2o}$};
  \node[right] at (3,0) {$j'$};
\end{scope}

\draw[->, thick] (6.75, -0.75) -- (7.75, -0.75);

\begin{scope}[xshift=9.5cm]
    \draw[dashed] (-1,-0.4) arc[start angle=90, end angle=-90,x radius=0.15cm, y radius=0.35cm];
  \draw (-1,-1.1) arc[start angle=270, end angle=90,x radius=0.15cm, y radius=0.35cm];
  \draw[dashed] (0,-0.75) ellipse (0.15 and 0.35);
  \draw (1,-0.75) ellipse (0.15 and 0.35);
  \draw (-1,-0.4) -- (1,-0.4);
  \draw (-1,-1.1) -- (1,-1.1);
  
  \draw[cyan, thick] (-1.15,-0.65) -- (-0.15,-0.85);
  \draw[cyan, thick] (-0.15,-0.85) -- (0.85,-0.65);
  
    \draw[orange, thick] (-1.15,-0.85) -- (-0.15,-0.65);
    \draw[orange, thick] (-0.15,-0.65) -- (0.85,-0.85);
  
  \node[left] at (-1.15,-0.34) {$U_1$};
  \node[right] at (1.15,-0.34) {$U_3$};
  \node[left] at (-1.12,-0.62) {$i$};
  \node[right] at (0.83,-0.64) {$l'$};
    \node[left] at (-1.12,-0.88) {$j$};
  \node[right] at (0.83,-0.86) {$k'$};
\end{scope}

\end{tikzpicture}
\caption{From left to right: cylinder with a pair of crossing open Wilson lines; its plaquette decomposition used to construct the state \eqref{eq:state_intersecting_lines}; representation of the reduced density matrix before tracing out gauge degrees of freedom $U_1$ and $U_3$.}
\label{fig:cylinder_crossed_open_lines}
\end{figure}

Let us now consider the case of a pair of intersecting Wilson lines ending at opposite ends of a cylinder, as shown in Figure~\ref{fig:cylinder_crossed_open_lines}.

The wavefunction of this state can be written as
\begin{equation}\label{eq:state_intersecting_lines}
\begin{split}
    \psi&(U_{1i},U_{1o}, U_{1o},U_{2o};\gamma_1,\gamma_2)_{ii';jj'} \ = \ \sum_{\alpha,\beta,\alpha'} d_\alpha d_\beta d_{\alpha'} e^{-C_2(\alpha)\varrho_{1i}/2 - C_2(\beta)\varrho_o/2 - C_2(\alpha')\varrho_{2i}/2}\\
    &\times\int dA\ dB\ dC\ dD\ \chi_\alpha(U_{1i}AB^{-1}) \chi_\beta(U_{1o}BCU_{2o}D^{-1}A^{-1})\chi_{\alpha'}(U_{2i}C^{-1}D) \gamma_1(AC)_{ii'}\gamma_2(BD)_{jj'}\,.
\end{split}
\end{equation}

Tracing out the gauge degrees of freedom, we obtain a reduced density matrix of the Wilson lines:
\begin{equation}\label{eq:red_dens_matrix_intersecting_lines}
\begin{split}
    \Tr_{gauge}(\rho)_{il';jk'} &= \sum_{\alpha,\beta,\alpha'} d_\alpha d_{\alpha'} e^{-C_2(\alpha)\varrho_{1i} - C_2(\beta)\varrho_o - C_2(\alpha')\varrho_{2i}} \int dA\ dB\ dC\ dD\ dE\ dF\ dG\ dH\ \\
    &\quad\quad\times \chi_\alpha(AB^{-1}G^{-1}H)\chi_\beta(BCFG)\chi_\beta(H^{-1}E^{-1}D^{-1}A^{-1}) \chi_{\alpha'}(C^{-1}DEF^{-1})\\
    &\quad\quad\times \gamma_1(ACFH)_{il'}\gamma_2(BDEG)_{jk'}\,,
\end{split}
\end{equation}
where $l'$ and $k'$ label the endpoints of Wilson lines in representations $\gamma_1$ and $\gamma_2$ obtained after gluing the reflected cylinder to the original state; see the drawing on the right in Figure \ref{fig:cylinder_crossed_open_lines}.

Let us now consider the $SU(2)$ case for simplicity. In order to compute the integrals, we write the characters as products of matrix elements and calculate the eight integrals using equations~\eqref{eq:ortho3j} and~(\ref{6jfrom3j}), expressing the result in terms of Wigner 3j and 6j symbols. The unnormalized density matrix of the Wilson lines is then given by

\begin{align}
    \begin{split}
        \rho(\gamma_1,\gamma_2)_{il';jk'} &= \sum_{\alpha,\beta,\alpha'} d_\alpha d_{\alpha'}\, e^{-C_2(\alpha)\varrho_{1i} - C_2(\beta)\varrho_o - C_2(\alpha')\varrho_{2i}}\left\{\begin{array}{ccc}
            \beta & \gamma_2 & \alpha' \\
            \beta & \gamma_1 & \alpha
        \end{array}\right\}^2\\
        &\times\left(\begin{array}{ccc}
            \alpha & \gamma_1 & \beta \\
            a & i & k
        \end{array}\right) 
        \left(\begin{array}{ccc}
            \alpha & \gamma_1 & \beta \\
            a & l' & k
        \end{array}\right)
        \left(\begin{array}{ccc}
            \beta & \gamma_1 & \alpha' \\
            g & t & o
        \end{array}\right)\left(\begin{array}{ccc}
            \beta & \gamma_1 & \alpha' \\
            g & t & o
        \end{array}\right)\\
        &\times \left(\begin{array}{ccc}
            \beta & \gamma_2 & \alpha \\
            e & j & c
        \end{array}\right) \left(\begin{array}{ccc}
            \beta & \gamma_2 & \alpha \\
            e & k' & c
        \end{array}\right)   \left(\begin{array}{ccc}
            \alpha' & \gamma_2 & \beta \\
            q & w & m
        \end{array}\right) \left(\begin{array}{ccc}
            \alpha' & \gamma_2 & \beta \\
            q & w & m
        \end{array}\right) .
    \end{split}
\end{align}

This expression can be simplified using the orthogonality of the 3j symbols (equation~\eqref{eq:3j_orthogonality_1}):
\be
\rho(\gamma_1,\gamma_2)_{il';jk'} = \frac{\delta_{il'}\delta_{jk'}}{d_{\gamma_1}d_{\gamma_2}}\sum_{\alpha,\beta,\alpha'} d_\alpha d_{\alpha'} e^{-C_2(\alpha)\varrho_{1i} - C_2(\beta)\varrho_o - C_2(\alpha')\varrho_{2i}}\left\{\begin{array}{ccc}
            \beta & \gamma_2 & \alpha' \\
            \beta & \gamma_1 & \alpha
        \end{array}\right\}^2.
\ee
The triangular deltas that appear in the use of the orthogonality relations are enforced by the 6j symbols and are therefore omitted.

In other words, we reproduce the same structure as in the previous examples, e.g. in the case of nonintersecting lines~(\ref{eq:IIsetupDM}), corresponding to separable particles in a maximally mixed state. The normalization factor is the partition of a pair of intersecting loops on a torus, cf. the result of Section~\ref{sec:intersectingloops}.

Overall, the analysis of Section~\ref{sec:MixedParticles} showed that the reduced density matrices of Wilson lines anchored on different boundaries reproduce the same structure of a pair of uncorrelated particles in a maximally mixed state. While this seems rather natural for lines connecting different boundaries, for lines connecting the same boundary this result may appear less intuitive. Nevertheless, this structure of the density matrices is consistent with the gauge invariance, as explained before.


\section{Free energy and confinement}
\label{sec:confinement}

In the examples of the previous section, we observed that the reduced density matrix of the degrees of freedom associated with the endpoints of Wilson lines (particles) has a factorized structure
\be
\label{eq:Gen2partRDM}
\rho(\gamma,\ldots)_{ij;\ldots} \ \sim \ \frac{\delta_{ij}\cdots}{d_\gamma^2\cdots}\, Z(\gamma,\ldots)\,,
\ee
where $\gamma$ is the representation of the Wilson line connecting the particles and $Z$ is a gauge invariant partition function (on $T^2$ or $S^2$, for example, with a Wilson loop in representation $\gamma$ and other features). In other words, this density matrix describes nonentangled particles, each in a maximally mixed state. 

In this section we will supplement our analysis with a study of the properties of confinement of the same states. Despite the fact that placing a pair of particles in the environment of gauge modes makes them lose completely their mutual correlations (vanishing mutual information), the particles, nevertheless, remain confined. In particular, we will analyze the behavior of the free energy of the particles to demonstrate the confining behavior as well as to understand its large-volume properties. We will also discuss the analog of a hadron force between pairs of meson-like configurations of particles.

Confinement is a well-known property of 2D YM and the original discussion of its properties in the two-dimensional theory can be found in~\cite{tHooft:1974pnl,Callan:1975ps,Migdal:1975zg,Makeenko:1979pb,Rusakov:1990rs,Gross:1992tu,Gross:1993hu}.

\subsection{Two particles}

Computing the free energy of a state with a Wilson line with its endpoints on a boundary circle allows one to see the linear dependence of the potential of the pair of particles on their separation. The free energy is captured by the piece in~(\ref{eq:Gen2partRDM}) that is irrelevant for the entanglement -- the closed manifold partition function factor $Z$.

\begin{figure}[h!]
\centering
\begin{subfigure}{0.45\textwidth}
\centering
    \includegraphics[width=7cm]{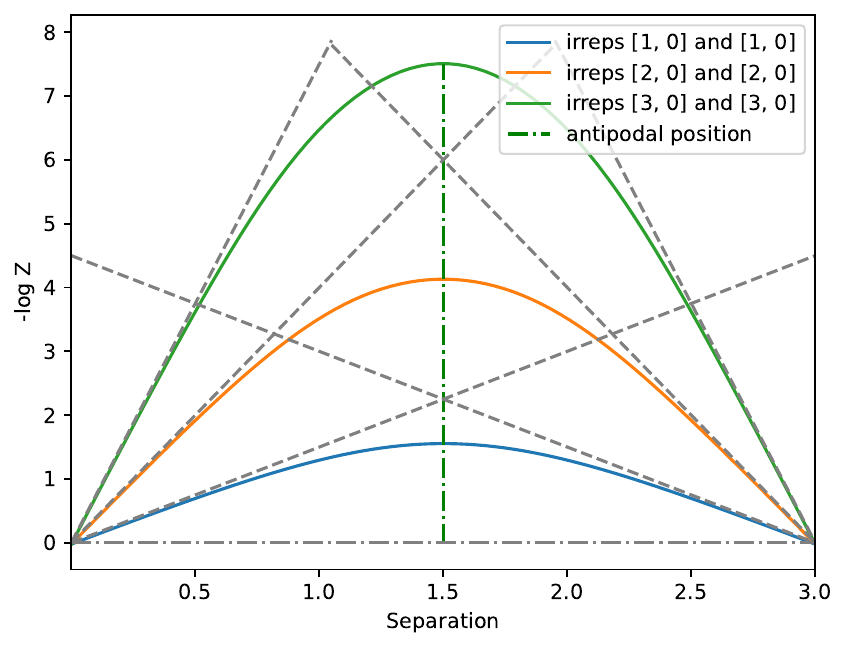}
\subcaption{}
\label{subfig:cyl1WLsu2repa}
\end{subfigure}
\hfill
\begin{subfigure}{0.45\textwidth}
\centering
    \includegraphics[width=7cm]{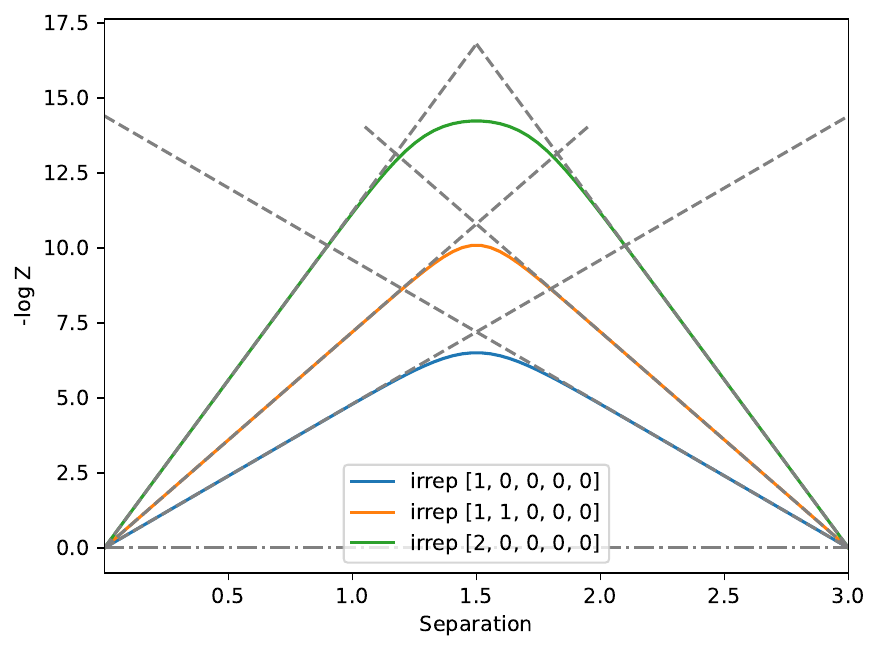}
\subcaption{}
\label{subfig:cyl1WLsu5repb}
\end{subfigure}
\caption{Free energy of a pair of quarks in conjugate representations carried by endpoints of parallel Wilson lines on a cylinder (Figure~\ref{fig:cylinderparallel}) in three different representations of $SU(2)$ (\subref{subfig:cyl1WLsu2repa}) and $SU(5)$ (\subref{subfig:cyl1WLsu5repb}). Dashes lines illustrate the leading linear behavior for small separations.}
\label{fig:free_energy_2quarks_su2}
\end{figure}

In Figure~\ref{fig:free_energy_2quarks_su2} we show the free energy of a pair of particles in state~(\ref{IIsetupState}), with parallel lines in conjugate representations on a cylinder, choosing $SU(2)$ and $SU(5)$ as examples. In this setup, there is an obvious notion of the separation of the particles in terms of the length of the cylinder (inverse temperature) and the areas bounded by the lines. For small separations, the free energy is linear in distance, while finite size effects alter this behavior at separations comparable to the size of the circle. 

\begin{figure}[h!]
\centering
\begin{subfigure}{0.45\textwidth}
\centering
    \includegraphics[width=7cm]{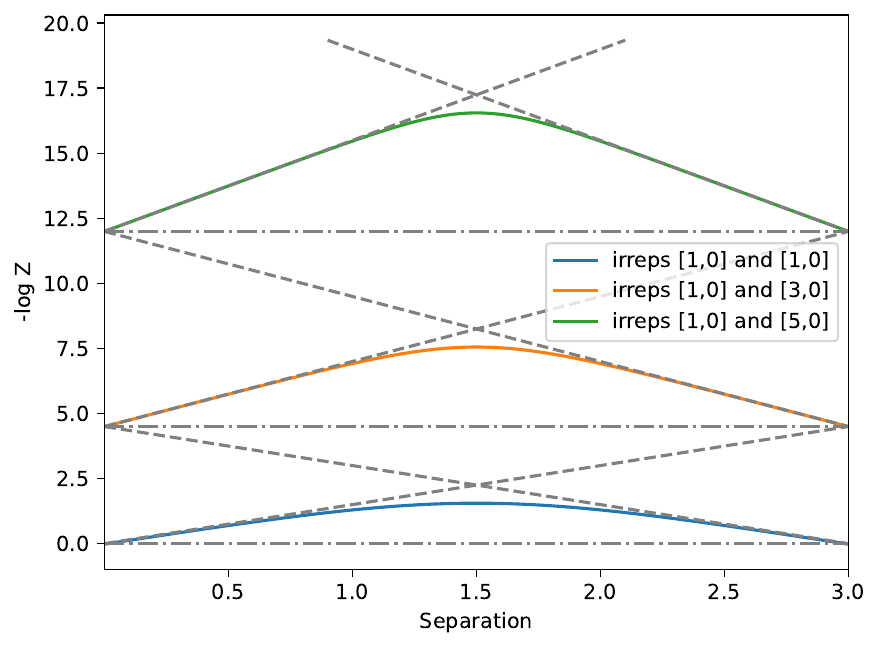}
\subcaption{}
\label{subfig:free_energy_2quarks_su2_2a}
\end{subfigure}
\hfill
\begin{subfigure}{0.45\textwidth}
\centering
    \includegraphics[width=7cm]{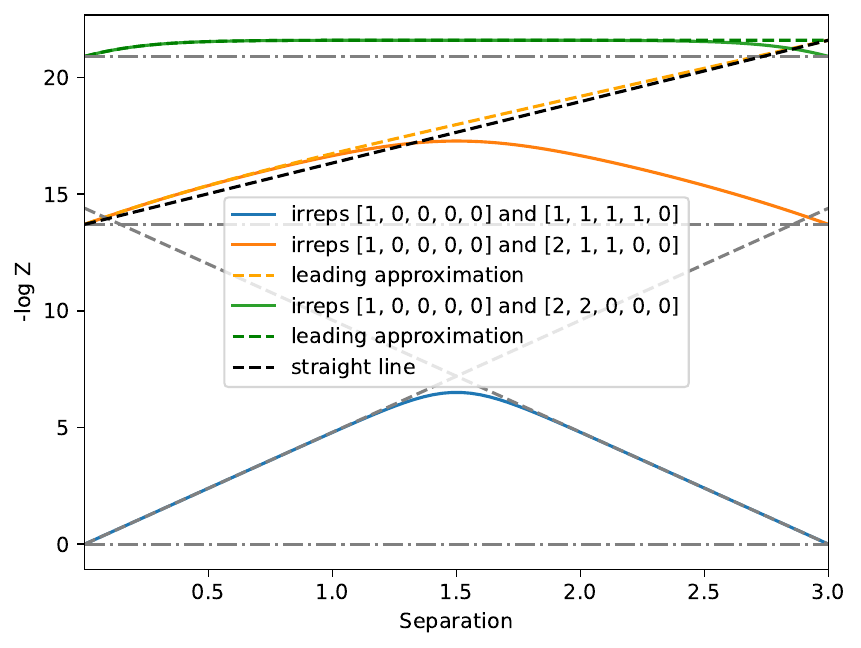}
\subcaption{}
\label{subfig:free_energy_2quarks_su2_2b}
\end{subfigure}
\caption{Free energy of a pair of quarks in non-conjugate representations of $SU(2)$ (\subref{subfig:free_energy_2quarks_su2_2a}) and $SU(5)$ (\subref{subfig:free_energy_2quarks_su2_2b}). The horizontal dash-dotted lines scale as $C_2(|\gamma_1-\gamma_2|/2)$ in the units of the total area. The dashed lines show the dominant small-area linear behavior of the free energy. The black dashed line indicates contribution of the dominant representations $\sigma_d$ and $\lambda_d$.}
\label{fig:free_energy_2quarks_su2_2}
\end{figure}

For vanishing separation, $\varrho_1\ll \varrho_2$, the partition function~(\ref{eq:ZofT2w2WLoops}) defining the free energy reduces to 
\begin{align}
\label{ZIIanalytic}
\begin{split}
Z(T^2;&\gamma_1,\gamma_2;\varrho_1,\varrho_2) = \ \sum_{\lambda,\sigma}N^{\lambda}_{\gamma_1\sigma} N^{\sigma}_{\lambda\gamma_2}e^{- C_2(\lambda)\varrho_1- C_2(\sigma)\varrho_2} \\
& = \ \sum_{\lambda}\left(N^{\lambda}_{\gamma_10} N^{0}_{\lambda\gamma_2} + N^{\lambda}_{\gamma_1\Box} N^{\Box}_{\lambda\gamma_2}e^{ -C_2(\Box)\varrho_2} + \ldots\right)e^{- C_2(\lambda)\varrho_1 } \ = \ \delta_{\gamma_1\gamma_2^\ast}e^{- C_2(\gamma_1)\varrho_1 } + \ldots\,.
\end{split}
\end{align}
Thus, the free energy is close to zero at leading order for $\varrho_1\ll \varrho_2$ if the representations are compatible, $\gamma_1=\gamma_2^\ast$. The slope of the curve in this regime is controlled by $C_2(\gamma_1)$, as illustrated in Figure~\ref{fig:free_energy_2quarks_su2}. If $\gamma_1$ and $\gamma_2$ are not compatible, one has to go to the next order. We can do this exercise for $SU(2)$:
\be
Z(T^2;\gamma_1,\gamma_2;0,\varrho_2) \ = \  \sum_{\lambda}\left( N^{\lambda}_{\gamma_1\Box} N^{\Box}_{\lambda\gamma_2}e^{-\frac{3}{2}\,\varrho_2} + \ldots\right) \ = \ (\delta_{\gamma_1,\gamma_2+2} + \delta_{\gamma_1,\gamma_2-2})e^{-\frac{3}{2}\,\varrho_2} + \ldots
\ee
Hence, the leading zero-area value of the free energy, divided by $\rho_2$, will be defined by the quadratic Casimir of the dominant representation $\gamma=|\gamma_1-\gamma_2|/2$. The general form of the leading contribution to the $SU(2)$ partition function at small area ($\varrho_1\ll \varrho=\varrho_1+\varrho_2$) has the structure (assuming $\gamma_1\geq \gamma_2$)
\be
Z(T^2;\gamma_1,\gamma_2;\varrho_1,\varrho_2) \ \simeq \ \delta_{\gamma_1,\gamma_2+2\gamma}\, e^{-(C_2(\gamma_1-\gamma)-C_2(\gamma))\varrho_1 - C_2(\gamma)\varrho}\,.
\ee
The corresponding asymptotic behavior of the zero-area values and the slopes is illustrated in Figure~\ref{subfig:free_energy_2quarks_su2_2a}. 

The situation is qualitatively similar for $SU(5)$, but there are some interesting features due to a more complex hierarchy of quadratic Casimirs. First, only pairs $\gamma_1$ and $\gamma_2$ for which the total number of boxes in the two Young diagrams is a multiple of $N=5$ give a nonvanishing free energy. The zero-area free energy is proportional to the Casimir of the dominant representation $\sigma$, when $\varrho_1\to 0$, or $\lambda$, when $\varrho_2\to 0$. The dominant $\sigma$ can be found as the smallest Casimir representation that shares a common term in the expansions of both $\sigma\otimes\sigma^\ast$ and $\gamma_1\otimes\gamma_2$. Unless $\sigma_d=\sigma_d^\ast$ there will be two exponential terms contributing to free energy, each with some dominating $\lambda_d$. Consequently, one does not expect a linear behavior for very small $\varrho_1$. As $\varrho_1$ increases, the lowest of the two $\lambda_d$ dominates and the linear slope is recovered. This is illustrated by examples in Figure~\ref{subfig:free_energy_2quarks_su2_2b}. The dominating $\lambda_d$ is found as the lowest Casimir representation in the fusion of $\sigma_d$ and $\gamma_1$. The linear slope is given by $C_2(\lambda_d)-C_2(\sigma_d)$. 

In examples of Figure~\ref{subfig:free_energy_2quarks_su2_2b}, the case $\gamma_1=[1,0,0,0,0]$ and $\gamma_2 = [2,1,1,0,0]$ corresponds to either $\sigma_d=[1,0,0,0,0]$ or $\sigma_d=[1,1,1,1,0]$. The pair of corresponding $\lambda_d$ is $[1,1,0,0,0]$ and $[2,1,1,1,0]$. The first $\lambda_d$ wins for intermediate areas. Consequently,
\begin{equation}
    Z \simeq e^{-(C_2(\scalebox{0.25}{\yng(1,1)}) - C_2(\scalebox{0.25}{\yng(1)}))\varrho_1 - C_2(\scalebox{0.25}{\yng(1)})\varrho}\,, \qquad \text{for} \qquad \varrho_1\ll \varrho\,. 
\end{equation}
Note that for $\varrho_2\ll \varrho$ different pairs of representations dominate: $\sigma_d=[1,0,0,0,0]$, $\lambda_d=[2,1,1,1,0]$ and $\sigma_d=[1,1,1,1,0]$, $\lambda_d=[1,1,0,0,0]$. Nevertheless, the free energy curve remains symmetric due to the degeneracy of the quadratic Casimir for conjugate representations. 

For $\gamma_2 = [2,2,0,0,0]$, an example of the dominating pair is $\sigma_d=[1,1,0,0,0]$ and $\lambda_d=[1,1,1,0,0]$. Note that these two representations are conjugate and one expects a vanishing slope, as indeed happens in Figure~\ref{subfig:free_energy_2quarks_su2_2b}.

\begin{figure}[h!]
\centering
\begin{subfigure}{0.45\textwidth}
\centering
    \includegraphics[width=7cm]{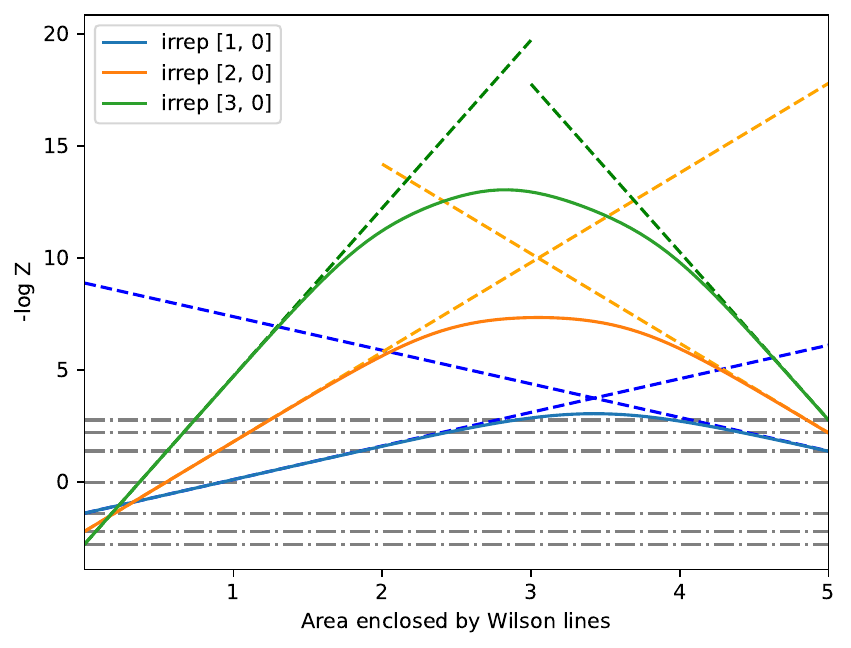}
\subcaption{}
\label{subfig:23a}
\end{subfigure}
\hfill
\begin{subfigure}{0.45\textwidth}
\centering
    \includegraphics[width=7cm]{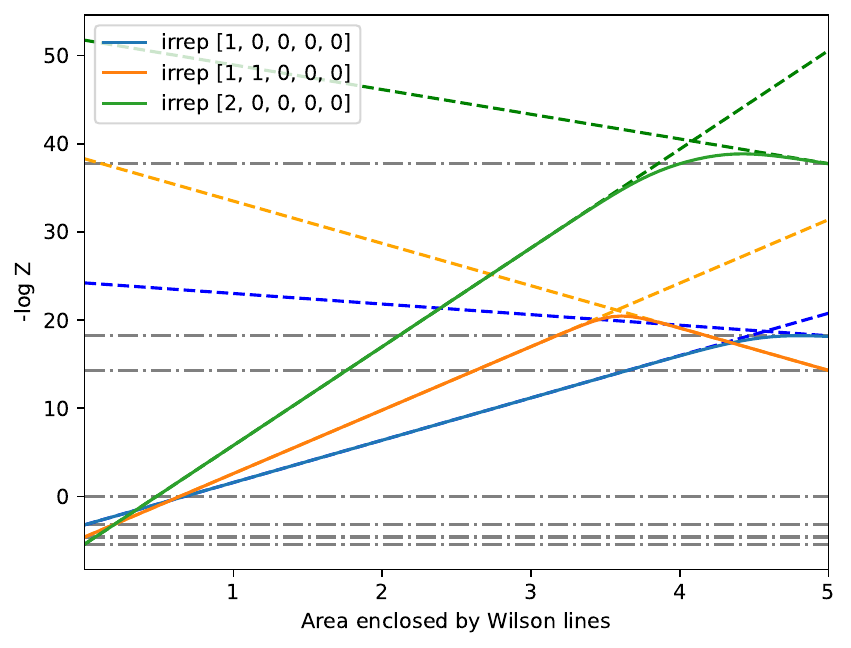}
\subcaption{}
\label{subfig:23b}
\end{subfigure}
\caption{Free energy of a pair of quarks in three different representations of $SU(2)$ (\subref{subfig:23a}) and $SU(5)$ (\subref{subfig:23b}) in configuration~(\ref{eq:UUsetup}). The plot takes the two Wilson lines in conjugate representations and assumes equal areas of the regions bounded by the lines. Dashed lines show the leading order linear approximation for either small, or large areas. The horizontal dash-dotted lines indicate the variation of the boundary values of the curves.}
\label{fig:free_energy_2quarks_UU}
\end{figure}

For Wilson lines in configuration~(\ref{eq:UUsetup}) the relation between the separation and the areas bounded by the lines is less obvious. However, for the discussion of confinement, one only needs to see that the effective potential is a linear function of the area. The corresponding plots for groups $SU(2)$ and $SU(5)$ are shown in Figure~\ref{fig:free_energy_2quarks_UU}. The curves have a qualitatively similar behavior to the case of parallel Wilson lines, compatible with confinement up to finite-size effects. In this case the small-area and large-area expansions are dominated by different representations producing asymmetric plots.

In the zero area limit the partition functions reduces to
\be
Z(T^2;\gamma_1,\gamma_2;0,0,\varrho_e) \ = \  d_{\gamma_1} d_{\gamma_2} Z(T^2;\varrho_e)\,.
\ee
Consequently, the initial points of the curves are controlled by the dimensions of the representations of the Wilson lines. The slopes are dominated by the Casimirs of these representations, since the fusion channels in~(\ref{ZinUUconfig}) are dominated by the trivial representation, that is $N_{\alpha\gamma_1}^0N_{\lambda\gamma_1^\ast}^0=\delta_{\alpha,\gamma_1^\ast}\delta_{\lambda,\gamma_1}$:
\be
Z(T^2;\gamma_1,\gamma_2;\varrho/2,\varrho/2,\varrho_e-\varrho) \simeq d_{\gamma_1}^2e^{- C_2(\gamma_1)\varrho}.
\ee

In the $\varrho_e\to 0$ limit, the leading exponential term is controlled by allowed representations $\alpha$ and $\lambda$ in $N_{\alpha\gamma_1}^\beta N_{\lambda\gamma_1^\ast}^\beta$ with the lowest sum $C_2(\alpha)+C_2(\lambda)$. For $SU(2)$ this corresponds to $\alpha=\lambda=0$ and $\beta=\gamma_1$, so that the $\varrho_e=0$ values of the free energy and of the slopes are negatives of the small-area values above. For generic $SU(N)$ the dominant $\alpha$ and $\lambda$ should be searched in the fusion channel $\gamma_1+\gamma_1^\ast\to\beta\to \alpha+\lambda$. We show the characteristic asymmetric plots in Figure~\ref{fig:free_energy_2quarks_UU} without further details.  

\begin{figure}[h!]
\centering
\begin{subfigure}{0.45\textwidth}
\centering
    \includegraphics[width=7cm]{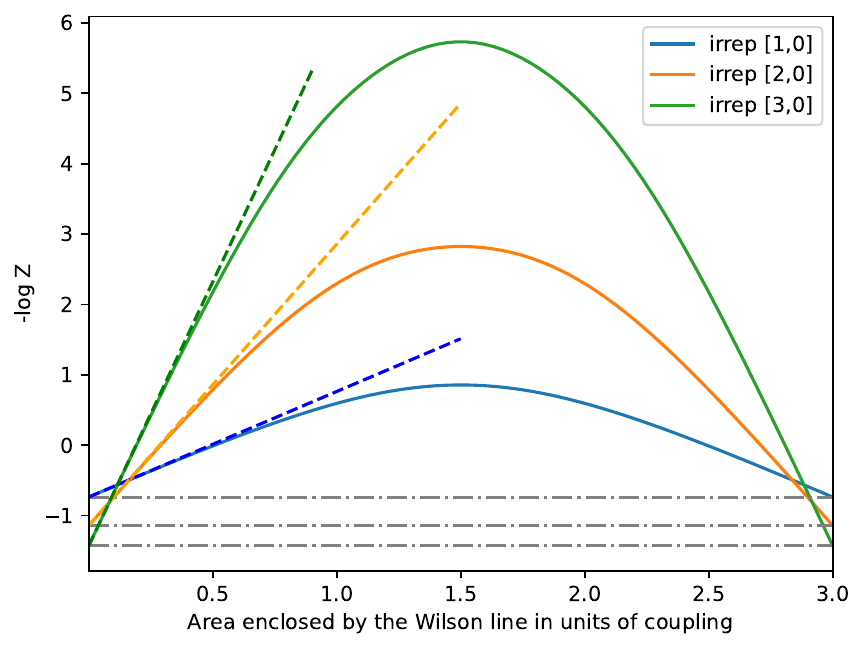}
\subcaption{}
\label{subfig:24a}
\end{subfigure}
\hfill
\begin{subfigure}{0.45\textwidth}
\centering
    \includegraphics[width=7cm]{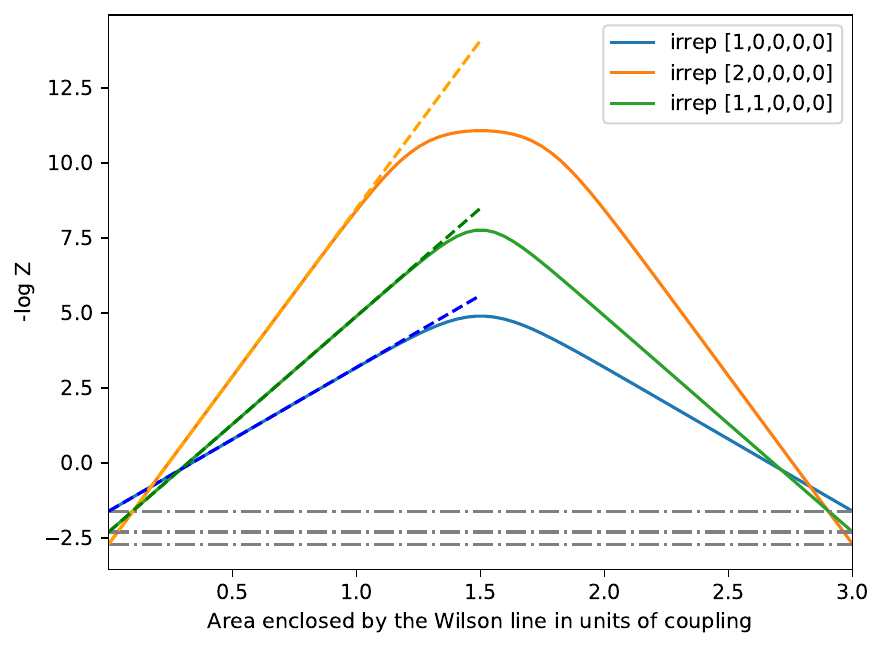}
\subcaption{}
\label{subfig:24b}
\end{subfigure}
\caption{Free energy of a pair of quarks in three different representations of $SU(2)$ (\subref{subfig:24a}) and $SU(5)$ (\subref{subfig:24b}) in configuration~(\ref{eq:reddiskline}) on a disk. The horizontal lines show the zero-area values of the free energy calculated using equation~(\ref{eq:0LoopDisk}). Colored dashed lines show the leading order linear approximation.}
\label{fig:free_energy_disk}
\end{figure}

One can also compare the previous results with the case of a single Wilson line on a disk. In this case, the free energy is controlled by the partition function~(\ref{eq:ZofS2wWLoop}) of a loop $\gamma$ on $S^2$. The plots as functions of area $\varrho_2$ are shown in Figure~\ref{fig:free_energy_disk}. The curves are symmetric in this case, which resembles more the situation of a pair of parallel Wilson lines on a cylinder, with the exception of the dependence of the zero-area values of the representation. These values are
\be
\label{eq:0LoopDisk}
Z(S^2;\gamma;0,\varrho_2) \ = \ d_\gamma Z(S^2;\varrho_2)\,.
\ee
In particular, the ratios of the zero-area values for different representations must be equal to the ratios of the dimensions of the representations running in the loop. The asymptotic linear slopes are given by the Casimirs of the loop representations. This behavior is illustrated in the plot. 

All the studied examples are compatible with the confining behavior at relatively small areas. At areas comparable with the area of the manifold finite size effects come into play. The effects can be controlled by either increasing $N$ or increasing the total area of the 2D surface. However, new effects come into play at large areas, as we shall discuss.

\subsection{Large-volume effects}
\label{sec:ConfinementLargeVolume}

\begin{figure}[h!]
\centering
\begin{subfigure}{0.45\textwidth}
\centering
    \includegraphics[width=7cm]{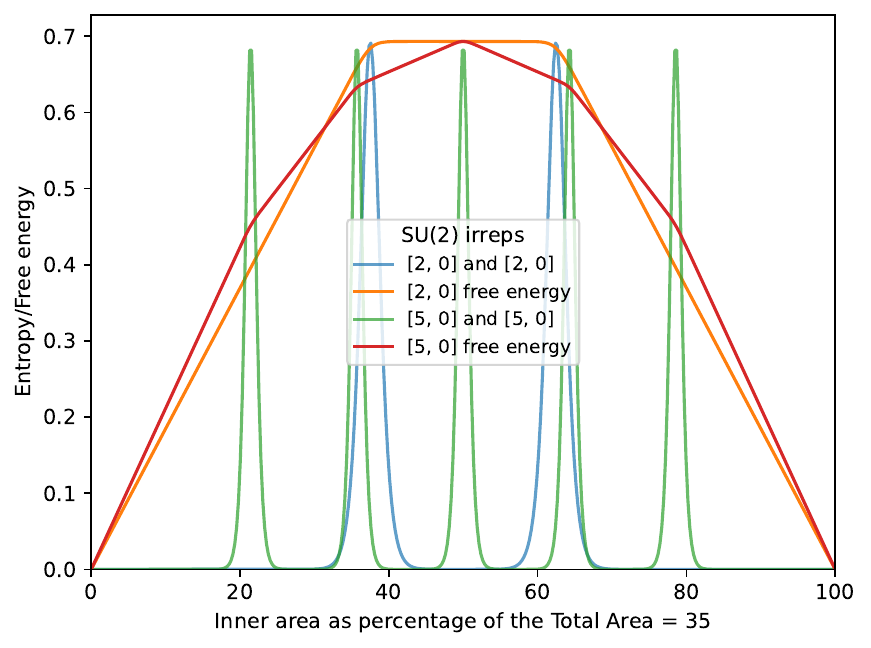}
\subcaption{}
\label{subfig:25a}
\end{subfigure}
\hfill
\begin{subfigure}{0.45\textwidth}
\centering
    \includegraphics[width=7cm]{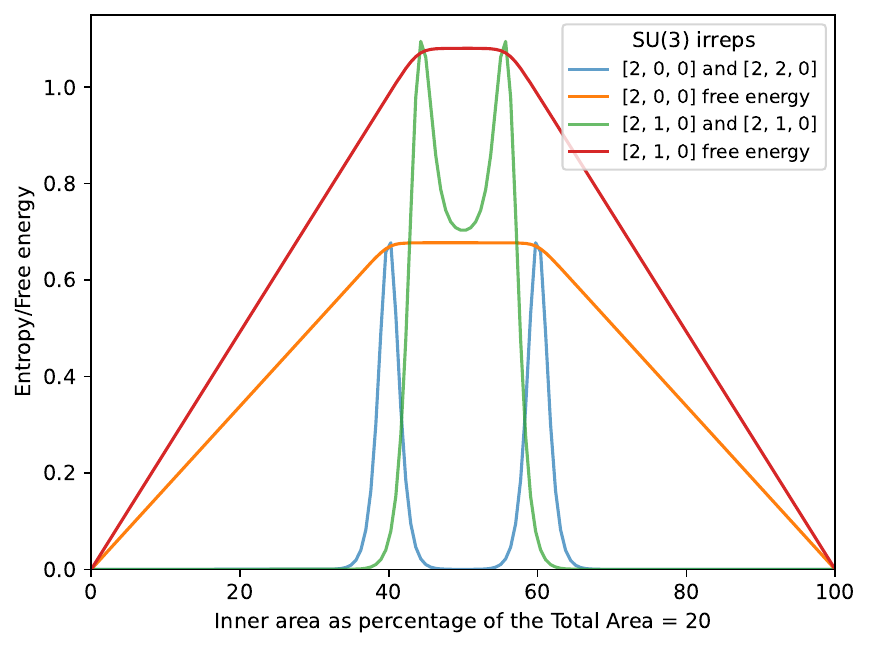}
\subcaption{}
\label{subfig:25b}
\end{subfigure}
\caption{(\subref{subfig:25a}) Free energy of two parallel Wilson lines on a cylinder in representations $[2,0]$ and $[5,0]$ of $SU(2)$ plotted together with peaks of entanglement entropy. The maximum of the free energy is scaled to the height of the peaks. (\subref{subfig:25b}) Free energy and entropy peaks for several representations of $SU(3)$.}
\label{fig:entropy_free_energy}
\end{figure}

So far, our analysis of confinement focused on relatively small areas. It is natural to ask whether additional phenomena emerge in the large-area regime. In the study of entanglement entropy presented in Sections~\ref{sec:WilsonLoops} and~\ref{sec:WilsonLines}, such effects were associated with a change in the dominant contribution to the partition function. In an analogous way, this change of dominance should also manifest itself in the behavior of the free energy, in particular in the slopes near the optimal fractions. In Figure~\ref{fig:entropy_free_energy}, we display the free energy together with the corresponding peaks of the entanglement entropy at large total area. We confirm the anticipated behavior in the plots.

For parallel Wilson lines on a cylinder, both the positions of the entropy peaks and the slopes of free energy are determined by equations~(\ref{eq:rngamma1gamma2}) and~(\ref{eq:AsymptEntropyII}). Upon crossing an entropy peak, the slope decreases, reaching its minimum near the symmetric configuration in which the two areas separating the Wilson lines are equal (i.e. at a 50\% ratio). In certain cases, such as integer-spin representations of $SU(2)$, this minimal slope vanishes. 

Recall that at finite total area, the entropy peaks are slightly shifted relative to the optimal fractions given in~(\ref{eq:rngamma1gamma2}). As illustrated in Figure~\ref{fig:entropy_free_energy}, the actual peaks accurately track the locations at which the slopes of the free-energy curve change. For finite areas, these are smooth crossover transitions, which we expect to sharpen into genuine phase transitions in the infinite-area limit. In this sense, the entanglement entropy provides a useful diagnostic of these transitions.

It is important to emphasize that the entanglement entropy considered here does not quantify correlations between quarks, which remain unentangled, as discussed in Section~\ref{sec:MixedParticles}. Rather, it measures the entropy of the mixed state associated with one end of the cylinder. At the transition points, this state becomes maximally mixed, thereby signaling (and effectively driving) the transition.

A similar staircase structure in the free energy also appears in the state obtained from two copies of a single Wilson line on a disk, whose partition function is given in~(\ref{eq:ZofS2wWLoop}). In that case, the state is pure, yet the transitions and the corresponding slopes are again governed by the same equations as for parallel Wilson lines on a cylinder.

\subsection{Four particles}
\label{sec:hadronforce}

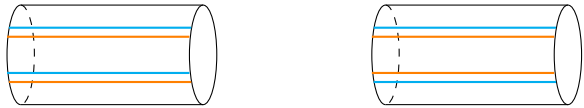
\begin{figure}[h!]
\centering
\begin{tikzpicture}[scale=1.2, every node/.style={font=\small}]

\begin{scope}[xshift=-2cm]
    \draw[dashed] (-1,-0.2) arc[start angle=90, end angle=-90,x radius=0.15cm, y radius=0.55cm];
  \draw (-1,-1.3) arc[start angle=270, end angle=90,x radius=0.15cm, y radius=0.55cm];
  \draw (1,-0.75) ellipse (0.15 and 0.55);
  \draw (-1,-0.2) -- (1,-0.2);
  \draw (-1,-1.3) -- (1,-1.3);

  \draw[cyan, thick] (-1.13,-0.45) -- (0.87,-0.45);
  \draw[orange, thick] (-1.15,-0.55) -- (0.85,-0.55);

  \draw[orange, thick] (-1.13,-1.05) -- (0.87,-1.05);
  \draw[cyan, thick] (-1.15,-0.95) -- (0.85,-0.95);
  
\end{scope}

\begin{scope}[xshift=2cm]
    \draw[dashed] (-1,-0.2) arc[start angle=90, end angle=-90,x radius=0.15cm, y radius=0.55cm];
  \draw (-1,-1.3) arc[start angle=270, end angle=90,x radius=0.15cm, y radius=0.55cm];
  \draw (1,-0.75) ellipse (0.15 and 0.55);
  \draw (-1,-0.2) -- (1,-0.2);
  \draw (-1,-1.3) -- (1,-1.3);
  
  \draw[cyan, thick] (-1.13,-0.45) -- (0.87,-0.45);
  \draw[orange, thick] (-1.15,-0.55) -- (0.85,-0.55);

  \draw[cyan, thick] (-1.13,-1.05) -- (0.87,-1.05);
  \draw[orange, thick] (-1.15,-0.95) -- (0.85,-0.95);

\end{scope}

\end{tikzpicture}
\caption{Cylinders with two pairs of parallel Wilson lines in meson-like configuration with two possible ``dipole'' orientations. Different colors illustrate conjugate representations.}
\label{fig:cylinder4parallel}
\end{figure}

In this section, we study the free energy of a pair of meson-like configurations and observe a basic example of the hadronic force. For this, we place two quark-anti-quark pairs on the circle, fixing the distance between the quark and the anti-quark, and study the variation of the energy with the distance between the pairs. We only consider the setup with the quarks represented by the endpoints of parallel Wilson lines on the cylinder; that is, mixed state quarks, as shown in Figure~\ref{fig:cylinder4parallel}. 

\vspace{0.2cm}

For four parallel Wilson lines, the partition function is given by
\begin{equation}
\label{eq:ZofT2w4WLoops}
    Z(\varrho_p,\gamma_q) = \dfrac{\delta_{ii'}\delta_{jj'}\delta_{kk'}\delta_{ll'}}{\prod_{q=1}^4 d_{\gamma_q}} \sum_{\alpha_1,\alpha_2,\alpha_3,\alpha_4} \exp\left(-\sum_{p=1}^4 C_2(\alpha_p)\varrho_p\right) N_{\alpha_1,\gamma_1}^{\alpha_2}N_{\alpha_2,\gamma_2}^{\alpha_3}N_{\alpha_3,\gamma_3}^{\alpha_4}N_{\alpha_4,\gamma_4}^{\alpha_1}\,.
\end{equation}
where $\gamma_i$ is the $SU(N)$ representation of line $i$.\footnote{Comparing with~(\ref{eq:ZofT2w2WLoops}) one can see that the result has an obvious generalization to an arbitrary number of lines. Also, due to modular invariance, the piece of equation~(\ref{eq:ZofT2w4WLoops}) corresponding to the torus partition function with four Wilson lines has the same structure as the partition function~(\ref{eq:ZofT2w2WLoops0}).}

We can choose $\gamma_2 = \gamma_1^*$ and $\gamma_4 = \gamma_3^*$, where we assume that the labels of the Wilson lines are naturally ordered. Once the cylinder's length is held fixed, the plaquette areas $\varrho_i$ can be understood as distances between quarks. Assuming that the quarks in the pair $(\gamma_1,\gamma_1^*)$ are close to each other (which means $\varrho_2 \ll \varrho_t$ for the total cylinder area $\varrho_t$) and assuming the same for the pair $(\gamma_3,\gamma_3^*)$, we can test the behavior of the free energy for different representation choices.

\begin{figure}[H]
    \centering
    \begin{subfigure}{0.45\textwidth}
    \centering
    \includegraphics[width=7cm]{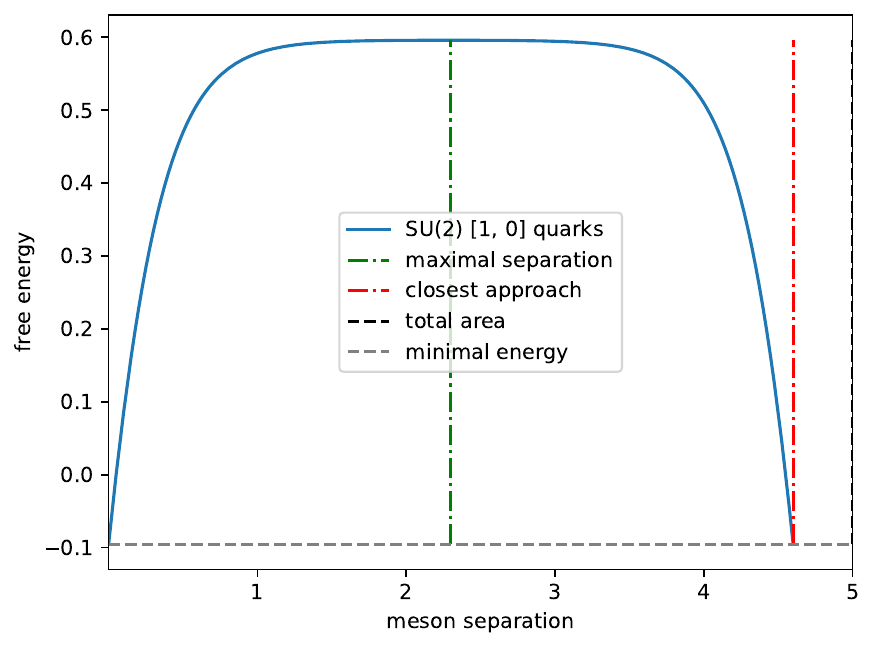}
    \subcaption{}
    \label{subfig:free_energy_4_particles_su3_udud-a}
    \end{subfigure}
    \hfill
    \begin{subfigure}{0.45\textwidth}
    \centering
    \includegraphics[width=7cm]{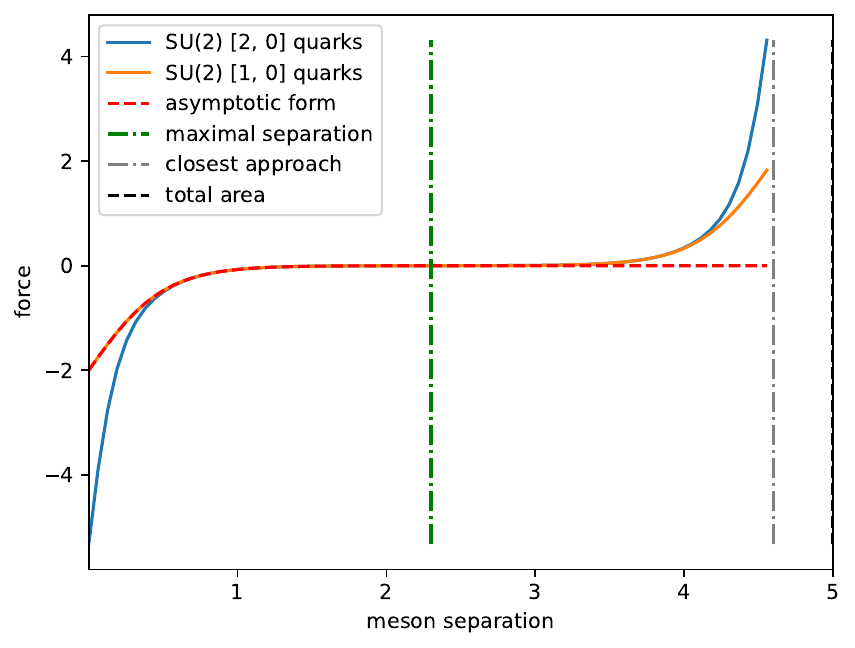}
    \subcaption{}
    \label{subfig:free_energy_4_particles_su3_udud-b}
    \end{subfigure}
    \caption{Free energy in the case $\gamma_3 = \gamma_1$ for $SU(2)$ quarks in the fundamental representation (\subref{subfig:free_energy_4_particles_su3_udud-a}) and force for fundamental and symmetric irreps of $SU(2)$ (\subref{subfig:free_energy_4_particles_su3_udud-b}). Red dashed line shows the analytic approximation~(\ref{eq:forceSU2}) for the fundamental representation.}
    \label{fig:free_energy_4_particles_su3_udud}
\end{figure}

For $SU(2)$ and all quarks in the fundamental representation, we obtain the free energy shown in Figure~\ref{subfig:free_energy_4_particles_su3_udud-a}. For areas that are not too close to the total area of the cylinder, the partition function, as a function of $\varrho=\varrho_1$, for example, can be well approximated by truncating the $\alpha_3$ sum on the trivial representation. This restricts the expression to a single sum:
\be
\label{eq:approxZ4lines}
Z \ \simeq \ e^{-C_2(\gamma_1)\varrho_2-C_2(\gamma_3)\varrho_4}\sum_{\alpha_1} N_{\alpha_1,\gamma_1}^{\gamma_1}N_{\gamma_3,\gamma_3^\ast}^{\alpha_1}e^{-C_2(\alpha_1)\varrho_1}\,.
\ee
For example, for $\gamma_1=\gamma_3$  and the fundamental representation, the sum only has two terms.

The approximation \eqref{eq:approxZ4lines} may be used to obtain the analytic behavior of the force in any representation $[n,0]$ valid up to area ranges sensitive to finite size effects:
\be
\label{eq:forceSU2}
F \ \simeq \ - \frac{\sum_{j=0}^{n}C_2([2j,0])e^{-C_2([2j,0])\varrho}}{\sum_{j=0}^{n}e^{-C_2([2j,0])\varrho}}\,.
\ee
When area $\varrho$ is not too small, the force has a universal form for all representations:
\be
\label{eq:leadingforce}
F \ \sim \ - C_2(\alpha_d)e^{-C_2(\alpha_d)\varrho}\,, 
\ee
that is, it rapidly decays with the distance with the exponent dominated by the first symmetric representation, $\alpha_d=[2,0]$. This behavior of the force is illustrated in Figure~\ref{subfig:free_energy_4_particles_su3_udud-b}, where the analytic approximation~(\ref{eq:forceSU2}) for the fundamental representation is also shown.

\begin{figure}[h!]
    \centering
    \begin{subfigure}{0.45\textwidth}
    \centering
    \includegraphics[width=7cm]{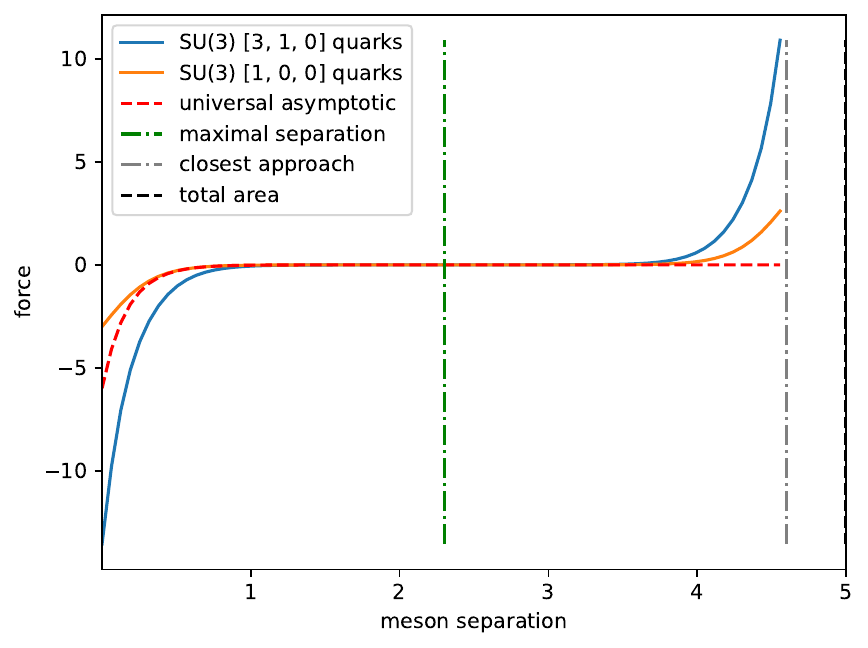}
    \subcaption{}
    \label{fig:free_energy_4_particles_su4_udud-a}
    \end{subfigure}
    \hfill
    \begin{subfigure}{0.45\textwidth}
    \centering
    \includegraphics[width=7cm]{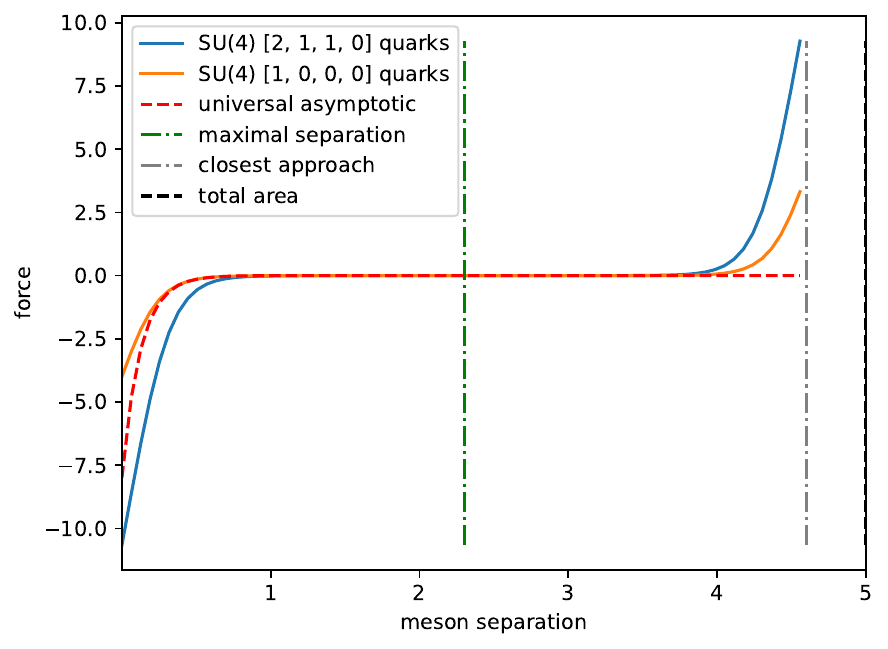}
    \subcaption{}
    \label{fig:free_energy_4_particles_su4_udud-b}
    \end{subfigure}
    \caption{(\subref{fig:free_energy_4_particles_su4_udud-a}) Force between $SU(3)$ ``mesons'' for quarks in selected representations. (\subref{fig:free_energy_4_particles_su4_udud-b}) Same for $SU(4)$. Red dashed line shows the leading decay law~(\ref{eq:leadingforce}).}
    \label{fig:free_energy_4_particles_su4_udud}
\end{figure}

For $SU(N)$ the behavior is very similar. Approximation~(\ref{eq:approxZ4lines}) remains valid if $\gamma_1=\gamma_3=\gamma_2^\ast=\gamma_4^\ast$. The leading representation is universal: $\alpha_d =[2,1,\ldots,1,0]$, with $N-2$ single-box rows -- the adjoint representation of $SU(N)$. The leading expression for the force in the intermediate regime is again~(\ref{eq:leadingforce}). This behavior is shown in Figure~\ref{fig:free_energy_4_particles_su4_udud} for $SU(3)$ and $SU(4)$. 

The behavior is similar for quark-anti-quark pairs in different representations.


\section{Conclusions and discussion}
\label{sec:conclusions}

In this work, we investigated a class of states in 2D YM defined by Euclidean path integrals over Riemann surfaces, primarily cylinders, with various configurations of closed and open Wilson lines. Our central goal was to understand how the topology and the geometry of the configurations affect entanglement and related physical properties. We found that the most nontrivial and sharp statements arise in the regime where the domains cut by Wilson lines have large areas. In this limit, the states prepared by the path integral behave as quasi-projectors onto finite-dimensional sectors of the otherwise infinite-dimensional Hilbert space, with the dimension controlled by the rank of the gauge group, the representations of the Wilson lines, and related group-theoretic data.

From the entanglement perspective, the large-area states are generically entangled but not maximally entangled when viewed as states in the full infinite dimensional Hilbert space. Within the finite-dimensional subspace, such states can apparently be engineered to have an arbitrary amount of entanglement, limited by the dimension of the subspace. In particular, they can be interpreted as finite-temperature mixed states with entropy bounded by $\log N$ for $SU(N)$. Nonvanishing entanglement typically occurs at special ratios, or finite intervals of ratios, of the areas partitioning the Riemann surface. However, we also identified configurations where the entropy remains finite for arbitrary partitions, even in the infinite-area limit. The latter situation is a consequence of nontrivial multiplicities of representations appearing in the tensor product, as illustrated by the example in Figure~\ref{subfig:cyl2parallelWLsu3}.

The $\log N$ bound is not strict and we found examples of states with entropy above it. The mentioned case of nontrivial multiplicity is one of them. In some cases the violation of the bound can happen without nontrivial multiplicities, as in the case of nested $SU(2)$ loops, not studied here. In general we expect more complex configurations, with multiple loops and intersection to have a much richer entanglement structure.

If one views Wilson loops as defects inserted in 2D surfaces then the large-area limit supports the picture of the entanglement as a measure of the space's connectedness. Defects reduce the connectedness and the entanglement entropy reduces consistently with the number of defects. One can see this as a variant of the Ryu-Takayanagi formula~\cite{Ryu:2006bv} at the topological (rather than geometrical) level. It would also be interesting to have a more quantitative geometric version of this formula for states of the 2D YM.

For cylindrical geometries without open Wilson lines, the density matrices obtained in the large area limit take a diagonal thermofield double structure. While noncontractible loops generically produce nondiagonal density matrices at finite area, the off-diagonal components decay rapidly as the area increases. It would be interesting to identify geometric constructions that yield genuinely nondiagonal finite-dimensional matrices in the large-area limit. More broadly, developing a systematic toolkit for engineering general finite-dimensional operators, such as unitaries, invertible maps, and projectors, from Riemann surfaces with defects remains an open direction.

Open Wilson lines ending at the boundaries lead to a richer structure. The associated density matrices are intrinsically nondiagonal due to the presence of representation matrices. Their detailed behavior, particularly in the large-area regime, deserves further investigation. A systematic classification of such states as quantum resources would be especially desirable.

Finally, the large-area limit of 2D YM theory defines families of vacuum sectors with varying degrees of degeneracy. An important question is how these sectors influence observable physics. In our analysis of confinement, we found that large-area effects can modify the confining force, although such modifications appear only at high energies, or zero-temperture limit of finite-size systems. Whether finite-energy signatures persist in the infinite-area limit remains an open problem.


\subsubsection*{Acknowledgments}

JO and VP acknowledge the support of the Brazilian Federal Agency for Support and Evaluation of Graduate Education (CAPES) in the realization of this work. The work of DM and MT was supported by the Simons Foundation through award number 1023171-RC. The work was also partially supported by grants from the Brazilian National Council for Scientific and Technological Development (CNPq) number 308580/2022-2 and 404274/2023-4 (DM) as well as number 445944/2024-2 (MT).

\begin{appendix}

\section{List of useful formulae}
\label{sec:usefulformulae}

In this appendix we collect some group-theoretic formulae used in the calculation of the amplitudes of this paper.

Let $\mathcal{G}$ be a group, $\alpha$ and $\beta$ irreducible finite-dimensional representations of $\mathcal{G}$. For $\mathcal{G}=SU(N)$ irreducible representations are labeled by the Young diagrams, or equivalently, partitions $(n_1,\ldots n_N)$, with $n_1\geq n_2\geq\cdots\geq n_N=0$. In terms of the partitions the dimension of a representation $\alpha$ is given by
\begin{equation}\label{eq:dimension_formula_su_n}
    d_\alpha = \sum_{1\leq i < j\leq N} \left(1 + \dfrac{n_i - n_j}{j-i}\right).
\end{equation}

Eigenvalues $C_2(\alpha)$ of the quadratic Casimir operator corresponding to irrep $\alpha$, can be obtained from
\begin{equation}\label{eq:quadratic_casimir_formula_su_n}
    C_2(\alpha) = \sum_{i=1}^{N-1} n_i(n_i - 2i + N + 1) - \dfrac{n^2}{N}\,,
\end{equation}
where $n = \sum_i n_i$ is the total number of boxes on the diagram. 

For any $\mathcal{G}$, matrix elements of the representation satisfy the following orthogonality relation:
\begin{equation}\label{eq:orthogonality_mat_elem}
    \int dU\ \alpha(U)_{ij}\beta(U^{-1})_{kl} = \dfrac{\delta_{\alpha,\beta}}{d_\alpha}\delta_{il}\delta_{jk}\,,
\end{equation}
where $U$ is a group element of $\mathcal{G}$ and the integral is calculated with the Haar measure $dU$.

Formula~\eqref{eq:orthogonality_mat_elem} can be used to prove the following results for the characters of the group elements. Namely, $\forall\, V,W\in\mathcal{G}$ we have
\begin{eqnarray}
    \int dU\ \chi_\alpha(U) \chi_\beta(U^{-1}) & = & \delta_{\alpha,\beta}\,, \\
    \int dU\ \chi_\alpha(VU) \chi_\beta(U^{-1}W) & = &  \dfrac{\delta_{\alpha,\beta}}{d_\alpha} \chi_\alpha(VW)\,,\\
    \int dU\ \chi_\alpha(UVU^{-1}W) &  = & \dfrac{1}{d_\alpha}\chi_\alpha(V)\chi_\alpha(W)\,.
\end{eqnarray}

Multiplicities defined by equation~(\ref{eq:multiplicities}) are encoded by the products of the characters. Using the properties of the sum and multiplication of characters, one can obtain that $\forall\, U\in \mathcal{G}$
\begin{equation}\label{eq:product_characters_property}
    \chi_\alpha(U)\chi_\beta(U) = \sum_\sigma N^\sigma_{\alpha\beta} \chi_\sigma(U)\,,
\end{equation}
where $N^\sigma_{\alpha\beta}$ is the  multiplicity of irrep $\sigma$ in the expansion of the tensor product $\alpha\otimes\beta$.

Equation~\eqref{eq:product_characters_property} can be used with the previous formulae to prove that
\begin{equation}\label{eq:multiplicity_fusion_number_integral}
    N^\gamma_{\alpha\beta} \ = \ \int dU\ \chi_\alpha(U)\chi_\beta(U)\chi_\gamma(U^{-1})\,.
\end{equation}
The analog of this formula for matrix elements yields 3j symbols:
\be\label{eq:ortho3j}
\int dU\ \alpha_{am}(U)\beta_{bn}(U)\gamma_{cp}(U^{-1}) \ = \ 
\left(
\begin{array}{ccc}
   \alpha  & \beta & \gamma \\
   a  & b & p
\end{array}
\right)
\left(
\begin{array}{ccc}
   \alpha  & \beta & \gamma \\
   m  & n & c
\end{array}
\right).
\ee
Consequently, contractions of four 3j symbols over the lower row indices produce a 6j symbol:
\be
\label{6jfrom3j}
\left\{
\begin{array}{ccc}
   \alpha  & \gamma_1 & \gamma_2 \\
   \beta  & \gamma_4 & \gamma_3
\end{array}
\right\}
\ = \ \sum\limits_{a,b,c,m,n,p}
\left(
\begin{array}{ccc}
   \gamma_2  & \gamma_1 & \alpha \\
    c & b & a
\end{array}
\right)
\left(
\begin{array}{ccc}
   \gamma_3  & \gamma_4 & \alpha \\
    m & n & a
\end{array}
\right)
\left(
\begin{array}{ccc}
   \gamma_1  & \gamma_3 & \beta \\
    b & m & p
\end{array}
\right)
\left(
\begin{array}{ccc}
   \gamma_4  & \gamma_2 & \beta \\
    n & c  & p
\end{array}
\right)
\ee

For $\mathcal{G}=SU(2)$, contractions of four of the six entries of two 3j symbols leads to the following orthogonality relation:
\begin{equation}\label{eq:3j_orthogonality_1}
    \sum_{a,b} \left(
    \begin{array}{ccc}
       \alpha & \gamma & \beta \\
        a & b & m
    \end{array}\right) \left(
    \begin{array}{ccc}
       \alpha & \gamma & \beta' \\
        a & b & n
    \end{array}\right) = \dfrac{\delta_{\beta,\beta'}\,\delta_{m,n}}{d_\beta} \, \{\alpha\quad \gamma\quad \beta\}\,,
\end{equation}
where $\{\alpha\quad \beta\quad \gamma\}$ are triangular delta symbols, which for $SU(2)$ coincide with $N_{\alpha\beta}^\gamma$. Then the 6j symbols also have an orthogonality relation:
\begin{equation}\label{eq:6j_orthogonality}
    \sum_{\beta} d_{\beta}\, \left\{\begin{array}{ccc}
        \alpha & \gamma_1 & \beta \\
        \lambda & \gamma_2 & \sigma
    \end{array}\right\} \left\{\begin{array}{ccc}
        \alpha & \gamma_1 & \beta \\
        \lambda & \gamma_2 & \sigma'
    \end{array}\right\} = \dfrac{\delta_{\sigma,\sigma'}}{d_\sigma} \, \{\alpha\quad \gamma_2\quad \sigma\} \, \{\lambda\quad \gamma_1\quad \sigma\}\,.
\end{equation}

\end{appendix}

\bibliographystyle{JHEP}
\bibliography{refs}

\end{document}